\documentclass{aa}
\usepackage[varg]{txfonts}
\usepackage{graphicx}
\usepackage{float}
\usepackage{subcaption}
\usepackage{lscape}
\usepackage{longtable}
\usepackage{supertabular}

\usepackage{natbib}
\bibpunct{(}{)}{;}{a}{}{,}

\usepackage{bm}

\usepackage{amsmath}
\usepackage{amsfonts}
\usepackage{amssymb}

\usepackage[colorlinks=true,citecolor=blue]{hyperref}

\begin{document}

\title{A homogeneous spectroscopic analysis of a {\it Kepler\/} legacy sample of dwarfs for gravity-mode asteroseismology}

\author{Sarah Gebruers\inst{\ref{KUL},\ref{MPIA}} 
\and Ilya Straumit\inst{\ref{KUL},\ref{Ohio}} 
\and Andrew Tkachenko\inst{\ref{KUL}} 
\and Joey S. G. Mombarg\inst{\ref{KUL}} 
\and May G. Pedersen\inst{\ref{KUL},\ref{Kavli}} 
\and Timothy Van Reeth\inst{\ref{KUL}} 
\and Gang Li\inst{\ref{SIfA},\ref{Aarhus}} 
\and Patricia Lampens\inst{\ref{ROB}} 
\and Ana Escorza \inst{\ref{KUL},\ref{ESO}} 
\and Dominic M. Bowman\inst{\ref{KUL}} 
\and Peter De Cat\inst{\ref{ROB}} 
\and Lore Vermeylen\inst{\ref{ROB}}  
\and Julia Bodensteiner\inst{\ref{KUL}} 
\and Hans-Walter Rix\inst{\ref{MPIA}} 
\and Conny Aerts\inst{\ref{KUL},\ref{MPIA},\ref{Radboud}}}

\institute{Institute of Astronomy, KU Leuven, Celestijnenlaan 200D, B-3001 Leuven, Belgium \\ \email{sarah.gebruers@kuleuven.be}\label{KUL} 
\and Max Planck Institute for Astronomy, K\"onigstuhl 17, 69117 Heidelberg, Germany \label{MPIA}
\and Department of Astronomy, The Ohio State University, Columbus, OH 43210, USA \label{Ohio}
\and Kavli Institute for Theoretical Physics, University of California, Santa Barbara, CA 93106, USA\label{Kavli}
\and Sydney Institute for Astronomy (SIfA), School of Physics, University of Sydney, NSW 2006, Australia\label{SIfA} 
\and Stellar Astrophysics Centre, Department of Physics and Astronomy, Aarhus University, Ny Munkegade 120, DK-8000 Aarhus C, Denmark\label{Aarhus} 
\and Koninklijke Sterrenwacht van België, Ringlaan 3, B‐1180 Brussel, België\label{ROB} 
\and European Southern Observatory, Alonso de C\'{o}rdova 3107, Vitacura, Casilla 190001, Santiago, Chile\label{ESO}
\and Department of Astrophysics, IMAPP, Radboud University Nijmegen, PO Box 9010, 6500 GL Nijmegen, The Netherlands\label{Radboud}}

\date{Received XXX / Accepted XXX}

\abstract {Asteroseismic modelling of the internal structure of main-sequence stars born with a convective core has so far been based on homogeneous analyses of space photometric {\it Kepler\/} light curves of 4 years duration, to which most often incomplete in-homogeneously deduced spectroscopic information was added to break degeneracies.}
{Our goal is twofold: 1) to compose an optimal sample of gravity-mode pulsators observed by the {\it Kepler\/} space telescope for joint asteroseismic and spectroscopic stellar modelling, and 2) to provide spectroscopic parameters for its members, deduced in a homogeneous way.}
{We assembled HERMES high-resolution optical spectroscopy at the 1.2-m Mercator telescope for 111 dwarfs, whose {\it Kepler\/} light curves allowed for determination of their near-core rotation rates. Our spectroscopic information offers additional observational input to also model the envelope layers of these non-radially pulsating dwarfs.}
{We determined stellar parameters and surface abundances from atmospheric analysis with spectrum normalisation based on a new machine learning tool. Our results suggest a systematic overestimation of [M/H] in the literature for the studied F-type dwarfs, presumably due to normalisation limitations caused by the dense line spectrum of these rotating stars. CNO-surface abundances were found to be uncorrelated with the rotation properties of the F-type stars. For the B-type stars, we find a hint of deep mixing from C and O abundance ratios; N abundances have too large uncertainties to reveal a correlation with the rotation of the stars.}
{Our spectroscopic stellar parameters and abundance determinations allow for future joint spectroscopic, astrometric ({\it Gaia\/}), and asteroseismic modelling of this legacy sample of gravity-mode pulsators, with the aim to improve our understanding of transport processes in the core-hydrogen burning phase of stellar evolution.}

\keywords{asteroseismology -- stars: variables: general -- stars: oscillations -- stars: fundamental parameters -- stars: abundances -- techniques: spectroscopic}

\titlerunning{A homogeneous spectroscopic analysis of a {\it Kepler\/} legacy sample of gravity-mode pulsating dwarfs}
\authorrunning{Gebruers et al.}
\maketitle

\section{Introduction}

Space asteroseismology has made stellar interiors accessible for observational investigation across stellar evolution. Its current status is broadly summarised in various review papers to which we refer for a wealth of information \citep{HekkerJCD2017,GarciaBallot2019,Corsico2019,Bowman2020c,Aerts2021}. As highlighted in these reviews, most focus of asteroseismic modelling was so far put on low-mass stars across the Hertzsprung-Russell diagram (HRD), because these types of stars were predominantly observed by time-series photometric space telescopes. Moreover, the internal structure of these stars resembles that of the Sun, implying that their pulsation mode excitation is the same. It concerns stochastic excitation of pressure (p) modes in the extensive convective envelopes of these stars. Low-mass stars are slow rotators due to magnetic braking such that the Coriolis and centrifugal forces are typically negligible restoring forces for asteroseismology. Hence, the basic diagnostic observables and modelling techniques developed in the context of helioseismology \citep{JCD2002} could easily be transformed for the stellar modelling of large samples of low-mass pulsators. In addition to the space photometry also spectroscopy was used to deliver the effective temperature and metallicity as constraints \citep[e.g.][]{Lebreton2014,Mazumdar2014,SilvaAguirre2017,Bellinger2019a,Bellinger2019b,Verma2019a,Angelou2020}.

The possibilities of gravity-mode (g-mode hereafter) asteroseismology of dwarf stars only became apparent after the first detection of period spacing patterns in B-type pulsators observed with the CoRoT space mission \citep{Auvergne2009} about a decade ago \citep{Degroote2010,Papics2012,Neiner2012}. While the CoRoT mission was a pioneer to show the potential of this branch of asteroseismology, it took until the release of the 4-year {\it Kepler\/} light curves before g-mode asteroseismic modelling saw its first light. Indeed, the frequency precision of the modes scales as the inverse of the total time base of the light curve \citep[see e.g.][]{Montgomery1999} and this has to be of order of years to achieve sufficient g-mode probing power in order to evaluate the theory of stellar interiors \citep[e.g.][]{Aerts2018}. 

The initial studies of stellar modelling based on {\it Kepler\/} g-mode asteroseismology were done for individual targets belonging to the two classes of g-mode pulsators: the late-A to early F-type dwarfs called $\gamma\,$Doradus ($\gamma\,$Dor) stars \citep{Kurtz2014,Saio2015,Murphy2016,Ouazzani2017} and the Slowly Pulsating B (SPB) stars \citep{Moravveji2015,Moravveji2016,Szewczuk2018,WuLi2019-HD50230,Wu2020}. These pulsators all have a convective core and reveal high-radial order low-degree g~modes excited by the flux blocking mechanism for $\gamma\,$Dor stars with a thin convective envelope \citep{Guzik2000,Dupret2005}, and by the heat-engine mechanism acting in the iron-nickel opacity bump for SPB stars with a radiative envelope \citep{Dziembowski1993,Szewczuk2017}.
The $\gamma\,$Dor stars have masses between 1.3 and 1.9\,M$_{\sun}$ and effective temperatures between 6700-7900\,K, while the dwarf SPB stars cover the mass range from 3 to 9\,M$_{\sun}$ and have effective temperatures between 11\,000 and 22\,000\,K \citep[see Chap. 3,][for a general review of their properties]{Aerts2010}. Both these types of g-mode pulsators show non-radial g modes with pulsation periods from 0.5 to about 5\,d and amplitudes below 15\,mmag. Nominal 4-year {\it Kepler\/} light curves of these stars have allowed for the detection of tens of g~modes with amplitudes below 1\,mmag, several of which are members of g-mode period spacing patterns. Such patterns are critical observables as they allow us to achieve mode identification, which is a prerequisite for asteroseismic modelling \citep{Aerts2021}.

Dwarfs with g-mode pulsations cover the entire range of rotational velocities between zero and the critical break-up velocity, due to the absence of a convective dynamo and magnetic braking. Their high-radial order g~modes have low frequencies and eigenfunctions with dominant horizontal vector components \citep[cf.][for a general review]{Aerts2021}. Thanks to these properties, the g~modes are well described by the Traditional Approximation of Rotation (TAR), taking the Coriolis acceleration into account in the equation of motion, while ignoring the vertical component of the rotation vector \citep{LeeSaio1987,Townsend2003,Mathis2009}. In fact, the Coriolis acceleration plays a key role in the modelling of the detected g~modes because almost all of them occur in the gravito-inertial regime \citep{Aerts2017}. This implies that the stellar rotation throughout the star is a key ingredient of g-mode asteroseismology of dwarfs. Assessment of the internal structure of those stars in terms of angular momentum and chemical element transport must thus come from ensemble studies covering the entire range of rotation rates and treating the rotational properties as dominant unknowns in the modelling process \citep{Aerts2019}. This makes g-mode asteroseismology considerably more challenging than the one of the slowly rotating solar-like p-mode pulsators \citep{GarciaBallot2019}. Particularly, g-mode asteroseismology is subject to strong correlations among the properties of the convective core, the convective boundary mixing, and the rotational and pulsational behaviour in the radiative envelope. 
Moreover, uncertainties for the theoretical predictions occur, as the Sun cannot be used as calibrator for the most important physical ingredients of the models. 
While \citet{Aerts2018} developed a method to take these uncertainties and correlations into account in asteroseismic modelling, the addition of spectroscopic information helps to break degeneracies. 

Few ensemble studies with g-mode asteroseismology of $\gamma\,$Dor and SPB stars are available. So far, three such studies have been done, covering 37 $\gamma\,$Dor stars and 26 SPB stars \citep{VanReeth2016,Mombarg2019,Pedersen2021}. 
For none of them surface abundances and surface rotation rates have been included as inputs in the stellar modelling, due to lack of this information or the in-homogeneous treatments to derive those quantities in the literature. \citet{Pedersen2018} showed the potential of adding the surface nitrogen abundance for the assessment of mixing in the radiative envelope of SPB stars. These authors found that a combination of g-mode asteroseismology and high-precision spectroscopic measurements of the atmospheric parameters and surface chemical composition offers a powerful route to calibrate stellar interiors in terms of element transport processes. Moreover, \citet{Mombarg2020} investigated the occurrence of microscopic atomic diffusion, including radiative levitation, from g-mode pulsations in two F-type stars. In this study, the spectroscopic effective temperature, surface gravity, and surface abundances were used as extra constraints aside from the g~modes in the modelling of two slowly rotating $\gamma\,$Dor pulsators. The authors found strong evidence for signatures of radiative levitation for one of the stars, but not in the other one.
 
Clearly, a systematic homogeneous spectroscopic analysis of g-mode pulsating dwarfs selected based on their asteroseismic potential would be highly beneficial for ensemble asteroseismic modelling of these stars. This would allow to maximally exploit the power of the g~modes to probe the deep stellar interior and of the surface abundances to assess the mixing profiles in the radiative envelope.
With the exceptions of \citet{Tkachenko2013}, \citet{Papics2017}, and \citet{Pedersen2021}, spectroscopic studies of g-mode pulsators were done prior to the availability of any asteroseismic modelling results and only used the brightness of the pulsators or the availability of detected frequencies as selection criterion to compose the samples \citep[e.g.][]{Lehmann2011,Tkachenko2012,Niemczura2015,Niemczura2017,Lampens2018}. All these previous studies showed that $\gamma\,$Dor and SPB stars have spectroscopic properties that are not different from those of non-pulsators of the same spectral type.
Here, we took a different approach, by first selecting the most promising g-mode pulsators from the point of view of asteroseismic modelling capacity. We focus on g-mode pulsators whose internal rotation frequency ($\Omega_{\rm{core}}$) was estimated from period spacing patterns thanks to identified g~modes of consecutive radial order deduced from the {\it Kepler\/} light curves \citep{VanReeth2016,Papics2017,Li2020,Pedersen2021}.

For this carefully selected legacy sample of intermediate-mass stars with g-mode pulsations, we add homogeneously analysed high-resolution spectroscopy as complementary information to space photometry to aid future asteroseismic ensemble modelling of {\it Kepler\/} g-mode pulsators.
We do so by using high-resolution spectroscopy from one optimally suited spectrograph \citep[HERMES,][]{Raskin2011} and by analysing it using a machine learning approach. The spectra are analysed with {\it The Payne\/} \citep{Ting2019}, a spectrum interpolator based upon a neural network that is able to predict stellar parameters of a given spectrum, and that was adapted to also include continuum normalisation. It has as advantage that spectra can be processed fast and homogeneously, which is important for both this work and (future) samples from large spectroscopic surveys. 
In addition to spectroscopy, also {\it Gaia\/} distances were considered to provide constraints on the luminosity of the stars. Here we improve estimation of the luminosity from eDR3, while \citet{Pedersen2020} and \citet{Li2020} relied on DR2.
We present our results for a {\it Kepler\/} asteroseismic legacy sample of g-mode pulsators. As discussed in the next section, it consists of 91 $\gamma\,$Dor and 20 SPB {\it Kepler\/} stars. 

In Sect.~\ref{sec:sample} we define the sample and describe the available observational data and their reduction process. Section~\ref{sec:analysis} is dedicated to the atmospheric analysis. It also includes a few tests for a new machine learning technique for stellar parameter determination. The stellar parameters and surface abundances for all stars in the sample are given in Sect.~\ref{sec:results} together with a literature comparison. The results are placed in a physical context in Sect.~\ref{sec:discussion} and we conclude in Sect.~\ref{sec:conclusion}.

\section{Sample selection and spectroscopic observations} \label{sec:sample}

We searched for an optimal sample of g-mode pulsators, defined as stars with period spacing patterns composed of modes with consecutive radial order and identified spherical degree and azimuthal order. Moreover, our selection required the availability of an asteroseismic estimate of the near-core rotation frequency $\Omega_{\rm{core}}$. For g-mode pulsators meeting these two stringent selection criteria, we aim to determine their spectroscopic stellar parameters (effective temperature, surface gravity, metallicity, projected rotational velocity and microturbulent velocity) and individual surface abundances in a homogeneous way for the whole selected sample. In this way, future combined spectroscopic asteroseismic modelling will allow to investigate if the addition of spectroscopic parameters for the whole ensemble improves the current state-of-the-art asteroseismic modelling. We focused on $\gamma\,$Dor and SPB stars with {\it Gaia\/} luminosities and published {\it Kepler\/} period spacing patterns and $\Omega_{\rm{core}}$ values by \citet{VanReeth2016}, \citet{Li2020}, and \citet{Pedersen2020,Pedersen2021}.
All of these g-mode pulsators were detected from 4-year {\it Kepler\/} light curves. The first two studies concern some 650 $\gamma$\,Dor stars, 37 of which having as well asteroseismic masses, ages, and core properties from ensemble modelling by \citet{Mombarg2019}. 
The study of \citet{Pedersen2020} concerns 32 SPB pulsators for which period spacing patterns have been detected. Twenty-six of them were asteroseismically modelled, including estimation of convective boundary mixing as well as envelope mixing. 

Out of these nearly 700 g-mode pulsators with asteroseismic parameters available in the literature, we distilled a g-mode legacy sample. We selected stars for which high-resolution spectroscopic observations were subsequently assembled with intermediate to high signal-to-noise-ratios (S/N $\ge$ 20), allowing for a homogeneous analysis and study of the atmospheric properties of the sample stars. For this reason, we focused on spectroscopic data taken exclusively with the HERMES {\'e}chelle spectrograph \citep{Raskin2011} mounted on the 1.2-m Mercator telescope at Observatorio del Roque de los Muchachos (La Palma, Canary Islands, Spain). HERMES combines two important characteristics for our aim, namely high spectral resolution (R$\sim$ 85\,000) and full coverage of the optical wavelength range from 377 to 900~nm. With the above requirements, we ended up with a total of 127 g-mode pulsators in our sample, including spectroscopic binaries (see Sect.~\ref{sec:RV} for the final sample of 111 stars that only consists of single stars and single-lined binaries). Of these, 104 are $\gamma$\,Dor stars taken from \citet{Li2020} including 40 that overlap with \citet{VanReeth2016}, and 23 SPB stars that were studied by \citet{Pedersen2020}. An overview of the spectroscopic observations and the S/N values of the combined exposures of each target is given in Tables~\ref{tab:gd_log} and \ref{tab:SPB_log} for the $\gamma\,$Dor and SPB stars, respectively.

The spectra were reduced with version 7.0 of the HERMES reduction pipeline which includes bias subtraction, cosmic ray removal, wavelength calibration, barycentric correction, order merging and flat field correction. Every spectrum was also subjected to additional outlier rejection, such that cosmic ray hits are removed. For this step, we compared the local flux at every point in the spectrum to the median flux within a surrounding window of 50 pixels ($\sim$\,1.5\AA) for $\gamma\,$Dor stars or 100 pixels ($\sim$\,3\AA) for SPB stars, where we took a smaller window for the $\gamma\,$Dor stars because these spectra contain many more spectral lines. If the difference between the local and median flux was higher than four times the standard deviation within this window, we replaced the local flux with the median value.

\section{Atmospheric analysis} \label{sec:analysis}

The atmospheric analysis was split into two parts. We used a machine learning framework, {\it The Payne\/} \citep[][see Sect.~\ref{sec:parameters}]{Ting2019}, to determine stellar parameters while the surface abundances were obtained with the spectrum analysis code called Grid Search in Stellar Parameters \citep[GSSP,][]{Tkachenko2015}. Both the stellar parameters and surface abundances are computed using local thermodynamic equilibrium (LTE) models. For the atmospheric analysis and test described in this section we considered the wavelength range between 4200 and 5800 \AA, unless stated otherwise. This includes two Balmer lines and a region of metal lines.

\subsection{Radial velocity signal-based classification} \label{sec:RV}

We started with the classification of our sample of stars according to the radial velocity (RV) information present in their spectra. We computed a discrete 1D cross-correlation function \citep[CCF,][]{Tonry1979} where the observed spectrum, preliminary pseudo-normalised by convolution with a Gaussian kernel, is superimposed on a list of delta functions (also termed line mask) and on a grid of RV values, resulting in a correlation function between the observations and the line mask. A line mask represents a set of (typical) atomic lines at their laboratory wavelength's reference frame and the corresponding line depths predicted from spectral synthesis. In our case, several of those line masks were computed with the GSSP software package \citep[][discussed further in  Sects.~\ref{sec:training_sample} and \ref{sec:abundances}]{Tkachenko2015}. We computed such masks for a range of effective temperatures (6500\,-8000\,K with a step of 500\,K for $\gamma\,$Dor stars and 10\,000\,-20\,000\,K with a step of 2000\,K for SPB stars), surface gravities (3.0\,-5.0\,dex with a step of 0.5\,dex) and metallicities ($-$0.8 to +0.8\,dex with a step of 0.2\,dex). For each star the line mask closest to its atmospheric parameters found in the literature was chosen or, when no literature data were available, we used the line mask for an effective temperature of 7000\,K respectively 16\,000\,K for $\gamma\,$Dor and SPB stars, a surface gravity of 4.0\,dex and solar metallicity as the default values. The corresponding cross-correlation function was obtained for each spectrum of the star separately. The number of spectra for each star are given in the second column of Tables~\ref{tab:gd_log} and \ref{tab:SPB_log}.

\begin{figure}
    \centering
    \resizebox{\hsize}{!}{\includegraphics{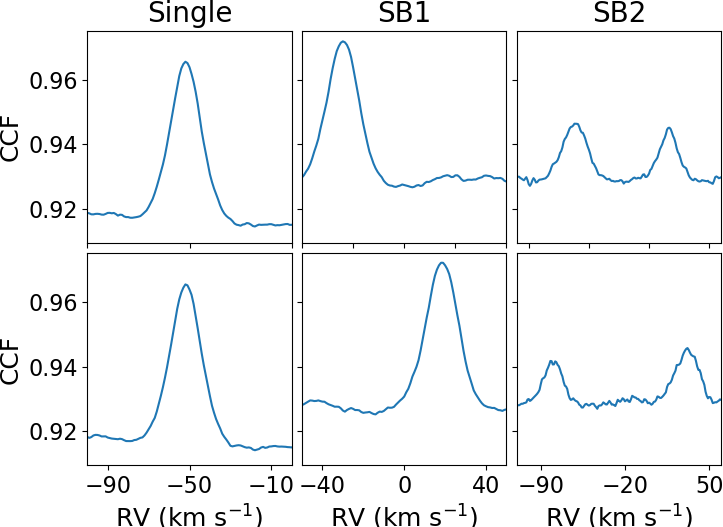}}
    \caption{From left to right: CCFs for a single star (KIC\,9751996), a single-lined binary (KIC\,6292398) and a double-lined binary (KIC\,10080943). Top and bottom rows show CCFs taken at two different observational epochs.}
    \label{fig:RV}
\end{figure}

The CCFs were subject to both visual inspection and inference of RVs, which allowed us to classify stars into three main categories: 1) single stars, 2) spectroscopic single-lined binaries (SB1), and 3) spectroscopic double- (SB2), triple-lined (SB3) or higher-order multiple systems. The last class can be separated from the other two based on visual inspection of the CCFs provided both (or all) components are of similar spectral type as demonstrated in Fig.~\ref{fig:RV}. Distinguishing between single stars and SB1 systems requires establishing a threshold beyond which RV variations are considered as significant, hence pointing to an SB1 binary nature.
We computed RVs by fitting a Gaussian to the upper half of the CCF. The maximum of the best-fitting Gaussian gave a first estimate of the RV. By shifting the Gaussian on a RV grid covering the range $\pm$\,50\,km\,s$^{-1}$ around this RV value and calculating the $\chi^2$ for each perturbed Gaussian, we obtained a $\chi^2$ distribution. For each spectrum we took the value with minimal $\chi^2$ as RV and the 3$\sigma \ \chi^2$-level of the distribution as its uncertainty.
The detected RV variations were considered as significant whenever the RV difference between two or more spectra exceeded the RV uncertainty. The uncertainty depends on the S/N of the spectrum and the projected rotational velocity, since lower S/N spectra or higher velocities broaden the CCF profile, resulting in larger uncertainties. Thus the effect of rotation on the RV accuracy is implicitly taken into account. g-mode pulsations might also distort the wings of line profiles, but our method of RV determination is not sensitive to these changes.

We find that ten of the $\gamma\,$Dor stars are SB2 or possibly higher-order binary systems, 12 stars are SB1s and the remaining are single stars or undetected systems. The RVs are given in the last column of Table~\ref{tab:gd_log}, where binaries are indicated appropriately according to their type. 
Most of the SPB stars are single stars except for two SB2s and one SB1. This is surprising since the binary fraction of OB-type stars in the Milky Way is found to be 0.3-0.5 \citep{Kobulnicky2014}. It is possible that RV variations of some systems are below our detection limit. Many of the SPB spectra have low S/N values which led to high RV uncertainties. There might also be a selection bias since \citet{Pedersen2020} excluded potential SPB stars from their sample if they showed periodicities in their light curves matching binary signatures or rotational modulation.
For two SPB stars there is only one epoch available, so they are currently treated as single stars. The RVs or binarity information of the SPB sample can be found in Table~\ref{tab:SPB_log}. Most of the $\gamma$ Dor and SPB binaries were already identified as such in the literature and corresponding references are included in Tables~\ref{tab:gd_log} and \ref{tab:SPB_log}.

The SB2 binaries, including possible higher-order multiple systems, were removed from the sample because they require disentangling methods or combined analysis of two components. We shifted all the spectra to zero radial velocity and for stars with multiple spectra we added the exposures to obtain a spectrum with a higher S/N value. For four targets, KIC\,5254203, KIC\,7746984, KIC\,9533489, and KIC\,9715425, the S/N is too low (S/N $<$ 20) for any atmospheric analysis and they were also excluded from the sample. This left us with 111 stars in total (91 $\gamma\,$Dor and 20 SPB pulsators).

\subsection{Stellar parameters}\label{sec:parameters}

{\it The Payne\/} is a machine learning method that allows for the simultaneous
determination of various stellar properties, such as atmospheric parameters and surface abundances, from an observed spectrum \citep{Ting2019}. It uses a neural network (NN) trained on a few thousand model spectra to create a spectral interpolator that can predict a spectrum and its stellar parameters in multi dimensional space. It can be used to fit an observed spectrum and derive the corresponding parameters. In this work, we use a modified version of the original {\it The Payne\/} algorithm; a brief summary of the modifications is provided in Sect.~\ref{sec:normalisation}, while a detailed description is given in Straumit et al. (in prep) (cf. short description later).

\subsubsection{The training sample} \label{sec:training_sample}

A training sample can consist of synthetic model \citep[e.g.][]{Kovalev2019} or observed spectra \citep[e.g.][]{Ness2015,Xiang2019}. Both approaches have their advantages and disadvantages. A data-driven training set contains realistic spectra with noise, but it is restricted to the spectra that are available and the accuracy of their determined parameters. On the other hand, the atmospheric structure and radiative transfer of synthetic spectra are consistently computed and these spectra can cover large parameter ranges. Both approaches are being used to derive atmospheric parameters and surface abundances, by relying on atmosphere models with specific input physics. The latter can have a variety of approximations, such as 1D versus 3D geometry, an LTE versus non-LTE approximation, a static versus dynamical atmosphere or a time-dependent versus time-independent treatment of convection.

Since there are not enough B- and F-type stars with accurate stellar parameters deduced from HERMES spectra in a homogeneous way available, we trained NNs with grids of synthetic spectra computed with GSSP. This is an LTE spectrum analysis software package that adopts the SynthV radiative transfer code \citep{Tsymbal1996} for calculation of synthetic spectra based on a grid of plane-parallel atmosphere models pre-computed with the LLmodels code \citep{Shulyak2004}. 
The $\gamma\,$Dor and SPB stars require two separate training grids because they span different temperature regimes. We created grids in five dimensions including the most important stellar parameters: effective temperature ($T_{\rm{eff}}$), surface gravity ($\log g$), metallicity ([M/H]), projected rotational velocity ($v\sin i$), and microturbulent velocity ($\xi$). Since our modified version of {\it The Payne\/} code is also aimed for general use, it is impractical to compute large, computationally expensive grids of synthetic spectra each time the NN needs to be re-trained. Instead, we implement a quasi-random grid, which has the advantage that it covers a whole parameter space with a fairly low amount of points while it still has some structure as opposed to a fully random grid that leads to clustering of points \citep[cf. Fig. 14 in][]{Bellinger2016}. The individual synthetic spectra were obtained by generating 5D quasi-random numbers between zero and one using Sobol numbers \citep{Sobol1967}, and mapping the parameter ranges defined in Table~\ref{tab:training_grid} onto these Sobol numbers. For each of these quasi-random points, a small grid of 2$^5$ (=32) models surrounding the grid point was computed with GSSP and linearly interpolated to get the synthetic spectrum at that point. 

\begin{table}
    \centering
    \caption{Parameter ranges of the training grids.}
    \label{tab:training_grid}
    \begin{tabular}{lcc}
        \hline \hline
        Parameter & $\gamma\,$Dor & SPB \\
        \hline 
        $T_{\rm{eff}}$ (K) & 6000 - 10\,000 & 10\,000 - 25\,000 \\
        $\log{g}$ (dex) & 3 - 5 & 3 - 5 \\
        $\mathrm{[M/H]}$ (dex) & -0.8 - 0.8 & -0.8 - 0.8\\
        $v\sin{i}$ (km s$^{-1}$) & 0 - 200 & 0 - 400 \\
        $\xi$ (km s$^{-1}$) & 0 - 5 & 2 - 20 \\
        \hline
    \end{tabular}
\end{table}

We tested how large the training sets for $\gamma\,$Dor and SPB stars should be so that the NN predicts spectra with sufficient precision. In previous studies, {\it The Payne\/} was mostly used for the analysis of low-resolution (LAMOST) spectra \citep[e.g.][]{Xiang2019}, where the relatively small number of some 1000 training spectra in $>$10D space proved sufficient because of the limited resolving power in metal line regions. For high-resolution HERMES spectra, however, the NN must be able to handle slowly rotating stars and resolve their many narrow spectral lines. The training sample should therefore be dense enough to be able to differentiate between spectra with quasi-similar parameters.
We trained multiple NNs, each with a training sample of different size. For the $\gamma\,$Dor stars, the NNs were trained for grids with 100, 200, 300, 500, 1000, 2000 and 3000 spectra and for SPB stars with 500, 1000, 2500, 4000, 5000 and 7000 spectra. We let each of these NNs predict spectra for certain stellar parameter values and compared them to synthetic spectra computed with GSSP for the same parameters. Comparison was done by means of a merit function based on the sum of the squared differences between the NN-predicted and GSSP-computed spectra, namely $\sum{\mathrm{(flux_{NN} - flux_{GSSP})}^2}$. We accepted as the optimal size of the training set the point where the above merit function reached a plateau, that is 1000 and 5000 models for $\gamma$\,Dor and SPB stars, respectively (see also Figs.~\ref{fig:size_sample_gDor} and \ref{fig:size_sample_SPB}). Smaller training samples result in less accurate NN predictions while larger samples require more computation time and only slightly improve the results.
The training sample is larger for the SPB stars than for the $\gamma\,$Dor stars because the former cover a much wider temperature range.

\subsubsection{Normalisation} \label{sec:normalisation}
Both grids were used to train a NN as described in \citet{Ting2019}. The result is a 5D spectrum interpolator that we used to fit the observed HERMES spectra. The optimal stellar parameters of an observed spectrum can be found by least squares fitting, which determines the best matching interpolated spectrum with its corresponding stellar parameters. This method can also incorporate the normalisation of the spectrum by adding extra free parameters to the fitting routine. This adaptation of {\it The Payne\/} that includes normalisation will be discussed in Straumit et al.\ (in prep.) and we give a brief summary here.

We assumed that the residual HERMES response function of a spectrum can be represented by a Chebyshev polynomial. We characterised Chebyshev polynomials with coefficients that are treated as free parameters alongside the atmospheric parameters. In each iteration of the fitting routine, the best fitting coefficients are used to construct a polynomial which is introduced into the synthetic spectrum that is computed for the best fitting atmospheric parameters of that same iteration. This synthetic spectrum, containing now a proxy for the residual response function is compared to the observed HERMES spectrum. The iterations continue until convergence is reached.
This way the normalisation process is not subject to human intervention and is therefore more objective than any kind of manual spectrum normalisation. The latter typically assumes a subjective selection of continuum points and fitting a polynomial or a spline function through them to define the pseudo-continuum of the star. The resulting normalisation, and consequently the stellar parameters, are therefore highly dependent on the choice of the points. Even small offsets in the selected points can change the depth and shape of spectral lines, with the effect being typically more important for broad hydrogen and helium lines. These are in turn important $T_{\rm{eff}}$ and $\log g$ diagnostic lines, hence (manual) spectrum normalisation can be a major source of uncertainty for the inferred atmospheric parameters and surface composition of the stars. Instead, our approach of combining stellar parameters and Chebyshev polynomial coefficients into a single vector of unknowns allows for a self-consistent inference of atmospheric properties and pseudo-continuum of the star. Details of our pseudo-normalisation procedure are presented in Straumit et al.\ (in prep.).

\subsubsection{Performance of the neural network} \label{sec:performance}

Before applying the machine learning procedure to observed data we verified its performance on artificial data and benchmark stars.  
We computed ten synthetic spectra, four with stellar parameters within the ranges of SPB stars and six with typical values of $\gamma\,$Dor stars. Noise was added to simulate a S/N $\sim$ 100 and a response function similar to that of a HERMES spectrum was introduced. For each spectrum we determined the stellar parameters in three different ways: 1) using the NN approach as described in the previous (sub-)sections, that is simultaneous optimisation of the stellar parameters and pseudo-continuum of the star; 2) making use of the inferred pseudo-continuum of the star, applying it to the original spectrum to obtain a normalised spectrum which is then analysed with GSSP; 3) manually normalising the spectrum by selecting continuum points, fitting a spline through them and analysing the normalised spectrum with GSSP. The parameters obtained with these three methods were compared to the real values and are given in Table~\ref{tab:testNN}. 
This test showed that the values for $T_{\rm{eff}}$, $v\sin\,i$, $\xi$ and [M/H] determined with methods 1) and 2) agree with the real values to within the uncertainties. The $\log\,g$ values for F-type stars deviate by 0.05 to 0.1\,dex for most of the spectra. It is well-known that hydrogen lines in the spectra of F-type stars are insensitive to (small to moderate) $\log\,g$ variations \citep[][chap. 13-14]{Gray2005}. This means that metal lines have to be used for the inference of $\log\,g$ of the star, hence degeneracy with [M/H] can be expected. To test the hypothesis that the deviations in the inferred $\log\,g$ values are due to a [M/H]-$\log\,g$ correlation, we fixed $\log\,g$ and $\xi$ to their real values and repeated the analysis with method 2). The resulting [M/H] values are found to be within the uncertainties of the previous ones, which allows us to conclude that the [M/H] determination is done correctly by the NN, but small deviations in the $\log\,g$ parameter is something to bear in mind. 
Most of the [M/H] and some of the other parameter results from method 3) are less consistent with the real values and we attribute this to difficulties in the placement of the pseudo-continuum (which is done manually in this case). Manual normalisation assumes that the apparent continuum is at unity. This is not necessarily the case,  especially for F-type stars with their dense metal line spectra, and even more so for moderate and fast rotators ($v\sin\,i \gtrsim$ 40 km\,s$^{-1}$). Thus manual normalisation inevitably biases the parameter determination \citep[e.g.][]{Blanco-Cuaresma2015,Giribaldi2019}. For B-type stars this effect is smaller because there are fewer metal lines present in the spectrum, making continuum normalisation easier.  

In addition to the tests with synthetic spectra, we also compared the parameters predicted by the NN with literature values for a couple of F- and B-type standard stars that have been observed with HERMES and for which stellar parameter values are available in the literature. The results are shown in Table~\ref{tab:benchmark}. These are real observations with real noise and artefacts from the Earth's atmosphere and the instrument, which makes it more difficult to derive precise parameters. In addition, the literature values were obtained with data from different instruments and various analysis techniques. When taking this into account, the predictions from {\it The Payne\/} are comparable to literature values. Especially $T_{\rm{eff}}$ and $\log\,g$ are well determined by {\it The Payne\/}. Deviations in $v\sin{i}$ and $\xi$ are the result of correlations between these two parameters and the absence of other broadening mechanisms in {\it The Payne\/}, such as macroturbulence. 
The NN predictions for [M/H] are generally lower than those in the literature and this couples back to the pseudo-continuum placement.

The tests with artificial data and benchmark stars prove that a good performance is achieved with our NN-based approach. Concretely, we note that: 1) our approach delivers reliable atmospheric parameters which can further be adopted for the inference of elemental abundances; 2) deviations in $\log\,g$ of F-type stars of 0.05 to 0.1~dex is something to keep in mind, yet they are smaller than the typical $\log\,g$ uncertainty of $\gtrsim$ 0.15~dex; 3) we expect to observe differences with literature values due to our superior approach of pseudo-normalisation, in particular for the [M/H] parameter.

\begin{table*}[ht]
    \centering
    \caption{Stellar parameters from literature and the values obtained with {\it The Payne\/} for F-type benchmark stars $\pi^3$\,Orionis and $\iota$\,Piscium and B-type benchmark stars $\eta$\,Ursae Majoris and HD\,21071.}
    \label{tab:benchmark}
    \renewcommand{\arraystretch}{1.2}
    \begin{tabular}{cccccc}
        \hline\hline
        Reference & $T_{\rm{eff}}$ & $\log{g}$ & [M/H] & $v\sin{i}$ & $\xi$ \\
        & (K) & (dex) & (dex) & (km s$^{-1}$) & (km s$^{-1}$) \\
        \hline
        \multicolumn{6}{c}{F-type: $\pi^3$ Orionis (S/N$_{4970} \sim$ 360)} \\
        \hline
        1 & 6420$^{+26}_{-41}$ & 4.24$^{+0.04}_{-0.07}$ & -0.02$^{+0.02}_{-0.01}$ & $-$ & $-$ \\
        2 & 6509 $\pm$ 81 & 4.38 & +0.17 $\pm$ 0.13 & 18.5 & 2.22 \\
        3 & 6448 $\pm$ 50 & 4.29 $\pm$ 0.02 & 0.00 $\pm$ 0.04 & $-$ & $-$ \\
        {\it The Payne\/} & 6536 $\pm$ 27 & 4.33 $\pm$ 0.06 & -0.03 $\pm$ 0.02 & 19.2 $\pm$ 1.6 & 1.96 $\pm$ 0.09 \\
        \hline
        \multicolumn{6}{c}{F-type: $\iota$ Piscium (S/N$_{4970} \sim$ 490)} \\
        \hline
        1 & 6206$^{+26}_{-60}$ & 4.11$^{+0.06}_{-0.07}$ & -0.18$^{+0.02}_{-0.03}$ & $-$ & $-$ \\
        2 & 6177 $\pm$ 79 & 4.08 & -0.11 $\pm$ 0.07 & 8.1 & 1.60 \\
        3 & 6192$\pm$ 50 & 4.12 $\pm$ 0.02 & -0.16 $\pm$ 0.04 & $-$ & $-$ \\
        {\it The Payne\/} & 6109 $\pm$ 35 & 3.95 $\pm$ 0.06 & -0.18 $\pm$ 0.02 & 3.71 $\pm$ 0.63 & 0.98 $\pm$ 0.12 \\
        \hline
        \multicolumn{6}{c}{B-type: $\eta$ Ursae Majoris (S/N$_{4980} \sim$ 2200)} \\
        \hline
        4 & 16494 & 4.17 & 0.0 & $-$ & 2.0 \\
        5 & 17783 & $-$ & $-$ & 161 & $-$ \\
        {\it The Payne\/} & 16629 $\pm$ 71 & 4.20 $\pm$ 0.02 & -0.14 $\pm$ 0.04 & 131.9 $\pm$ 4.6 & 2.64 $\pm$ 0.37 \\
        \hline
        \multicolumn{6}{c}{B-type: HD\,21071 (S/N$_{4980} \sim$ 210)} \\
        \hline
        6 & 14355 $\pm$ 99 & $-$ & $-$ & 54 & $-$ \\
        7 & 14768 & 4.30 & -0.2 & 58 & $-$\\
        5 & 13804 & $-$ & $-$ & 67 & $-$ \\
        {\it The Payne\/} & 14043 $\pm$ 79 & 4.09 $\pm$ 0.02 & -0.41 $\pm$ 0.05 & 51.0 $\pm$ 4.0 & 4.45 $\pm$ 0.53 \\
        \hline
    \end{tabular}
    \tablebib{
    (1)~\citet{Boeche2016}; (2) \citet{Luck2017}; (3) \citet{Aguilera2018}; (4) \citet{Gray2003};
    (5) \citet{Simon-Diaz2017}; (6) \citet{Zorec2012}; (7) \citet{Saffe2014}.
    }
\end{table*}

\subsection{Surface abundances} \label{sec:abundances}
\begin{figure*}[ht]
    \centering
    \begin{subfigure}{0.24\textwidth}
        \includegraphics[width=\textwidth]{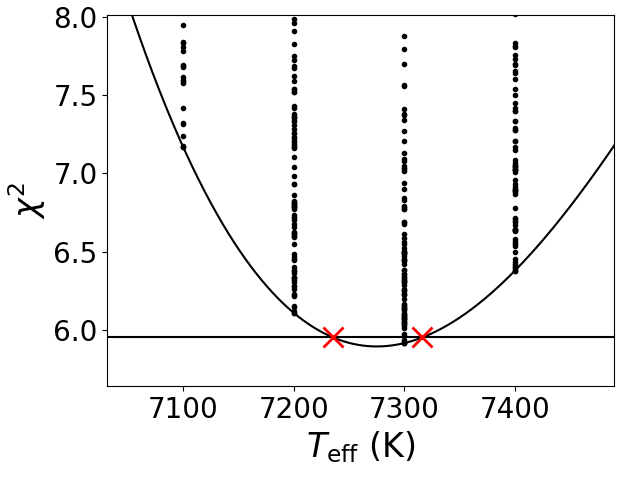}
    \end{subfigure}
    \begin{subfigure}{0.24\textwidth}
        \includegraphics[width=\textwidth]{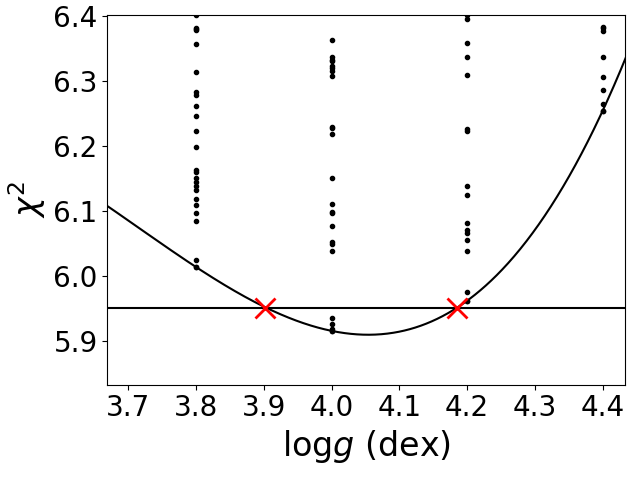}
    \end{subfigure}
    \begin{subfigure}{0.24\textwidth}
        \includegraphics[width=\textwidth]{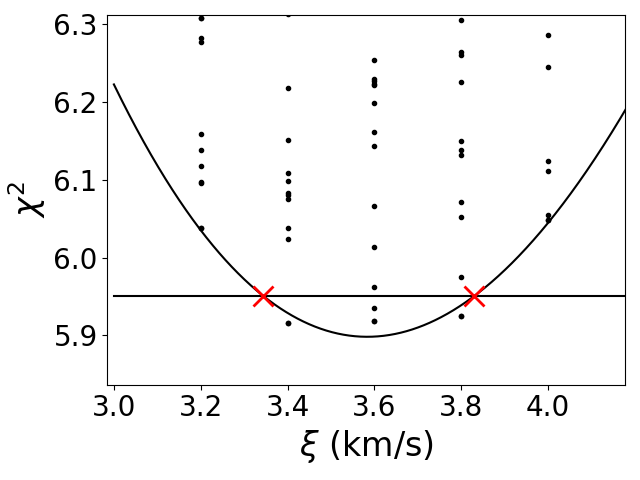}
    \end{subfigure}
    \begin{subfigure}{0.24\textwidth}
        \includegraphics[width=\textwidth]{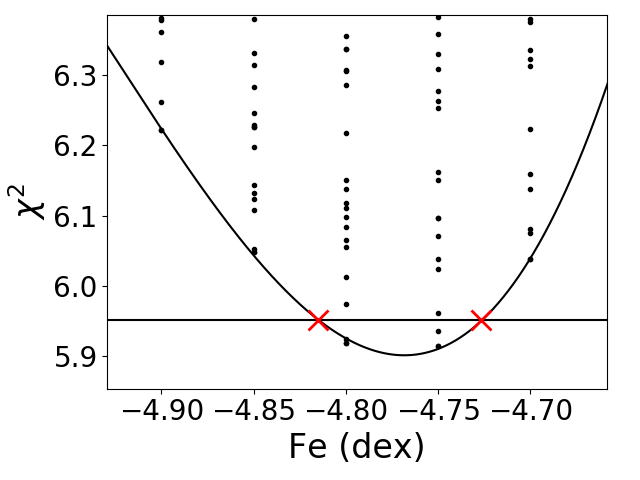}
    \end{subfigure}
    \caption{Projection of the $\chi^2$ hyper-surface with respect to $T_{\rm{eff}}$, $\log\,g$, $\xi$ and the Fe abundance for $\gamma\,$Dor star KIC\,7365537. The black dots are $\chi^2$ values for all the grid nodes, the black curve is a polynomial fit with as minimum the best-fit value and the red crosses are the intersection points of this fit with the 1$\sigma$ $\chi^2$ level (the horizontal line) and give the errors on the best-fit value.}
    \label{fig:parameterfit}
\end{figure*}

Currently, our version of {\it The Payne\/} that includes normalisation only delivers atmospheric parameters. We did not add surface abundances because it requires complex NNs that can resolve all high-resolution spectral features and we would need larger training samples which are very computationally demanding.
The surface abundances for all observed spectra were instead derived with GSSP \citep{Tkachenko2015}. This code calculates a grid of synthetic spectra and compares each one of them to the observed spectrum. The $\chi^2$ merit function is used as a measure of the goodness of fit. For every parameter or element abundance of the observed spectrum, the optimal value and its uncertainty are determined by projecting $\chi^2$ values of the entire grid on to the parameter or abundance in question. This allows us to take covariances between parameters/abundances into account when computing uncertainties. A polynomial of third or fourth order is fit to the lowest $\chi^2$ values to search for the global minimum. The errors are given by the intersection of this fit with the 1$\sigma$ limit in terms of $\chi^2$. This is illustrated in Fig.~\ref{fig:parameterfit} for KIC\,7365537.  

We use the GSSP software package to determine individual elemental abundances for our stars. For every star in the sample we normalised the observed spectrum by dividing it through the (residual) response function approximated with the Chebyshev coefficients obtained with {\it The Payne\/} algorithm in the previous step. We fixed [M/H] and $v\sin\,i$ to {\it The Payne\/} values and derived abundances for C, O, Na, Mg, Al, Si, S, Ca, Sc, Ti, V, Cr, Mn, Fe, Co, Ni, Cu, Zn, Sr, Y, Zr, Ba ($\gamma\,$Dor) and He, C, N, O, Ne, Mg, Si, S, Ca, Fe (SPB). Abundances of elements with many strong lines in the spectrum, such as Fe, Ni, Ti and Cr for $\gamma$ Dor stars and He for SPB stars, have to be optimised together with $T_{\rm{eff}}$, $\log\,g$ and $\xi$ atmospheric parameters \citep[][chap. 16]{Gray2005}. To take this into account, we allow for these parameters to vary in a small interval around the values found with {\it The Payne\/} algorithm. Furthermore, since elemental abundances are optimised one-by-one in GSSP, it is essential to perform abundance analysis in an iterative way to account for the fact that abundance of the element in question can be influenced by the presence of lines of other elements due to spectral line blending. This is relevant for chemical elements that show numerous lines in the spectrum and we found that typically 2-3 iterations are sufficient to achieve the convergence, that is when the abundance value does not deviate more than the typical uncertainty between two consecutive iterations. For chemical elements that display fewer lines in the spectrum, abundances were inferred using small wavelength ranges around positions of the corresponding spectral lines.

The abundance values returned by GSSP are given as $\log(n_{\rm{X}}/n_{\rm{tot}})$, with $n_{\rm{X}}$ the number density of element X and $n_{\rm{tot}}$ the total number density of all elements. However, abundances in the literature are mostly given relative to solar ones, that is [X/H]. The results from GSSP are converted to [X/H] as follows:

\begin{align}
    \mathrm{[X/H]} &= \log\left(\frac{n_{\rm{X}}}{n_{\rm{H}}}\right) - \log\left(\frac{n_{\rm{X}}}{n_{\rm{H}}}\right)_{\sun} \\ 
    &= \log\left(\frac{n_{\rm{X}}}{n_{\rm{tot}}}\right) - \log\left(\frac{n_{\rm{H}}}{n_{\rm{ tot}}}\right) - \log\left(\frac{n_{\rm{X}}}{n_{\rm{H}}}\right)_{\sun} \\ 
    &= \log\left(\frac{n_{\rm{X}}}{n_{\rm{tot}}}\right) - \log\left(\frac{n_{\rm{H}}}{n_{\rm{tot}}}\right) - \log\epsilon_{\rm{X,\sun}} + 12. \label{eq:conversion}
\end{align}

\noindent The first term in Eq.~(\ref{eq:conversion}) is the GSSP output, for the second term we use the same value for all stars \citep[$\log(n_{\rm{H}}/n_{\rm{tot}}) = -0.036$,][]{Asplund2005} since the sample has a relatively small [M/H] range, and the third term ($\log\epsilon_{X,\sun}$) is the solar value for element X (in 12-scale) from \citet{Asplund2005}. The [X/H] notation is the one used in the abundance tables of this paper. The values are converted to 12-scale $\left(\log\epsilon_X = \log{\left(\frac{n_{\rm{X}}}{n_{\rm{H}}}\right)} + 12\right)$ by adding the value for $\log\epsilon_{\rm{X,\sun}}$.

\begin{figure*}[ht]
    \centering
    \resizebox{\hsize}{!}{\includegraphics{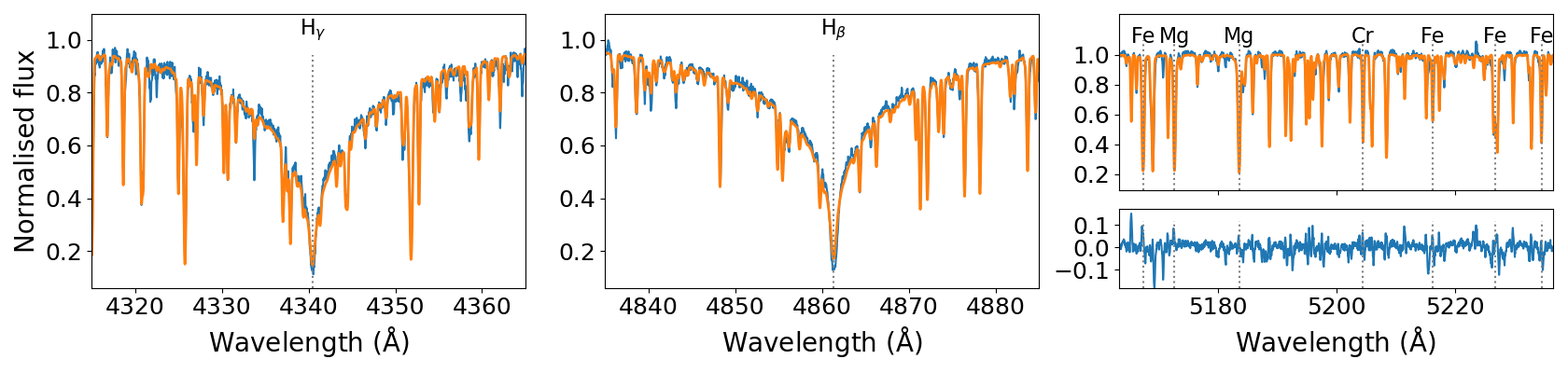}}
    \resizebox{\hsize}{!}{\includegraphics{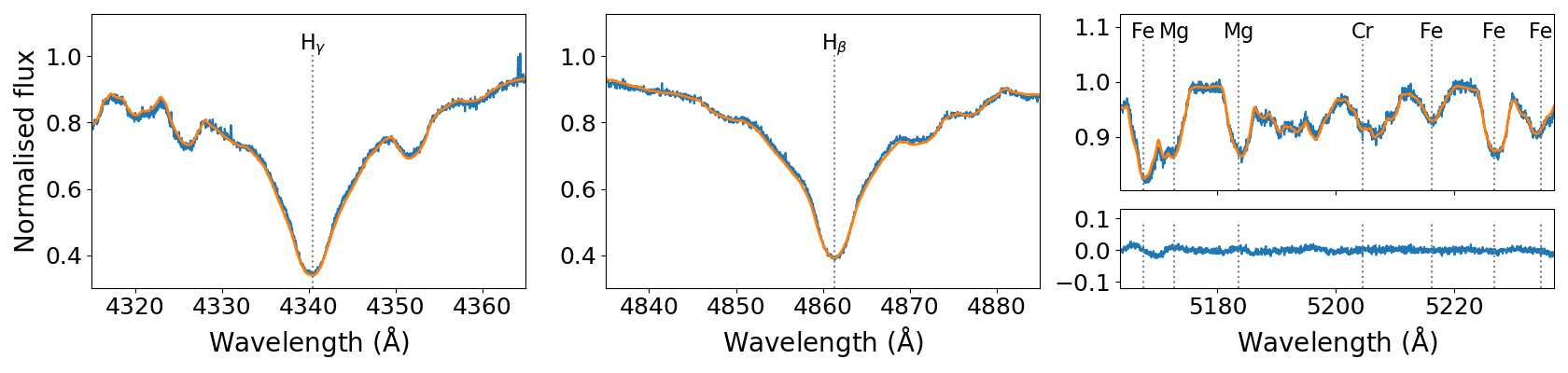}}
    \caption{{\it Top:} spectrum of $\gamma\,$Dor star KIC\,9751996 (S/N = 50) with $T_{\rm{eff}}=7139$\,K, $\log g=3.66$\,dex, $v\sin i=12.1$\,km\,s$^{-1}$, $\xi=3.38$\,km\,s$^{-1}$ and [M/H] = +0.12\,dex in blue and best fitting synthetic spectrum in orange. {\it Bottom:} the same for KIC\,7365537 (S/N = 170) with $T_{\rm{eff}}=7274$\,K, $\log g=4.02$\,dex, $v\sin i=148.3$\,km\,s$^{-1}$, $\xi=3.59$\,km\,s$^{-1}$ and [M/H] = $-$0.23\,dex. The bottom parts of the right panels show the residuals.}
    \label{fig:gd_spectrum}
\end{figure*}

\begin{figure*}[ht]
    \centering
    \resizebox{\hsize}{!}{\includegraphics{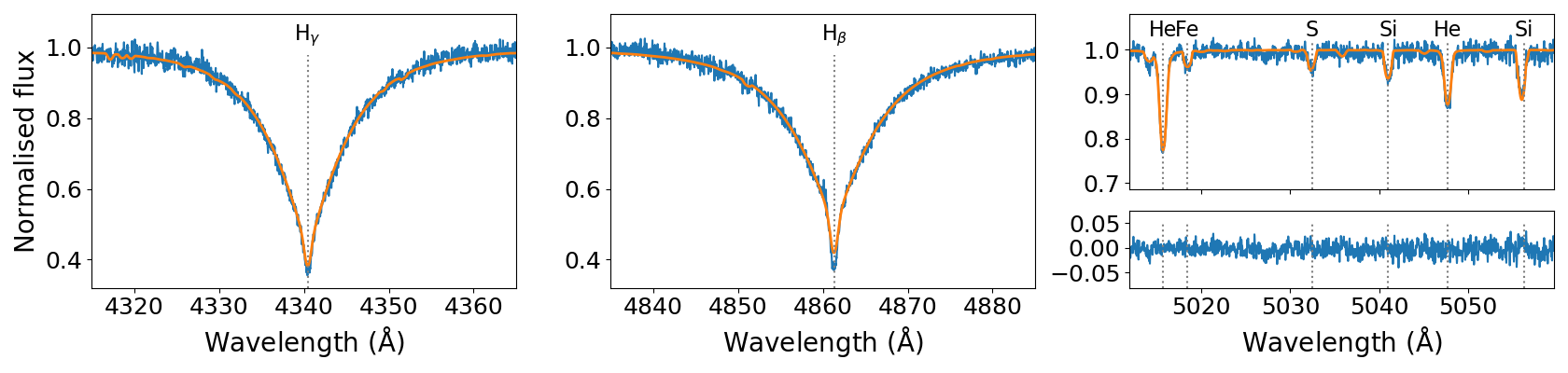}}
    \caption{Spectrum of SPB star KIC\,3756031 (S/N = 90) with $T_{\rm{eff}}=15947$\,K, $\log g=3.69$\,dex, $v\sin i=31.7$\,km\,s$^{-1}$, $\xi < 3.9$\,km\,s$^{-1}$ and [M/H] = $-$0.44\,dex in blue and best fitting synthetic spectrum in orange. The bottom part of the right panel shows the residuals.}
    \label{fig:spb_spectrum}
\end{figure*}

Figure~\ref{fig:gd_spectrum} shows part of the relatively low S/N spectrum (S/N = 60) of a slowly rotating $\gamma\,$Dor star in the top row and the high S/N spectrum (S/N = 170) of a fast rotating $\gamma\,$Dor star in the bottom row. In Fig.~\ref{fig:spb_spectrum} the spectrum of an SPB star with typical S/N of the SPB sample is plotted. All the spectra are over-plotted by their best fitting synthetic spectrum and zoomed in on the Balmer lines and a region with metal lines to demonstrate the quality of the fits we typically obtained for the whole sample. For every star in the sample we made similar diagnostic plots to visually check the quality of the fit.

\section{Results} \label{sec:results}
The stellar parameters and surface abundances we obtained for the $\gamma\,$Dor stars are given in Tables~\ref{tab:gd_params}, \ref{tab:gd_abund}, and \ref{tab:gd_abund2}. The results of the SPB stars can be found in Tables~\ref{tab:spb_params} and \ref{tab:spb_abund}. For all five parameters we report the values and uncertainties from the fitting routine with the NN in Cols.~2-6. Additionally in Cols.~7-9 we give the values for $T_{\rm{eff}}$, $\log\,g$ and $\xi$ determined with GSSP after optimising the abundances of elements with many spectral lines. The abundances in the tables are given in the [X/H] format and the uncertainties on the solar values of \citet{Asplund2005} are taken into account.

Typical uncertainties (although this depends on S/N) for the $\gamma\,$Dor sample are $\Delta T_{\rm{eff}} \sim 65\,K$, $\Delta \log\,g \sim 0.2$\,dex, $\Delta \rm{[M/H]} \sim 0.05$\,dex, $\Delta v\sin\,i \sim ~4 \,\rm{km\,s^{-1}}$ and $\Delta \xi \sim 0.3\,\rm{km\,s^{-1}}$. For the SPB stars the uncertainties are larger with typical values of $\Delta T_{\rm{eff}} \sim 700\,K$, $\Delta \log\,g \sim 0.15$\,dex, $\Delta  \rm{[M/H]} \sim 0.15$\,dex, $\Delta v\sin\,i \sim 13\,\rm{km\,s^{-1}}$ and $\Delta \xi \sim 3\,\rm{km\,s^{-1}}$. This is expected since most of the SPB spectra have lower S/N values and B-type stars have much fewer spectral lines (see Fig.~\ref{fig:spb_spectrum}), which complicates the parameter determination. Non-LTE effects on stellar parameters also become increasingly important at higher temperatures \citep{Przybilla2011} but these are not taken into account in the current analysis because we have specifically chosen to perform a homogeneous analysis for all the stars in this legacy sample, treating both F- and B-type stars in the same way.

\subsection{$\gamma\,$Dor stars} \label{sec:results_gDor}
\begin{figure*}[ht]
    \centering
    \includegraphics[width=0.3\textwidth]{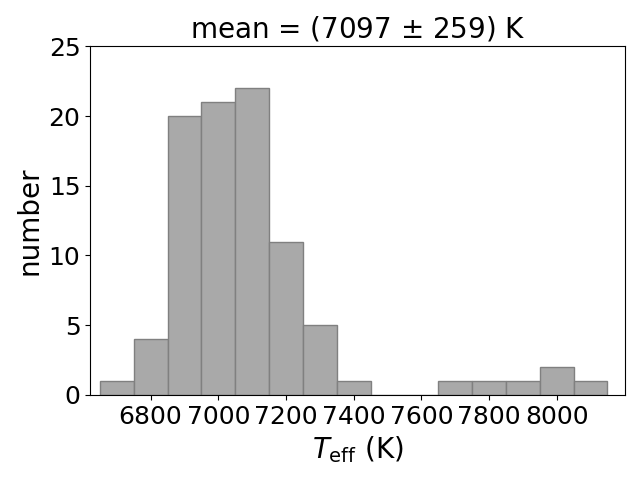}
    \includegraphics[width=0.3\textwidth]{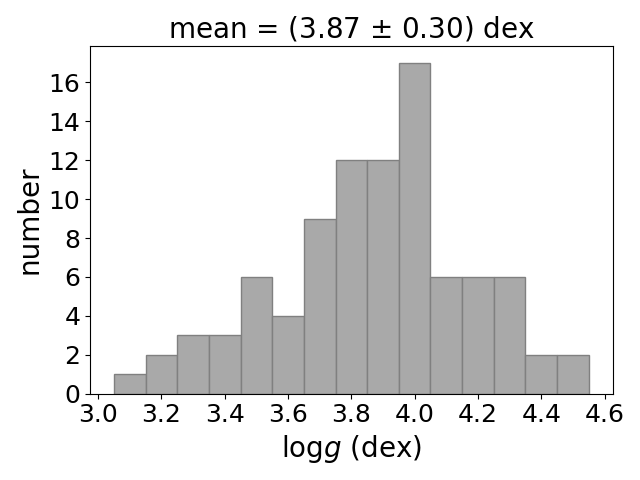}
    \includegraphics[width=0.3\textwidth]{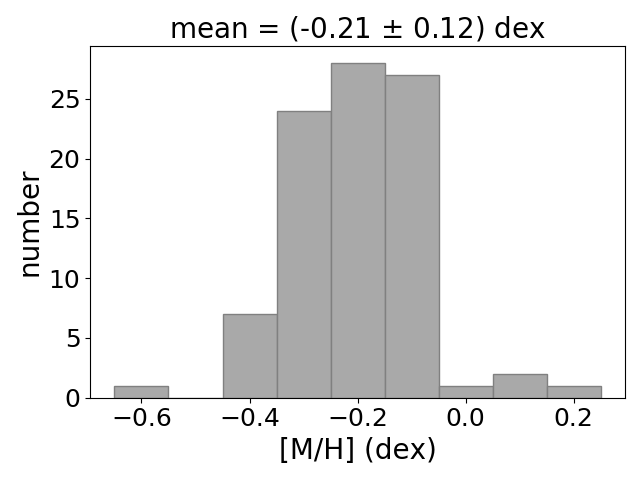}\\
    \includegraphics[width=0.3\textwidth]{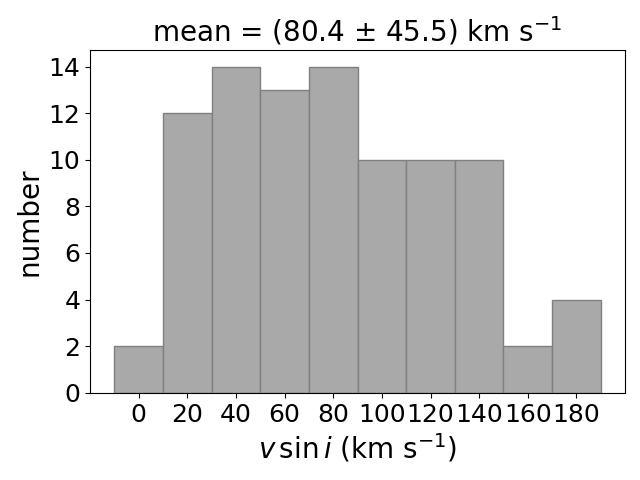}
    \includegraphics[width=0.3\textwidth]{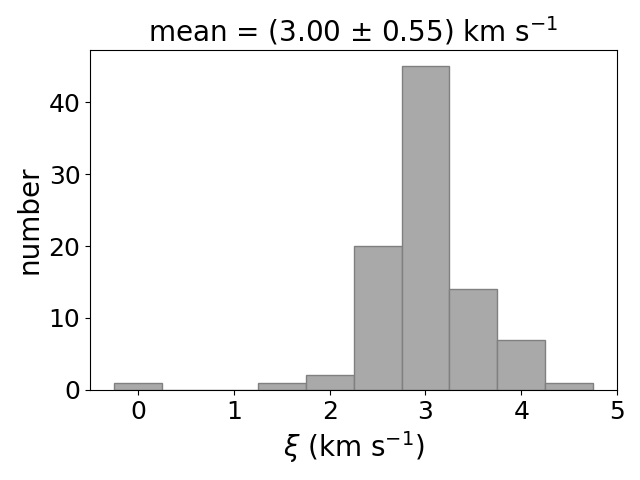}\\
    \includegraphics[width=0.3\textwidth]{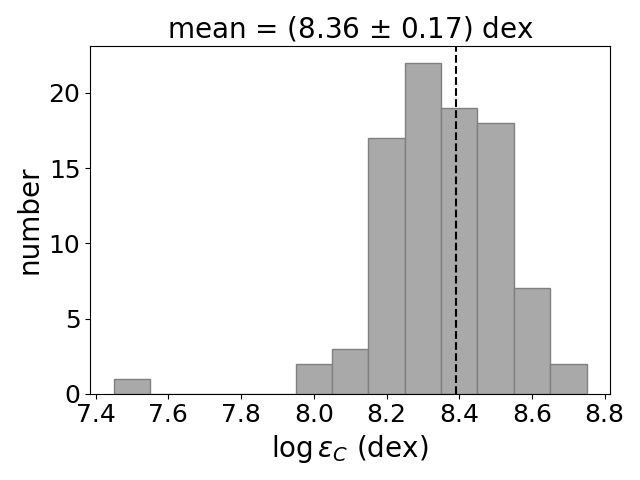}
    \includegraphics[width=0.3\textwidth]{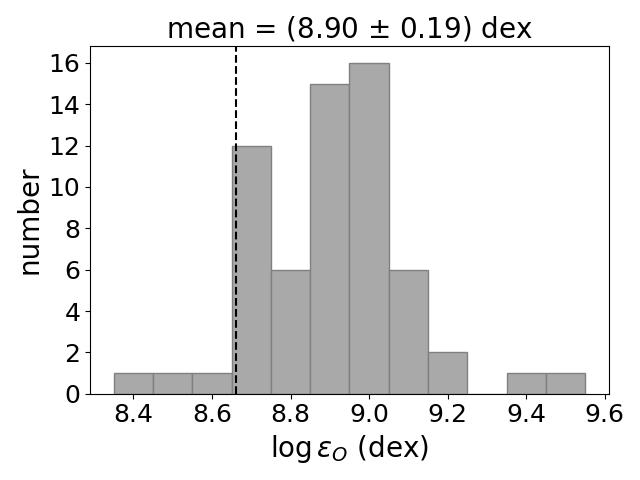}
    \caption{Histograms of $T_{\rm{eff}}$, $\log{g}$, [M/H], $v\sin{i}$, $\xi$, and the abundances of C and O in 12-scale for the sample of $\gamma\,$Dor stars. The [M/H] and $v\sin{i}$ values are the results from the NN, while $T_{\rm{eff}}$, $\log{g}$, $\xi$ and the C and O abundances are those obtained with GSSP.} Solar values of the abundances are indicated with a black dashed line.
    \label{fig:gDor_hist}
\end{figure*}

Distributions of the final stellar parameters and some element abundances for the $\gamma\,$Dor stars are shown in Fig.~\ref{fig:gDor_hist}. On top of each panel, the mean value and standard deviation of that parameter is given. Most of the $\gamma\,$Dor stars have a temperature within the expected range which is 6700-7900\,K \citep{Aerts2010}, but there are a few hotter stars with a temperature around 8000\,K. This is further discussed in Sect.~\ref{sec:discussionHRD}. There is a wide spread in $\log\,g$ and some targets have values close to 3 dex, implying that they are close to or even beyond the end of the main sequence. They cover a $v\sin i$ range from 0 to 200\,km\,s$^{-1}$ and $\xi$ is distributed around 3\,km\,s$^{-1}$. There is one outlier, KIC\,7215607, for which only an upper bound on $\xi$ could be determined so we fixed its value to 2\,km\,s$^{-1}$ in the further analysis. We note that the [M/H] and $v\sin i$ distributions shown in Fig.~\ref{fig:gDor_hist} are the results obtained with the NN, while the distributions of $T_{\rm{eff}}$, $\log{g}$, $\xi$ and the C and O abundances are those from the final GSSP analysis. This is due to the abundances being optimised along with $T_{\rm{eff}}$, $\log{g}$, and $\xi$ as explained in Sect.~\ref{sec:abundances}. 

\begin{figure*}
    \centering
    \includegraphics[width=17cm]{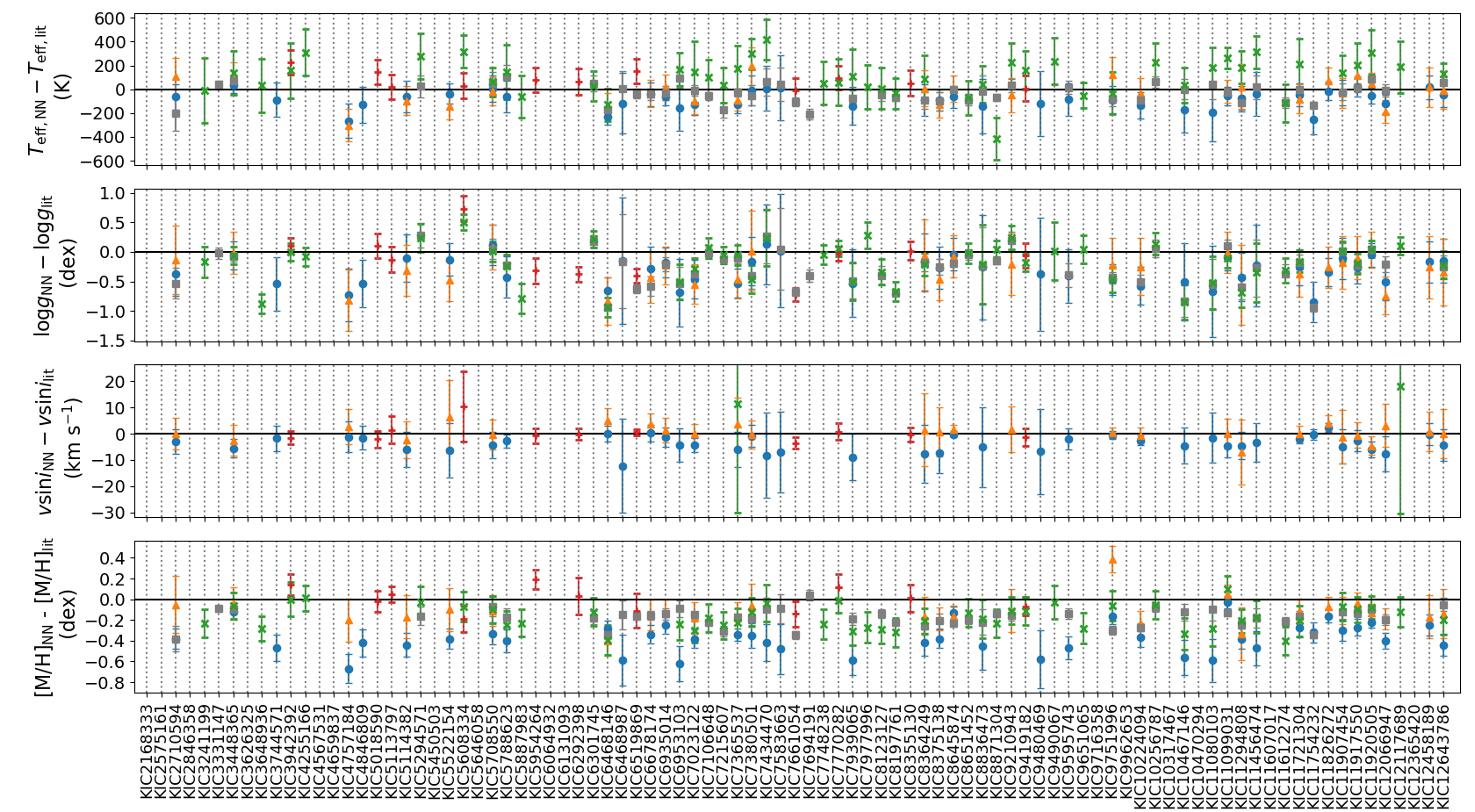}
    \caption{Comparison of $\gamma\,$Dor stellar parameters from the NN with literature values. From top to bottom the difference between the values in this paper and the ones from literature catalogues are shown for $T_{\rm{eff}}$, $\log{g}$, $v\sin{i}$ and [M/H]. Both the errors obtained from the NN and those from the literature are propagated to get the plotted error bars. The different symbols correspond to \citet{Tkachenko2013} (orange triangles), \citet{VanReeth2015} (blue circles), \citet{Niemczura2015} (red pluses), \citet{Frasca2016} (green crosses) and \citet{Qian2019} (grey squares).}
    \label{fig:gDor_lit}
\end{figure*}

The [M/H] distribution peaks around a value of $-0.21\pm0.12$\,dex with some outliers at both ends. This value is slightly sub-solar and is lower than what is found in other literature studies for this sample of $\gamma\,$Dor stars, as shown in Fig.~\ref{fig:gDor_lit}. In this figure we compare our NN values of $T_{\rm{eff}}$, $\log\,g$, $v\sin\,i$ and [M/H] with values from catalogue studies that determined spectroscopic stellar parameters for most of the stars in our sample. We use the NN values and not the final GSSP results in this comparison since they have been determined before the optimisation of individual abundances, as was the case in the literature studies. The error bars are the propagated uncertainties derived from the NN application and those reported in the literature.
\citet{Frasca2016} and \citet{Qian2019} used LAMOST spectra, which have a resolution of only\,$\sim$\,1800 \citep{Zhao2012}. The studies by \citet{Niemczura2015}, \citet{Tkachenko2013}, and \citet{VanReeth2015} analysed HERMES data and the latter two also used the GSSP code. For some stars our $T_{\rm{eff}}$ and $\log\,g$ deviate from these literature values and our $v\sin\,i$ is consistent when taking the uncertainties into account. But the [M/H] we find is systematically $\sim$0.2\,dex lower. We also compared our [M/H] distribution to that of a different sample of $\gamma\,$Dor stars from \citet{Bruntt2008}. The latter consists of 18 stars with a mean [M/H] of $-0.03\pm0.23\,$dex. The two distributions largely overlap within the uncertainties, but the majority of stars analysed here have [M/H] values at the lower end of the distribution in \citet{Bruntt2008}, with some stars being more metal poor. For most of these literature studies it is unclear which solar composition is adopted to compute [M/H], but the largest possible difference in solar metallicity is about 0.1\,dex thus the literature metallicities are still systematically lower than in this work. Other reasons for the [M/H] offset could be different model atmospheres used in the literature, different analysis methods or different quality of the data (high- versus low-resolution).
From our tests in Sect.~\ref{sec:performance} we conclude that the [M/H] offset between our results and literature values is not an underestimation by the NN, since it predicts correct metallicities and other parameters for synthetic spectra. Instead, we hypothesise that the $T_{\rm{eff}}$, $\log\,g$ and [M/H] differences are caused by the manual normalisation of the spectra done in mentioned literature papers, which resulted in different placement of the continuum. $T_{\rm{eff}}$ is determined by the shape and depth of the hydrogen lines, while $\log\,g$ only depends on their inner core strength and on metal lines. Small differences in normalisation of the hydrogen lines result in different $T_{\rm{eff}}$ and $\log\,g$ values. This has an effect on [M/H], since this parameter has to compensate for the $T_{\rm{eff}}$ changes and normalisation differences in the metal line regions. If the pseudo-continuum of the observed spectrum is placed higher, the spectrum is pulled down and this correspond to higher opacities in the stellar atmospheres and thus higher [M/H] values. The amount by which the parameters differ depends on the wavelength regions and the number of hydrogen lines used in the atmospheric analysis. Also, we expect larger deviations when the analysis includes global continuum scaling, which in the case of F-type stars shifts the continuum downwards, increasing [M/H].

\begin{figure}
    \centering
    \resizebox{\hsize}{!}{\includegraphics{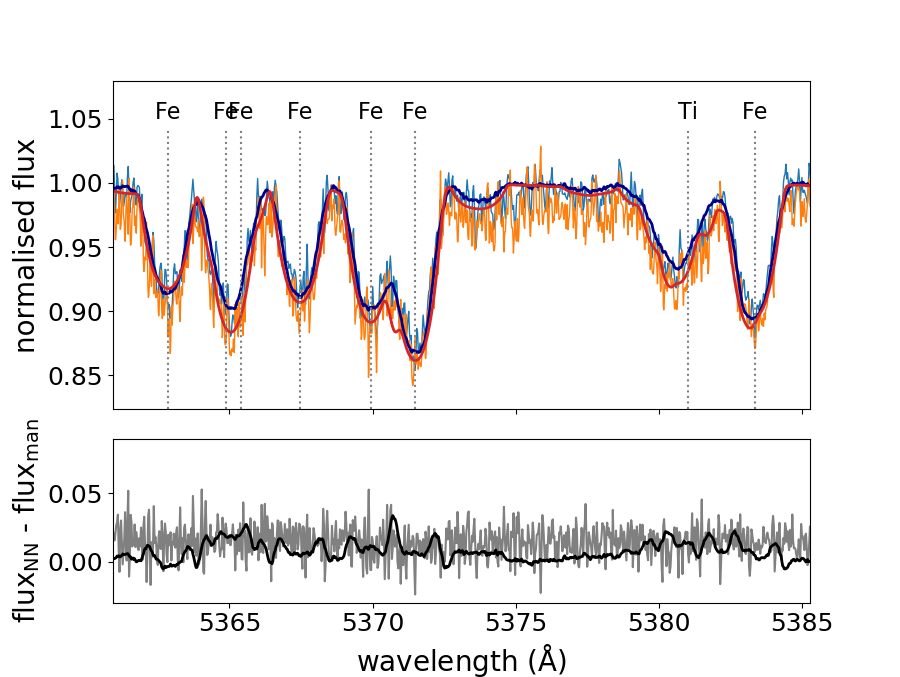}}
    \caption{Comparison of automated NN and manual normalisation for part of the spectrum of KIC\,7023122. The NN normalisation and best fitting predicted spectrum are shown in blue and dark blue. The manually normalised spectrum and a synthetic spectrum computed for the corresponding parameters from \citet{VanReeth2015} are plotted in orange and red. The lower panel shows the difference between the automated and manually normalised spectrum in grey, and the black line is the difference between the synthetic spectra.
    The most prominent element lines are indicated by dotted lines.}
    \label{fig:auto_vs_man}
\end{figure}

We were able to verify this normalisation statement by comparing our NN normalised spectra with manually normalised HERMES spectra that are available for the 37 stars in the sample of \citet{VanReeth2015}. These authors used a wavelength range between 4700 and 5800 \AA \ for the atmospheric analysis and normalised the spectra by fitting consecutively a cubic and a linear spline to the continuum.
We plotted both normalisations on top of each other for each of the 37 $\gamma\,$Dor stars and found that the manual normalisations are systematically shifted downwards with respect to the NN normalisations. We demonstrate this in Fig.~\ref{fig:auto_vs_man} for part of the spectrum of KIC\,7023122, where we over-plotted each spectrum with its best fitting synthetic spectrum. The [M/H] determined for this star by \citet{VanReeth2015} is $+0.04\pm0.08\,$dex while we found a value of $-0.34\pm0.03\,$dex. There is also a difference of more than 100\,K in $T_{\rm{eff}}$ and 0.46\,dex in $\log\,g$. 
Additionally, in Fig.~\ref{fig:gDor_lit} we observed similar patterns in the offsets of $T_{\rm{eff}}$, $\log\,g$ and [M/H], proving the connection between the derivation of these parameters. The various literature studies used different wavelength regions and some of them adopted a global continuum scaling. This explains why the values are different for all literature works. The largest discrepancies are found for the [M/H] values of \citet{VanReeth2015}, who used global continuum scaling and obtained higher [M/H] values than the other studies and the NN results.

This confirms that the small $T_{\rm{eff}}$ and $\log\,g$ offsets and the [M/H] inconsistency of $\sim$0.2\,dex is actually due to normalisation of the continuum level. We point out that this process is highly dependent on the positioning of points to which a polynomial or spline is fitted in order to define the continuum. Often points are selected in the upper part of the noise level while the NN normalisation fits a synthetic spectrum to the centre of the noise. From our analysis we are not able to tell which normalisation and corresponding parameters are correct. Independent, non-spectroscopic parameters are needed to lift the normalisation-($T_{\rm{eff}}$, $\log\,g$, [M/H]) degeneracy.

Apart from the global metallicity, the individual abundances are also affected. We derived the surface abundances of some elements from the manually normalised spectra with the same method as explained in Sect.~\ref{sec:abundances} and compared the results with the values in Tables~\ref{tab:gd_abund} and \ref{tab:gd_abund2}. The average differences and standard deviations are given in the header of those tables. These differences give an estimate of how much normalisation can affect abundance determinations. The differences are around a few tenths of dex for most elements and even reach 0.7~dex for Mn and Ba. This is normal for Mn and Ba since there are only a few lines of these elements present in the spectra and most of them are blended.

The mean abundance uncertainty is different for every element.
Fe and Mg could be determined with a precision below 0.1 dex, C and Ti have uncertainties of about 0.15 dex, the errors of Na, Ca, Sc, Cr, Mn, Ni and Ba lie around 0.2 dex, Y has a mean uncertainty of 0.3 dex, and that of Sr is $\sim$ 0.6 dex. 
For the remaining elements often only an upper limit could be found (this is indicated as such with the less than ($<$) symbol in Tables~\ref{tab:gd_abund} and \ref{tab:gd_abund2}) or uncertainties larger than 2 dex were obtained. The largest abundance uncertainties correspond to spectra with low S/N values. The metal lines in these spectra disappear into the noise, making it impossible to determine lower bounds on the abundances.  

\begin{figure*}
    \centering
    \includegraphics[width=12cm]{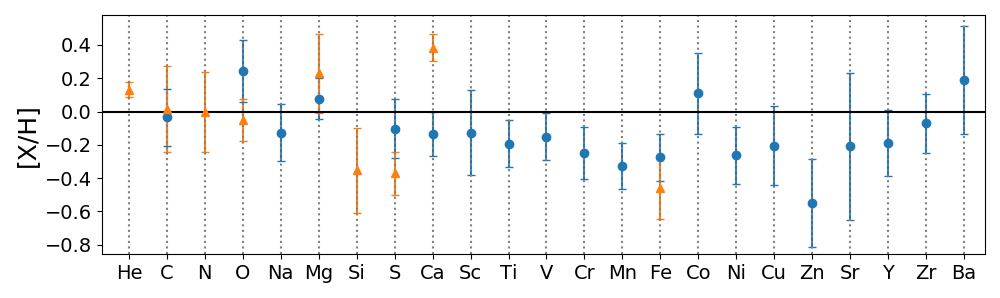}
    \caption{Average surface abundances with standard deviation of the $\gamma\,$Dor (blue circles) and SPB (orange triangles) samples with respect to the solar composition of \citet{Asplund2005}.}
    \label{fig:average_abund}
\end{figure*}

The abundances in Tables~\ref{tab:gd_abund} and \ref{tab:gd_abund2} are given with respect to the Sun and the average values over the whole sample are shown in Fig.~\ref{fig:average_abund}. As expected from the [M/H] values, most of the $\gamma\,$Dor stars have negative abundances and thus a sub-solar composition, except for Ba that is over-abundant in the majority of stars. There are only two Ba lines present in the wavelength range used in the analysis and both are heavily blended with other spectral lines. On top of that, one of the Ba lines (next to the H$_{\beta}$ hydrogen line) is a resonant line, so the results for Ba should be taken with caution. The average abundances of O and Co are also metal-rich. These elements only have a few weak lines in the F-type temperature range and for most of the stars only an upper bound or large errors on the abundance could be found.
KIC\,4757184, KIC\,6292398, KIC\,9751996, and KIC\,11099031 have systematically lower or higher abundances than the rest of the sample. These are also the stars with metallicities that deviate the most from the sample distribution, with KIC\,4757184 being metal poor ([M/H] = $-$0.65\,dex). The other three are the only $\gamma\,$Dor stars with positive [M/H].
We checked the abundance patterns of all stars for chemical peculiarities. Most of the stars have near-solar composition within the (sometimes large) error bars. Only one star, KIC\,6292398, is found to be a metallic line A star (Am star). It has underabundances of C, Ca and Sc while the heavy metals are enhanced \citep{Preston1974}. This star had already been classified as Am star by \citet{Niemczura2015}.
By ordering the elements according to their mean uncertainties it follows that Fe, Mg, C, Ti and to lesser extent Na, Ca, Sc, Cr, Mn, Ni and Ba are the best candidates to study and constrain interior physical processes. 

\subsection{SPB stars} 

\begin{figure*}[ht]
    \centering
    \includegraphics[width=0.3\textwidth]{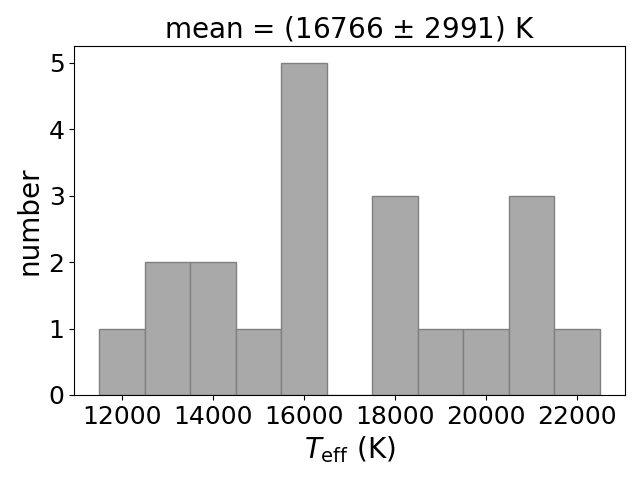}
    \includegraphics[width=0.3\textwidth]{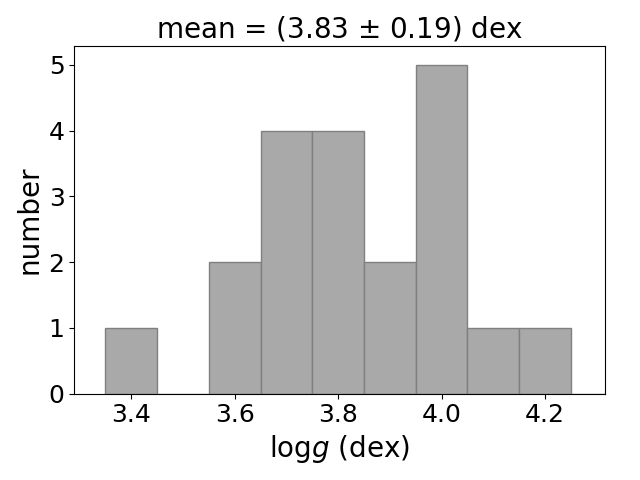}\\
    \includegraphics[width=0.3\textwidth]{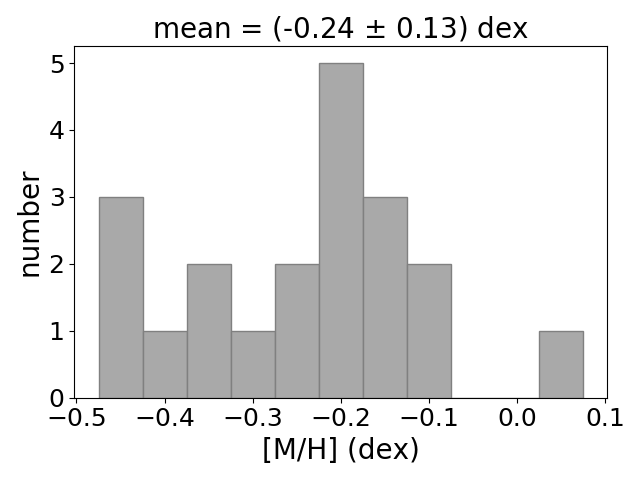}
    \includegraphics[width=0.3\textwidth]{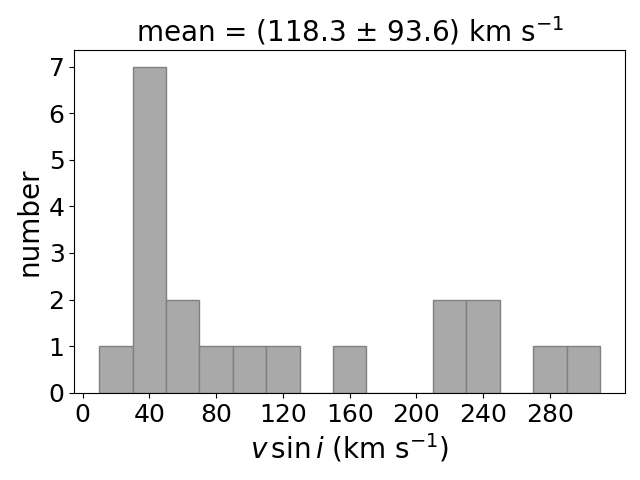}\\
    \includegraphics[width=0.3\textwidth]{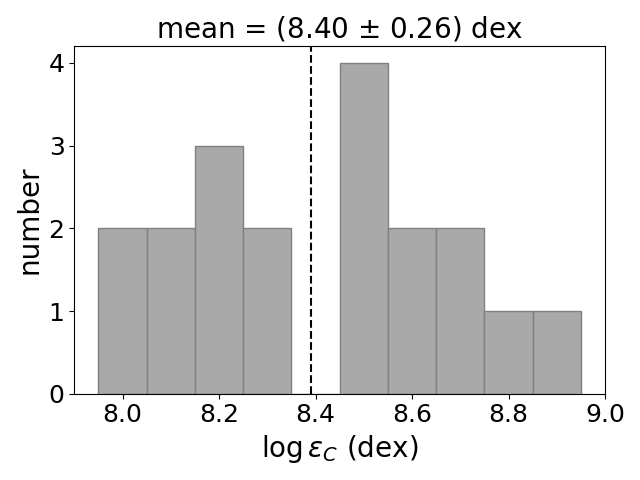}
    \includegraphics[width=0.3\textwidth]{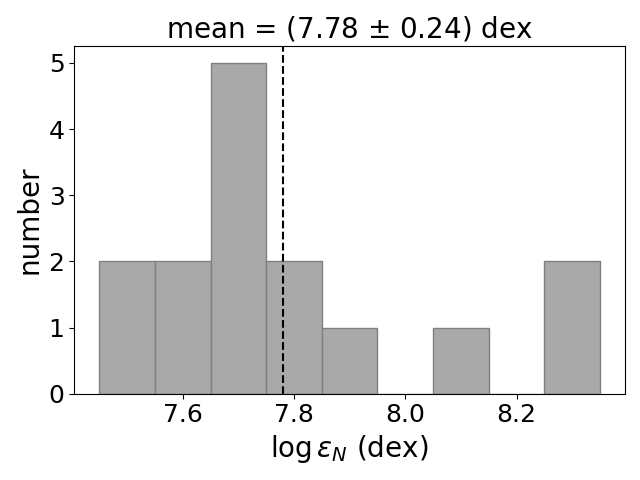}
    \includegraphics[width=0.3\textwidth]{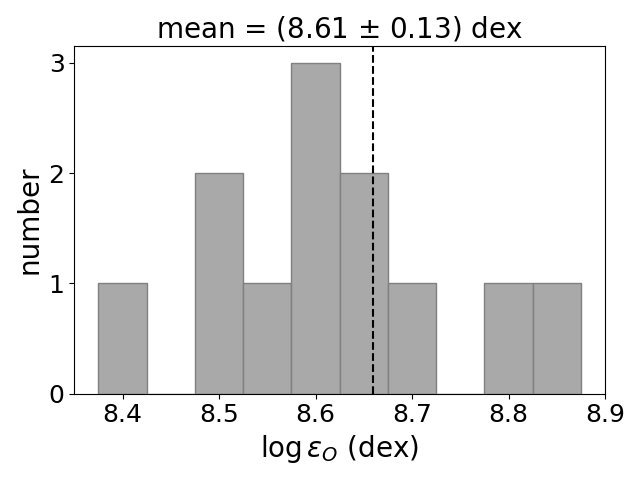}
    \caption{Histograms of $T_{\rm{eff}}$, $\log{g}$, [M/H], $v\sin{i}$ and C, N and O abundances in 12-scale for the sample of SPB stars. The [M/H] and $v\sin i$  values are the results from the NN, while $T_{\rm{eff}}$, $\log{g}$, and the C, N, and O abundances are those obtained with GSSP.} Solar values of the abundances are indicated with a black dashed line.
    \label{fig:SPB_hist}
\end{figure*}

The distributions of the final SPB parameters and a few abundances are shown in Fig.~\ref{fig:SPB_hist}. The temperatures are scattered over the full range typical for SPB stars which is 11\,000-22\,000\,K \citep{Aerts2010}. Most of the stars have $\log\,g$ values around 3.8 dex but there are a few with lower values and especially KIC\,8766405 seems to be more evolved. Just as for the $\gamma\,$Dor stars, the [M/H] distribution is centred around $-0.2$~dex. The SPB stars have projected rotational velocities over a wide range of values from 20 to 300\,km\,s$^{-1}$. It is difficult to determine $\xi$ as these stars only have a few metal lines and these are broadened by the fast rotation. For four SPB stars it was difficult to constrain $\xi$. In that case the value was fixed to 2\,km\,s$^{-1}$ which is a typical value for unevolved, core-hydrogen burning stars of intermediate mass \citep[][ chap. 17]{Gray2005} and which is also adopted in many model atmosphere codes \citep[e.g.][]{Kurucz1992,Lanz2007}. This value agrees with the NN determinations within the 1$\sigma$ uncertainties that, for these stars, reach over 60\% of the $\xi$ values themselves.

\begin{figure*}
    \centering
    \includegraphics[width=12cm]{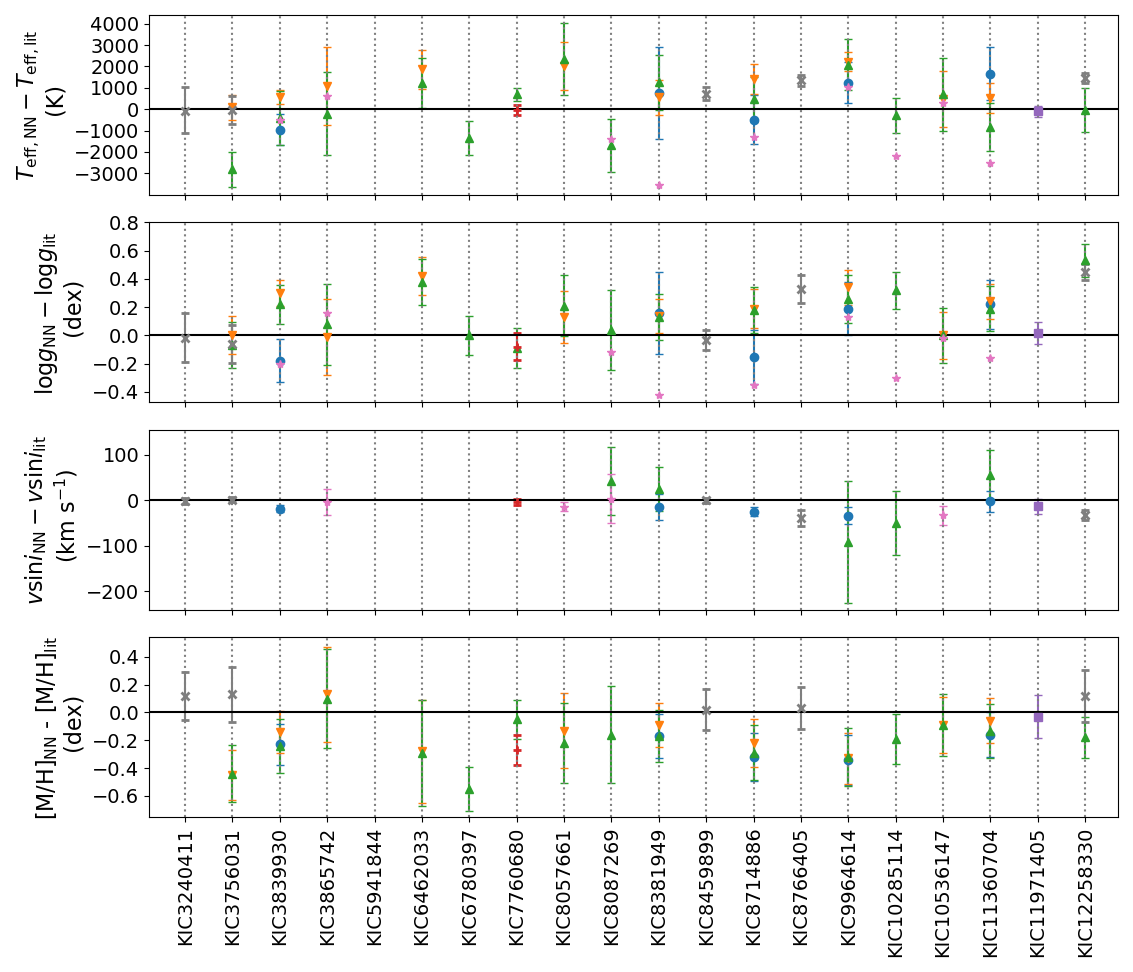}
    \caption{Comparison of SPB stellar parameters from the NN with literature values. From top to bottom the difference between the values in this paper and the ones from literature catalogues are shown for $T_{\rm{eff}}$, $\log{g}$, $v\sin{i}$ and [M/H]. Both the errors obtained from the NN and those from the literature are propagated to get the shown error bars. The different symbols correspond to \citet{Lehmann2011} (grey crosses), \citet{Balona2011a} (pink stars), \citet{Papics2015} (red pluses), \citet{Frasca2016} (green triangles), \citet{Papics2017} (purple squares), \citet{Zhang2018} (orange reversed triangles) and \citet{Hanes2019} (blue circles).}
    \label{fig:SPB_lit}
\end{figure*}

We compare our NN results to values in the literature in Fig.~\ref{fig:SPB_lit}. There are far fewer SPB stars than $\gamma\,$Dor stars known and fewer catalogue studies are available. Therefore we also include papers that only studied one or a few SPB stars. \citet{Lehmann2011} used spectra obtained with the Coude-Echelle spectrograph attached to the 2-m telescope at Th\"uringer Landessternwarte that has a resolution of 32\,000. The data were analysed with an early version of GSSP and apart from stellar parameters also surface abundances were determined. In \citet{Balona2011a} low resolution (R $\sim$ 550) spectra taken with the B\&C spectrograph on the Bok telescope at Kitt Peak observatory were analysed with LTE models. The largest SPB samples are found in \citet{Frasca2016} and \citet{Zhang2018} which exploit LAMOST data. Two stars have been studied in \citet{Papics2015,Papics2017} using HERMES data and the GSSP code. Only six SPB stars, with spectra from the RC spectrograph at KPNO 4-m Mayall telescope (R $\sim$ 7200), were analysed with a non-LTE method by \citet{Hanes2019}. Most of the parameters are consistent with these literature values, especially considering the large uncertainties for these hotter stars and the different methods used in all of the studies. We compare the [M/H] distribution of our sample with that of another sample of SPB stars \citep{Niemczura2009}. The mean and standard deviation of both samples are ($-0.21 \pm 0.12$)\,dex and ($-0.21 \pm 0.46$)\,dex respectively, so the SPB [M/H] values in this paper are typical for this kind of stars.

The SPB sample contains spectra that in general have low S/N for this kind of stars. Therefore, the abundances determination of the SPB stars is more difficult than for the $\gamma\,$Dor sample. The SPB stars are hotter and therefore have fewer metal lines. Most of them are also fast rotators which means that the few lines that are present are often largely broadened and sometimes disappear into the noise. This is reflected in the precision of the abundances that were estimated. For many elements such as Ne and Ca only an upper limit could be determined or estimated values with very large errors are obtained. Only He has an uncertainty below 0.1\,dex. C, Mg, Si and S have precisions between 0.15 and 0.25\,dex and the typical uncertainty on O, N and Fe is around 0.45~dex. This is partially caused by the low S/N of the spectra, given that the {\it Kepler\/} SPB stars are faint for the 1.2-m Mercator telescope. Because of the large uncertainties it is difficult to find trends in the abundances. As expected from the metallicities and illustrated in Fig.~\ref{fig:average_abund}, Fe, Si and S are generally below the solar value. We also find that the SPB stars are somewhat helium rich. 
However, this result should be taken with caution since non-LTE effects will strengthen the cores of He lines \citep{Nieva2007} leading to lower He abundances closer to solar value. Most of the abundances are also comparable with those of the cosmic B star standard \citep{Przybilla2008,Nieva2012} except for Si and Fe which have lower abundances.

\section{Discussion} \label{sec:discussion}

\subsection{Location in the Hertzsprung-Russell diagram} \label{sec:discussionHRD}

\begin{figure}
    \centering
    \resizebox{\hsize}{!}{\includegraphics{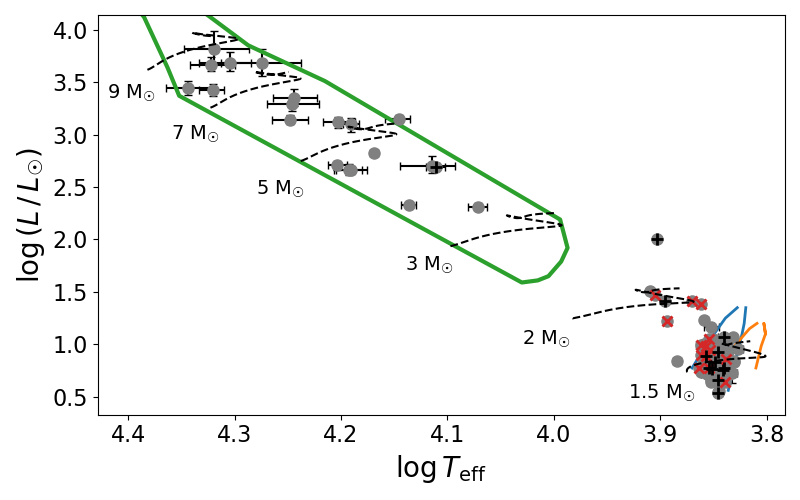}}
    \caption{Hertzsprung-Russell diagram containing the $\gamma\,$Dor and SPB stars studied in this work. The $\gamma\,$Dor instability strips are from \citet{Dupret2005}, for $\alpha_{\rm{mlt}}=1.5$ (orange) and $\alpha_{\rm{mlt}}=2.0$ (blue), while the SPB instability strip (green) is taken from \citet{Szewczuk2017}. All these strips were computed for solar metallicity. The red crosses are hybrid $\gamma\,$Dor - $\delta\,$Sct stars and the black pluses are SB1 stars. Evolutionary tracks for different masses taken from \citep{Johnston2019} and covering from zero age- to the terminal-age main sequence are plotted as black dashed lines to guide the eye.}
    \label{fig:HRD}
\end{figure}

In Fig.~\ref{fig:HRD} the whole sample is plotted in an HRD, based on our $T_{\rm{eff}}, \log g$ and [M/H] determination. We computed luminosities with

\begin{equation}
    \log \frac{L}{L_{\sun}} = -0.4 \left( M_{S_{\lambda}} + \rm{BC}_{S_{\lambda}} - M_{\rm{bol},\sun} \right) 
\end{equation}
\noindent and
\begin{equation}
    M_{S_{\lambda}} = m_{S_{\lambda}} - 5 \log \frac{d}{10\,\rm{pc}} - R_{S_{\lambda}} E(B - V),
\end{equation}

\noindent with $m_{S_{\lambda}}$ the apparent magnitudes from SIMBAD in the Johnson V-band or in the {\it Gaia\/} G-band when the previous is not available. There are three $\gamma\,$Dor stars with neither V- nor G-band magnitudes available for which we used their Kepler (K) magnitudes from the MAST archive. $d$ are {\it Gaia\/} eDR3 distances from \citet{Bailer-Jones2020} and $E(B-V)$ are reddening values obtained with the 3D reddening map (Bayestar19) from \citet{Green2019}. The reddening vector R$_{S_{\lambda}}$ is equal to 3.089 (V- or K-band) or 3.002 (G-band) \citep{Pedersen2020}. 
For the $\gamma\,$Dor stars with V- or K-band magnitudes we computed bolometric corrections (BC$_{S_{\lambda}}$) with prescriptions from \citet{Flower1996} with $3.7 < \log T_{\rm{eff}} < 3.9$ and we used $M_{\rm{bol,\sun}}=4.73$ \citep{Torres2010b}. The BC$_{G}$ values for the $\gamma\,$Dor stars with G-band magnitudes were obtained with coefficients from \citet{Andrae2018} which has a corresponding $M_{\rm{bol,\sun}}=4.74$. For the SPB stars the prescriptions from \citet{Pedersen2020} (model 3, LTE+non-LTE) were adopted and $M_{\rm{bol, \sun}}=4.74$. The luminosity values are given in the last columns of Tables~\ref{tab:gd_params} and \ref{tab:spb_params}.

The orange and blue lines in Fig.~\ref{fig:HRD} are the theoretical $\gamma\,$Dor instability strips from \citet{Dupret2005} for a mixing length parameter ($\alpha_{\rm{mlt}}$)  of 1.5 and 2.0 respectively. The parameter $\alpha_{\rm{mlt}}$ determines the length scale over which convective blobs move in MLT \citep{Bohm-Vitense1958} before they dissolve in their surroundings and hence affects the size of the convective envelope in low- and intermediate-mass stars \citep{Viani2018}. Higher values of $\alpha_{\rm{mlt}}$ lead to larger convective envelopes and shift the red edge of the instability strip based on the flux-blocking excitation mechanism towards higher $T_{\rm{eff}}$ \citep{Dupret2005}. Almost all objects in our legacy sample fall within the instability region for $\alpha_{\rm{mlt}} = 2.0$. There are ten stars, including five hybrid $\gamma\,$Dor - $\delta\,$Sct pulsators and two SB1 systems, which have a higher temperature and luminosity than theoretically predicted by excitation models. These hot $\gamma\,$Dor (- $\delta\,$Sct) stars have been discussed in other studies as well, including the instability study based on the flux-blocking mechanism based on time-dependent convection by \citep{Dupret2005}. Stars with such high temperatures do not have sufficiently deep convective envelopes for this mechanism to be effective. Different explanations have been given in the literature, such as binarity where the $\gamma\,$Dor star has a hotter non-pulsating A- or B-type companion \citep{Balona2011b} or rapidly rotating SPB stars appearing to be cooler due to gravity darkening \citep{Salmon2014,Balona2015}. \citet{Antoci2014} showed that high-order non-radial pulsations might also be excited stochastically by turbulent pressure in the hydrogen ionisation zone. \citet{Grassitelli2015} included turbulent pressure in stellar evolution calculations and found that the region in the HRD that contains models with a high fraction of turbulent pressure coincides with the observational $\gamma\,$Dor instability strip, such that these stars can excite stochastic gravito-inertial modes. However, Fig.~1 in that paper reveals that the turbulent pressure fraction of the five hottest $\gamma\,$Dor stars is small ($<$ 0.02) implying that this excitation mechanism is not efficient for these hotter stars.
Only two of the outliers are found to be in a spectroscopic binary based upon the RV measurements of their spectra (KIC\,7694191, which has the highest luminosity of the sample, and KIC\,6292398). None of the hot $\gamma\,$Dor (- $\delta\,$Sct) stars have a large value of $v\sin\,i$ which suggests either slow to moderate rotation or low inclination angle $i$. In both those cases, we can reject the hypothesis of rapidly rotating SPB stars because gravity darkening is negligible in the former case, whereas the star is observed nearly pole-on in the latter case and the $T_{\rm{eff}}$ determination is not subject to the gravity darkening effect either. Similar conclusions were found by \citet{Balona2016} and \citet{Kahraman2020} who also did not find signs of binarity or rapidly rotating SPB stars in hot $\gamma$\,Dor samples. The existence of such hot $\gamma$\,Dor stars suggests that the mode excitation in F-type stars is not yet fully understood. The problem of the pulsation mode excitation is beyond the scope of this observational spectroscopy paper.

For the SPB stars we used the theoretical instability strip from \citet{Szewczuk2017}. All the B-type targets in our legacy sample lie within the theoretically predicted region.

\subsection{Signatures of mixing processes} \label{sec:discussionMixing}

Mixing processes in the envelopes of stars can reveal themselves by changing element abundances at the stellar surface.  All the stars in our legacy sample burn hydrogen into helium via the CNO cycle, although the pp-chain is also active in $\gamma\,$Dor stars. During the initial phase of the CNO cycle, most of the C is transformed into N, creating an excess of N while C is depleted and the O level stays almost constant. Efficient mixing processes can transport the elements. Hence surface abundances of C, N, and O and their ratios can thus be used as tracers of mixing \citep{Maeder2000, Langer2012}. Several studies have looked at N abundances as a function of $v\sin{i}$ \citep{Hunter2008, Hunter2009, Brott2011} and found a positive correlation. These studies also reported a group of slowly rotating stars that are nevertheless N enriched. This cannot be explained by the current theory of rotational mixing and is attributed to magnetic fields \citep[e.g.][]{Morel2008}. However, \citet{Aerts2014b} delivered an alternative explanation in terms of pulsational wave mixing, in view of the absence of any correlation between the surface N abundance and the rotation frequency in a sample of OB-type dwarfs studied from ground-based photometry, spectro-polarimetry, and high-precision spectroscopy. This study was based on direct measurements of the rotation frequency at the surface or in the envelope of stars, rather than using projected rotation velocities from spectroscopy, which are subject to uncertainty due to the unknown inclination angle.

\citet{Przybilla2010}, \citet{Maeder2014} and \citet{Martins2015} used the relations between N/C - N/O and between N/C - $\log\,g$ to study the mixing in OB-type stars. N/C and N/O track the evolution in terms of nucleosynthesis, while $\log\,g$ is a proxy for the evolution in the HRD. From the theory of nucleosynthesis it follows that N/C and N/O should be tightly correlated and the different studies find various theoretical limits depending on the mass of the star and the adopted solar abundances. A correlation between N/C and $\log\,g$ is also expected since the N abundance should increase with age, or decrease with $\log\,g$.

\begin{figure}
    \centering
    \resizebox{\hsize}{!}{\includegraphics{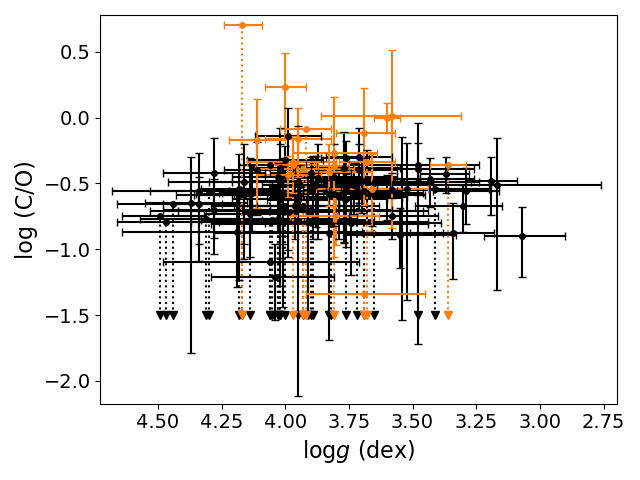}}
    \caption{log (C/O) as a function of $\log{g}$ for the sample of $\gamma\,$Dor stars in black and for the SPB stars in orange.}
    \label{fig:CO_logg}
\end{figure}

\begin{figure}
    \centering
    \resizebox{\hsize}{!}{\includegraphics{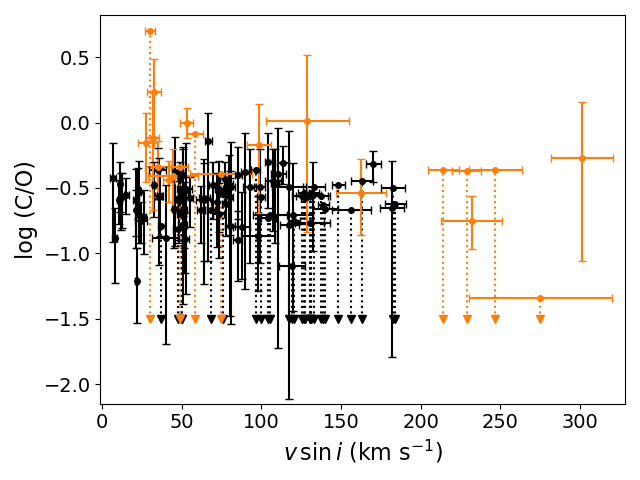}}
    \caption{log (C/O) as a function of $v\sin\,i$ for the sample of $\gamma\,$Dor stars in black and for the SPB stars in orange.}
    \label{fig:CO_vsini}
\end{figure}

The CNO cycle becomes more important with temperature. F-type stars have few N lines and we were not able to determine its abundance in any of the $\gamma\,$Dor stars. We did constrain the abundances of C and O and these can also be used as tracers of the CNO cycle, although to a lesser extent. In Figs.~\ref{fig:CO_logg} and \ref{fig:CO_vsini} the ratio between C and O is plotted as a function of $\log\,g$ and $v\sin\,i$, respectively. Since nucleosynthesis theory predicts that C gets depleted with time while O stays constant we expect to see a correlation with $\log\,g$. For faster rotating stars, mixing should be more efficient thus stars with a higher $v\sin\,i$ should have lower C/O ratios. We do not detect this from visual inspection nor from Pearson correlation coefficients, which are equal to $-$0.12 for $\log\,g$ and 0.12 for $v\sin\,i$ and imply the absence of a linear correlation between C/O and $\log\,g$ and between C/O and $v\sin\,i$. We do point out that for some stars the O abundance could not be determined precisely and this could be the reason that no correlation is found.

Microscopic atomic diffusion can also alter surface abundances in stars \citep{Michaud2015}. It influences the Mg/Fe ratio in such a way that it becomes higher when gravitational settling is dominant, because Mg with its lower atomic mass settles slower than Fe. For fast rotators it is assumed that the atomic diffusion process is counteracted by rotational mixing, but this assumption has been questioned lately, particularly for stars in the $\gamma\,$Dor mass regime. Indeed, for the hottest F-type dwarfs, radiative levitation comes into play as dominant mixing effect \citep{Deal2020}. This atomic diffusion process already occurs early on during the main sequence and affects the surface abundances at later evolutionary phases. This is important for chemical tagging of evolved stars, where surface abundances are used to identify members of dispersed clusters by assuming they were formed from the same cloud \citep{Freeman2002,Dotter2017}. Without internal mixing of the chemical elements, these stars must have the same composition. However, this is not the case when atomic diffusion or other mixing processes were at work in the stellar interior since they were born. 
As shown by \citet{Dotter2017}, the mixing  processes due to atomic diffusion must be included in stellar evolution models and calibrated with observed abundances to obtain the initial bulk abundances of  evolved stars. Such a calibration can be done from abundance studies of their progenitors, among which F-type dwarfs. 

\begin{figure}
    \centering
    \resizebox{\hsize}{!}{\includegraphics{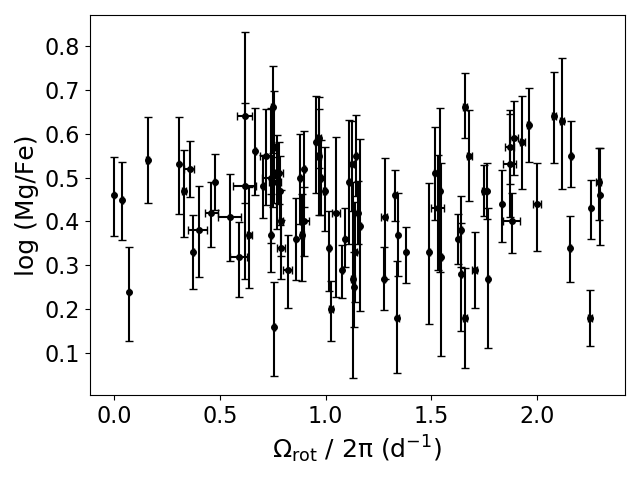}}
    \caption{log (Mg/Fe) as a function of $\Omega_{\rm{rot}}/2\pi$ for the sample of $\gamma\,$Dor stars.}
    \label{fig:FeMg_frot}
\end{figure}

\citet{Mombarg2020} compared measured asteroseismic parameters and surface abundances of two slowly rotating F-type dwarfs by means of models with and without atomic diffusion including radiative levitation. They found that precise abundance values with uncertainties below 0.1\,dex are needed to assess internal mixing due to atomic diffusion.  Only our values of Fe and Mg (and for some stars C and Ti) comply with this requirement.
In Fig.~\ref{fig:FeMg_frot} we plot the Mg/Fe ratio of the $\gamma\,$Dor stars as a function of $\Omega_{\rm{rot}}/2\pi$, with $\Omega_{\rm{rot}}$ the near-core angular rotation frequency determined from asteroseismic modelling by \citet{VanReeth2018,VanReeth2016} and \citet{Li2020}. 
This quantity is not dependent on the unknown inclination angle nor on the radius of the star. It offers a direct measurement of the internal rotation rate in the near-core convective boundary layer as estimated from high-order g~modes \citep{Aerts2019}.
We expect to see an enhancement of Mg/Fe for slow rotators in the absence of rotational mixing, but this is not observed. 
We thus do not find any observational evidence of mixing due to gravitational settling in our spectroscopic analysis of the $\gamma\,$Dor stars. 
This is in agreement with the low levels of deep mixing at the bottom of the radiative envelope, adjacent to the overshoot zone (hereafter called bottom envelope mixing), with values below 10\,cm$^2$\,s$^{-1}$ found from asteroseismology for these pulsators, which cover a wide range of rotational velocities \citep{VanReeth2016}.

\begin{figure}
    \centering
    \resizebox{\hsize}{!}{\includegraphics{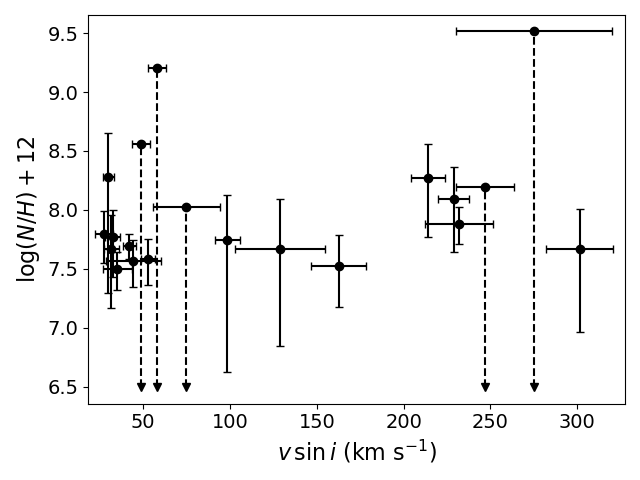}}
    \caption{N abundance of SPB stars in 12-scale plotted as a function of $v\sin\,i$.}
    \label{fig:SPB_Hunter_vsini}
\end{figure}

\begin{figure}
    \centering
    \resizebox{\hsize}{!}{\includegraphics{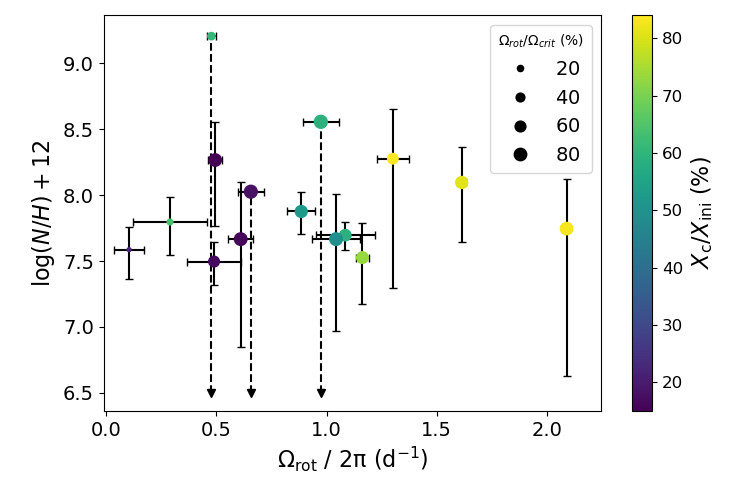}}
    \caption{N abundance of SPB stars in 12-scale plotted as a function of the asteroseismic $\Omega_{\rm{rot}}/2\pi$. The stars are coloured according to their ratio of central to initial hydrogen mass fraction, which is a proxy for their age. The size of the points depends on the fraction of the rotation rate with respect to the critical rate.}
    \label{fig:SPB_Hunter_omega_rot}
\end{figure}

\begin{figure}
    \centering
    \resizebox{\hsize}{!}{\includegraphics{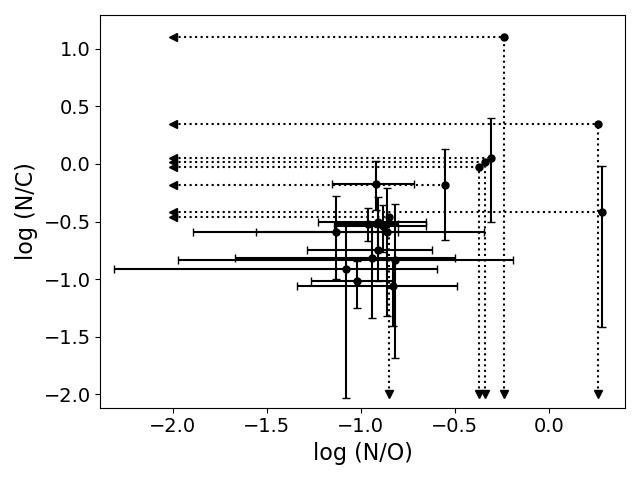}}
    \caption{log (N/C) as a function of log (N/O) for the sample of SPB stars.}
    \label{fig:SPB_CNO}
\end{figure}

\begin{figure}
    \centering
    \resizebox{\hsize}{!}{\includegraphics{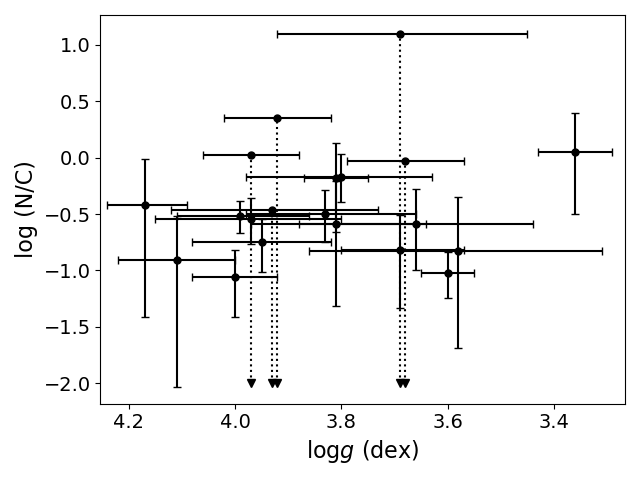}}
    \caption{log (N/C) as a function of $\log\,g$ for the sample of SPB stars.}
    \label{fig:SPB_CN_logg}
\end{figure}

For the SPB stars we did determine N abundances. In Fig.~\ref{fig:SPB_Hunter_vsini} we show the N abundance in 12-scale plotted as a function of $v\sin\,i$ and in Fig.~\ref{fig:SPB_Hunter_omega_rot} $v\sin\,i$ is replaced by the asteroseismic value of $\Omega_{\rm{rot}}/2\pi$.
These asteroseismic estimates are taken from \citet{Pedersen2021}. The Pearson correlation coefficients for both diagrams are 0.26 and $-$0.05 for $v\sin\,i$ and $\Omega_{\rm{rot}}/2\pi$, respectively. This confirms what can be seen by eye, namely there is no correlation between the N abundance at the stellar surface and the rotation frequency of the stars, in line with the earlier findings by \citet{Aerts2014b}. 
However, we do find evidence that deep mixing is active in the SPB stars from abundance ratios. As can be seen in Fig.~\ref{fig:SPB_CNO}, there is a positive trend between N/C and N/O. This is confirmed by the Pearson correlation coefficient of 0.66. This is compatible with nucleosynthesis and element transport of the resulting CNO products to the surface. The correlation with evolutionary state in Fig.~\ref{fig:SPB_CN_logg} is also present but less clear, with a Pearson correlation coefficient of 0.19. Our spectroscopic findings are in good agreement with the asteroseismic results of \citet{Pedersen2021}, who find a correlation between the level of mixing at the bottom of the radiative envelope and the star's $\Omega_{\rm rot}/\Omega_{\rm crit}$ at that position, with $\Omega_{\rm crit}$ being the critical rotation rate. The correlation coefficient between these two quantities measured from asteroseismology amounts to 0.60.

\begin{figure*}
    \centering
    \includegraphics[width=0.49\hsize]{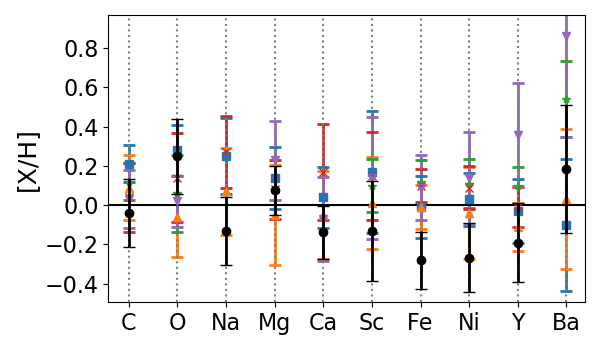}
    \includegraphics[width=0.49\hsize]{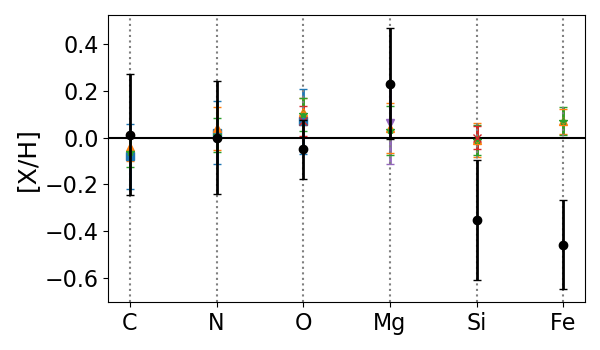}
    \caption{{\it Left}: Comparison of the mean element abundances from the $\gamma\,$Dor sample (black dots) with abundance values for non-pulsating stars from \citet{Varenne1999} (blue squares), \citet{Monier2005} (orange triangles), \citet{Gebran2008} (green stars), \citet{Gebran2010} (red crosses) and \citet{Kilicoglu2016} (purple reversed triangles). {\it Right}: Comparison of the mean element abundances from the SPB sample (black dots) with abundance values for non-pulsating stars from \citet{Lyubimkov2005} (purple reversed triangles), \citet{SimonDiaz2010} (red crosses), \citet{Nieva2011} (orange triangles), \citet{Nieva2012} (green stars) and \citet{Lyubimkov2013} (blue squares).}
    \label{fig:non_pulsators}
\end{figure*}

We also compared the abundances of the pulsating $\gamma\,$Dor and SPB stars with values for non-pulsating F- and B-type stars in Fig.~\ref{fig:non_pulsators}, where all the sources of the literature data are listed in the figure caption.
Except for Si and Fe which appear to be less abundant in the SPB stars, the abundances of non-pulsators and pulsators are in agreement within the uncertainties. Thus from their surface abundances, there is no evidence that pulsating stars experience more mixing than non-pulsating stars. This was also found by \citet{Kahraman2016}.

In conclusion, we find that F-type g-mode pulsators do not reveal any evidence of a connection between rotation and bottom envelope internal mixing, nor any sign of gravitational settling in their envelope. On the other hand, the surface abundances of B-type g-mode pulsators are compatible with CNO processed matter being transported to the surface by mixing process(es) that are mildly correlated with the internal rotation.

\section{Conclusion} \label{sec:conclusion}

We determined stellar parameters and surface abundances for a sample of 91 $\gamma\,$Dor stars, among which 14 are hybrid $\gamma\,$Dor - $\delta\,$Sct pulsators, and 20 SPB stars that were selected based on their asteroseismic properties. To analyse all the stars in a homogeneous way, we relied on spectra assembled with a single high-resolution spectrograph \citep[HERMES,][]{Raskin2011} and we used a machine learning tool \citep[{\it The Payne\/},][]{Ting2019}, for stellar parameter determination which includes pseudo-continuum normalisation of spectra. We performed tests for this method which was the first time it was applied to high-resolution spectroscopic data (R $\sim$ 85\,000). It was found that the size of the training sample is dependent on the parameter ranges that have to be covered, but is still reasonably small for 5D grids for $\gamma\,$Dor and SPB stars. Ideally, abundances must be optimised alongside stellar parameters to account for correlations between them, but this requires larger training samples that have almost impossibly long computation times. In future work we will explore ways to circumvent this and analyse stellar parameters and surface abundances simultaneously. The performance of the NN method was tested against artificial data and benchmark stars for which it determined accurate, precise stellar parameters that agree with the synthetic parameters and literature values. We applied the tested NN routine to the sample of $\gamma\,$Dor and SPB stars. $v\sin\,i$ was found to be consistent with values from literature studies, but $T_{\rm{eff}}$ and $\log\,g$ deviate a little and [M/H] is even systematically offset by $\sim$\,0.2\,dex for the F-type stars. We attribute the differences between values in this work and those in the literature for the F-type dwarfs to the different normalisation of the spectra. Normalisation in literature studies is typically done manually and therefore depends on the continuum points that are selected, while in this work normalisation is integrated into the fitting routine as additional coefficients that are optimised alongside the stellar parameters. The [M/H]$\sim$0.2\,dex difference is likely a manifestation of the degeneracy between ($T_{\rm{eff}}$, $\log\,g$, [M/H]) and normalisation, making it impossible to say with certainty which solution is correct. This must be investigated further for a large sample of F-type stars for which literature values from non-spectroscopic studies such as interferometry are available.

The surface abundances were determined individually with the LTE code GSSP \citep{Tkachenko2015}. Blending of lines was accounted for by iterating the elements with the most lines twice. The uncertainties obtained for the abundances depend on the S/N of the spectra. Results for spectra with S/N $<$ 50 should be taken with caution when studied individually, but they are reasonable in sample studies since they did not appear as outliers in the sample.
Our analysis does not take non-LTE effects into account. This becomes increasingly important for the stellar parameters of hotter B-type stars but was omitted here to have an analogous analysis for the F- and B-type pulsators. In future work we will increase the sample sizes and analyse the spectra based on a non-LTE approach.

We found no signatures of internal mixing from the CNO abundances for the F-type stars, nor did we detect consequences of gravitational settling. For the B-type stars, we do find consequences of internal mixing in the N/C and N/O ratios, but the abundance errors of N are too large to reveal effects of internal mixing. 
We also know from asteroseismology of a sample of 26 SPB stars, 20 of which were included here, that the level of internal mixing at the bottom of the radiative envelope is mildly correlated with the near-core rotation of the stars \citep{Pedersen2021}.
This correlation provides observational evidence that B stars have deep envelope mixing related to their rotation.
\citet{Pedersen2021} further investigated the internal mixing profiles and concluded that the envelope mixing profiles are stratified for the majority of stars, and of diverse nature for the 26 stars in their sample. 
Combining g-mode asteroseismic quantities that probe the bottom of the radiative envelope with high-precision surface abundances at the level of 0.1\,dex or better, constitutes a viable and promising route towards better probing of internal mixing in intermediate- and high-mass stars. It would also offer a great asset to better distinguish among various candidate envelope mixing theories of B-type stars. This requires higher S/N spectra to increase the abundance precision of as many chemical elements as possible to exploit the maximum power of joint g-mode asteroseismology and high-resolution spectroscopy.

\begin{acknowledgement}

This work is based on observations obtained with the HERMES spectrograph, which is supported by the Research Foundation - Flanders (FWO), Belgium, the Research Council of KU Leuven, Belgium, the Fonds National de la Recherche Scientifique (F.R.S.-FNRS), Belgium, the Royal Observatory of Belgium, the Observatoire de Genève, Switzerland and the Thüringer Landessternwarte Tautenburg, Germany. 
The computational resources and services used in this work were provided by the VSC (Flemish Supercomputer Centre), funded by the Research Foundation - Flanders (FWO) and the Flemish Government.
Part of the research leading to these results has received funding from the European Research Council (ERC) under the European Union’s Horizon 2020 research and innovation programme (grant agreement N$^\circ$670519: MAMSIE) and from the KU\,Leuven Research Council (grant C16/18/005: PARADISE). SG, TVR, and DMB gratefully acknowledge support from the Research Foundation Flanders (FWO) by means of a PhD Aspirant mandate and a Junior and Senior Postdoctoral Fellowship, under contracts No. 11E5620N, No. 12ZB620N, and No. 1286521N,
respectively. MGP acknowledges support from the National Science Foundation under Grant No. NSF PHY-1748958. JB acknowledges support from the FWO Odysseus program under project No. G0F8H6N.
We acknowledge the efforts of all observers who contributed to the gathering of the HERMES spectroscopy used in this paper and we are particularly grateful to have been given access to unpublished spectra of KIC\,6292398, KIC\,7748238, KIC\,7770282 and KIC\,2168333 from HERMES programs 52 and 86 prior to publication.
\end{acknowledgement}

\bibliographystyle{aa}
\bibliography{Gebruers.bib}

\begin{thebibliography}{134}
\expandafter\ifx\csname natexlab\endcsname\relax\def\natexlab#1{#1}\fi

\bibitem[{{Aerts}(2021)}]{Aerts2021}
{Aerts}, C. 2021, Reviews of Modern Physics, 93, 015001

\bibitem[{{Aerts} {et~al.}(2010){Aerts}, {Christensen-Dalsgaard}, \&
  {Kurtz}}]{Aerts2010}
{Aerts}, C., {Christensen-Dalsgaard}, J., \& {Kurtz}, D.~W. 2010,
  {Asteroseismology} (Springer Netherlands)

\bibitem[{{Aerts} {et~al.}(2019){Aerts}, {Mathis}, \& {Rogers}}]{Aerts2019}
{Aerts}, C., {Mathis}, S., \& {Rogers}, T.~M. 2019, \araa, 57, 35

\bibitem[{{Aerts} {et~al.}(2014){Aerts}, {Molenberghs}, {Kenward}, \&
  {Neiner}}]{Aerts2014b}
{Aerts}, C., {Molenberghs}, G., {Kenward}, M.~G., \& {Neiner}, C. 2014, \apj,
  781, 88

\bibitem[{{Aerts} {et~al.}(2018){Aerts}, {Molenberghs}, {Michielsen},
  {Pedersen}, {Bj{\"o}rklund}, {Johnston}, {Mombarg}, {Bowman}, {Buysschaert},
  {P{\'a}pics}, {Sekaran}, {Sundqvist}, {Tkachenko}, {Truyaert}, {Van Reeth},
  \& {Vermeyen}}]{Aerts2018}
{Aerts}, C., {Molenberghs}, G., {Michielsen}, M., {et~al.} 2018, \apjs, 237, 15

\bibitem[{{Aerts} {et~al.}(2017){Aerts}, {Van Reeth}, \&
  {Tkachenko}}]{Aerts2017}
{Aerts}, C., {Van Reeth}, T., \& {Tkachenko}, A. 2017, \textit{Astrophys. J.
  Lett.}, 847, L7

\bibitem[{{Aguilera-G{\'o}mez} {et~al.}(2018){Aguilera-G{\'o}mez},
  {Ram{\'\i}rez}, \& {Chanam{\'e}}}]{Aguilera2018}
{Aguilera-G{\'o}mez}, C., {Ram{\'\i}rez}, I., \& {Chanam{\'e}}, J. 2018, \aap,
  614, A55

\bibitem[{{Andrae} {et~al.}(2018){Andrae}, {Fouesneau}, {Creevey}, {Ordenovic},
  {Mary}, {Burlacu}, {Chaoul}, {Jean-Antoine-Piccolo}, {Kordopatis}, {Korn},
  {Lebreton}, {Panem}, {Pichon}, {Th{\'e}venin}, {Walmsley}, \&
  {Bailer-Jones}}]{Andrae2018}
{Andrae}, R., {Fouesneau}, M., {Creevey}, O., {et~al.} 2018, \aap, 616, A8

\bibitem[{{Angelou} {et~al.}(2020){Angelou}, {Bellinger}, {Hekker}, {Mints},
  {Elsworth}, {Basu}, \& {Weiss}}]{Angelou2020}
{Angelou}, G.~C., {Bellinger}, E.~P., {Hekker}, S., {et~al.} 2020,
  \textit{MNRAS}, 493, 4987

\bibitem[{{Antoci} {et~al.}(2014){Antoci}, {Cunha}, {Houdek}, {Kjeldsen},
  {Trampedach}, {Handler}, {L{\"u}ftinger}, {Arentoft}, \&
  {Murphy}}]{Antoci2014}
{Antoci}, V., {Cunha}, M., {Houdek}, G., {et~al.} 2014, \apj, 796, 118

\bibitem[{{Asplund} {et~al.}(2005){Asplund}, {Grevesse}, \&
  {Sauval}}]{Asplund2005}
{Asplund}, M., {Grevesse}, N., \& {Sauval}, A.~J. 2005, in Astronomical Society
  of the Pacific Conference Series, Vol. 336, Cosmic Abundances as Records of
  Stellar Evolution and Nucleosynthesis, ed. I.~{Barnes}, Thomas~G. \& F.~N.
  {Bash}, 25

\bibitem[{{Auvergne} {et~al.}(2009){Auvergne}, {Bodin}, {Boisnard}, {Buey},
  {Chaintreuil}, {Epstein}, {Jouret}, {Lam-Trong}, {Levacher}, {Magnan},
  {Perez}, {Plasson}, {Plesseria}, {Peter}, {Steller}, {Tiph{\`e}ne}, {Baglin},
  {Agogu{\'e}}, {Appourchaux}, {Barbet}, {Beaufort}, {Bellenger}, {Berlin},
  {Bernardi}, {Blouin}, {Boumier}, {Bonneau}, {Briet}, {Butler}, {Cautain},
  {Chiavassa}, {Costes}, {Cuvilho}, {Cunha-Parro}, {de Oliveira Fialho},
  {Decaudin}, {Defise}, {Djalal}, {Docclo}, {Drummond}, {Dupuis}, {Exil},
  {Faur{\'e}}, {Gaboriaud}, {Gamet}, {Gavalda}, {Grolleau}, {Gueguen},
  {Guivarc'h}, {Guterman}, {Hasiba}, {Huntzinger}, {Hustaix}, {Imbert},
  {Jeanville}, {Johlander}, {Jorda}, {Journoud}, {Karioty}, {Kerjean},
  {Lafond}, {Lapeyrere}, {Landiech}, {Larqu{\'e}}, {Laudet}, {Le Merrer},
  {Leporati}, {Leruyet}, {Levieuge}, {Llebaria}, {Martin}, {Mazy}, {Mesnager},
  {Michel}, {Moalic}, {Monjoin}, {Naudet}, {Neukirchner}, {Nguyen-Kim},
  {Ollivier}, {Orcesi}, {Ottacher}, {Oulali}, {Parisot}, {Perruchot},
  {Piacentino}, {Pinheiro da Silva}, {Platzer}, {Pontet}, {Pradines},
  {Quentin}, {Rohbeck}, {Rolland}, {Rollenhagen}, {Romagnan}, {Russ}, {Samadi},
  {Schmidt}, {Schwartz}, {Sebbag}, {Smit}, {Sunter}, {Tello}, {Toulouse},
  {Ulmer}, {Vandermarcq}, {Vergnault}, {Wallner}, {Waultier}, \&
  {Zanatta}}]{Auvergne2009}
{Auvergne}, M., {Bodin}, P., {Boisnard}, L., {et~al.} 2009, \aap, 506, 411

\bibitem[{{Bailer-Jones} {et~al.}(2020){Bailer-Jones}, {Rybizki}, {Fouesneau},
  {Demleitner}, \& {Andrae}}]{Bailer-Jones2020}
{Bailer-Jones}, C.~A.~L., {Rybizki}, J., {Fouesneau}, M., {Demleitner}, M., \&
  {Andrae}, R. 2020, arXiv e-prints, arXiv:2012.05220

\bibitem[{{Balona} {et~al.}(2015){Balona}, {Baran},
  {Daszy{\'n}ska-Daszkiewicz}, \& {De Cat}}]{Balona2015}
{Balona}, L.~A., {Baran}, A.~S., {Daszy{\'n}ska-Daszkiewicz}, J., \& {De Cat},
  P. 2015, \mnras, 451, 1445

\bibitem[{{Balona} {et~al.}(2016){Balona}, {Engelbrecht}, {Joshi}, {Joshi},
  {Sharma}, {Semenko}, {Pandey}, {Chakradhari}, {Mkrtichian}, {Hema}, \&
  {Nemec}}]{Balona2016}
{Balona}, L.~A., {Engelbrecht}, C.~A., {Joshi}, Y.~C., {et~al.} 2016, \mnras,
  460, 1318

\bibitem[{{Balona} {et~al.}(2011{\natexlab{a}}){Balona}, {Guzik},
  {Uytterhoeven}, {Smith}, {Tenenbaum}, \& {Twicken}}]{Balona2011b}
{Balona}, L.~A., {Guzik}, J.~A., {Uytterhoeven}, K., {et~al.}
  2011{\natexlab{a}}, \mnras, 415, 3531

\bibitem[{{Balona} {et~al.}(2011{\natexlab{b}}){Balona}, {Pigulski}, {De Cat},
  {Handler}, {Guti{\'e}rrez-Soto}, {Engelbrecht}, {Frescura}, {Briquet},
  {Cuypers}, {Daszy{\'n}ska-Daszkiewicz}, {Degroote}, {Dukes}, {Garcia},
  {Green}, {Heber}, {Kawaler}, {Lehmann}, {Leroy}, {Molenda-{\.Z}aaowicz},
  {Neiner}, {Noels}, {Nuspl}, {{\O}stensen}, {Pricopi}, {Roxburgh}, {Salmon},
  {Smith}, {Su{\'a}rez}, {Suran}, {Szab{\'o}}, {Uytterhoeven},
  {Christensen-Dalsgaard}, {Kjeldsen}, {Caldwell}, {Girouard}, \&
  {Sanderfer}}]{Balona2011a}
{Balona}, L.~A., {Pigulski}, A., {De Cat}, P., {et~al.} 2011{\natexlab{b}},
  \mnras, 413, 2403

\bibitem[{{Bellinger}(2019)}]{Bellinger2019a}
{Bellinger}, E.~P. 2019, \textit{MNRAS}, 486, 4612

\bibitem[{{Bellinger}(2020)}]{Bellinger2019b}
{Bellinger}, E.~P. 2020, \textit{MNRAS}, 492, L50

\bibitem[{{Bellinger} {et~al.}(2016){Bellinger}, {Angelou}, {Hekker}, {Basu},
  {Ball}, \& {Guggenberger}}]{Bellinger2016}
{Bellinger}, E.~P., {Angelou}, G.~C., {Hekker}, S., {et~al.} 2016, \apj, 830,
  31

\bibitem[{{Blanco-Cuaresma} {et~al.}(2015){Blanco-Cuaresma}, {Soubiran},
  {Heiter}, {Asplund}, {Carraro}, {Costado}, {Feltzing},
  {Gonz{\'a}lez-Hern{\'a}ndez}, {Jim{\'e}nez-Esteban}, {Korn}, {Marino},
  {Montes}, {San Roman}, {Tabernero}, \&
  {Tautvai{\v{s}}ien{\.{e}}}}]{Blanco-Cuaresma2015}
{Blanco-Cuaresma}, S., {Soubiran}, C., {Heiter}, U., {et~al.} 2015, \aap, 577,
  A47

\bibitem[{{Boeche} \& {Grebel}(2016)}]{Boeche2016}
{Boeche}, C. \& {Grebel}, E.~K. 2016, \aap, 587, A2

\bibitem[{{B{\"o}hm-Vitense}(1958)}]{Bohm-Vitense1958}
{B{\"o}hm-Vitense}, E. 1958, \zap, 46, 108

\bibitem[{{Bowman}(2020)}]{Bowman2020c}
{Bowman}, D.~M. 2020, Frontiers in Astronomy and Space Sciences, 7, 70

\bibitem[{{Brott} {et~al.}(2011){Brott}, {Evans}, {Hunter}, {de Koter},
  {Langer}, {Dufton}, {Cantiello}, {Trundle}, {Lennon}, {de Mink}, {Yoon}, \&
  {Anders}}]{Brott2011}
{Brott}, I., {Evans}, C.~J., {Hunter}, I., {et~al.} 2011, \aap, 530, A116

\bibitem[{{Bruntt} {et~al.}(2008){Bruntt}, {De Cat}, \& {Aerts}}]{Bruntt2008}
{Bruntt}, H., {De Cat}, P., \& {Aerts}, C. 2008, \aap, 478, 487

\bibitem[{{Christensen-Dalsgaard}(2002)}]{JCD2002}
{Christensen-Dalsgaard}, J. 2002, \textit{Rev. of Modern Phys.}, 74, 1073

\bibitem[{{C{\'o}rsico} {et~al.}(2019){C{\'o}rsico}, {Althaus}, {Miller
  Bertolami}, \& {Kepler}}]{Corsico2019}
{C{\'o}rsico}, A.~H., {Althaus}, L.~G., {Miller Bertolami}, M.~M., \& {Kepler},
  S.~O. 2019, \textit{Astron. Astrophys. Rev.}, 27, 7

\bibitem[{{Deal} {et~al.}(2020){Deal}, {Goupil}, {Marques}, {Reese}, \&
  {Lebreton}}]{Deal2020}
{Deal}, M., {Goupil}, M.~J., {Marques}, J.~P., {Reese}, D.~R., \& {Lebreton},
  Y. 2020, \aap, 633, A23

\bibitem[{{Degroote} {et~al.}(2010){Degroote}, {Aerts}, {Baglin}, {Miglio},
  {Briquet}, {Noels}, {Niemczura}, {Montalban}, {Bloemen}, {Oreiro},
  {Vu{\v{c}}kovi{\'c}}, {Smolders}, {Auvergne}, {Baudin}, {Catala}, \&
  {Michel}}]{Degroote2010}
{Degroote}, P., {Aerts}, C., {Baglin}, A., {et~al.} 2010, \nat, 464, 259

\bibitem[{{Dotter} {et~al.}(2017){Dotter}, {Conroy}, {Cargile}, \&
  {Asplund}}]{Dotter2017}
{Dotter}, A., {Conroy}, C., {Cargile}, P., \& {Asplund}, M. 2017, \apj, 840, 99

\bibitem[{{Dupret} {et~al.}(2005){Dupret}, {Grigahc{\`e}ne}, {Garrido},
  {Gabriel}, \& {Scuflaire}}]{Dupret2005}
{Dupret}, M.~A., {Grigahc{\`e}ne}, A., {Garrido}, R., {Gabriel}, M., \&
  {Scuflaire}, R. 2005, \aap, 435, 927

\bibitem[{{Dziembowski} {et~al.}(1993){Dziembowski}, {Moskalik}, \&
  {Pamyatnykh}}]{Dziembowski1993}
{Dziembowski}, W.~A., {Moskalik}, P., \& {Pamyatnykh}, A.~A. 1993, \mnras, 265,
  588

\bibitem[{{Flower}(1996)}]{Flower1996}
{Flower}, P.~J. 1996, \apj, 469, 355

\bibitem[{{Frasca} {et~al.}(2016){Frasca}, {Molenda-{\.Z}akowicz}, {De Cat},
  {Catanzaro}, {Fu}, {Ren}, {Luo}, {Shi}, {Wu}, \& {Zhang}}]{Frasca2016}
{Frasca}, A., {Molenda-{\.Z}akowicz}, J., {De Cat}, P., {et~al.} 2016, \aap,
  594, A39

\bibitem[{{Freeman} \& {Bland-Hawthorn}(2002)}]{Freeman2002}
{Freeman}, K. \& {Bland-Hawthorn}, J. 2002, \araa, 40, 487

\bibitem[{{Garc{\'\i}a} \& {Ballot}(2019)}]{GarciaBallot2019}
{Garc{\'\i}a}, R.~A. \& {Ballot}, J. 2019, \textit{Living Reviews in Solar
  Physics}, 16, 4

\bibitem[{{Gebran} {et~al.}(2008){Gebran}, {Monier}, \& {Richard}}]{Gebran2008}
{Gebran}, M., {Monier}, R., \& {Richard}, O. 2008, \aap, 479, 189

\bibitem[{{Gebran} {et~al.}(2010){Gebran}, {Vick}, {Monier}, \&
  {Fossati}}]{Gebran2010}
{Gebran}, M., {Vick}, M., {Monier}, R., \& {Fossati}, L. 2010, \aap, 523, A71

\bibitem[{{Giribaldi} {et~al.}(2019){Giribaldi}, {Ubaldo-Melo}, {Porto de
  Mello}, {Pasquini}, {Ludwig}, {Ulmer-Moll}, \&
  {Lorenzo-Oliveira}}]{Giribaldi2019}
{Giribaldi}, R.~E., {Ubaldo-Melo}, M.~L., {Porto de Mello}, G.~F., {et~al.}
  2019, \aap, 624, A10

\bibitem[{{Grassitelli} {et~al.}(2015){Grassitelli}, {Fossati}, {Langer},
  {Miglio}, {Istrate}, \& {Sanyal}}]{Grassitelli2015}
{Grassitelli}, L., {Fossati}, L., {Langer}, N., {et~al.} 2015, \aap, 584, L2

\bibitem[{{Gray}(2005)}]{Gray2005}
{Gray}, D.~F. 2005, {The Observation and Analysis of Stellar Photospheres}

\bibitem[{{Gray} {et~al.}(2003){Gray}, {Corbally}, {Garrison}, {McFadden}, \&
  {Robinson}}]{Gray2003}
{Gray}, R.~O., {Corbally}, C.~J., {Garrison}, R.~F., {McFadden}, M.~T., \&
  {Robinson}, P.~E. 2003, \aj, 126, 2048

\bibitem[{{Green} {et~al.}(2019){Green}, {Schlafly}, {Zucker}, {Speagle}, \&
  {Finkbeiner}}]{Green2019}
{Green}, G.~M., {Schlafly}, E., {Zucker}, C., {Speagle}, J.~S., \&
  {Finkbeiner}, D. 2019, \apj, 887, 93

\bibitem[{{Guzik} {et~al.}(2000){Guzik}, {Kaye}, {Bradley}, {Cox}, \&
  {Neuforge}}]{Guzik2000}
{Guzik}, J.~A., {Kaye}, A.~B., {Bradley}, P.~A., {Cox}, A.~N., \& {Neuforge},
  C. 2000, \apjl, 542, L57

\bibitem[{{Hanes} {et~al.}(2019){Hanes}, {Waskie}, {Labadie-Bartz}, {Wall},
  {Boyer}, \& {McSwain}}]{Hanes2019}
{Hanes}, R.~J., {Waskie}, S., {Labadie-Bartz}, J.~M., {et~al.} 2019, \aj, 157,
  129

\bibitem[{{Hekker} \& {Christensen-Dalsgaard}(2017)}]{HekkerJCD2017}
{Hekker}, S. \& {Christensen-Dalsgaard}, J. 2017, \textit{Astron. Astrophys.
  Rev.}, 25, 1

\bibitem[{{Hunter} {et~al.}(2009){Hunter}, {Brott}, {Langer}, {Lennon},
  {Dufton}, {Howarth}, {Ryans}, {Trundle}, {Evans}, {de Koter}, \&
  {Smartt}}]{Hunter2009}
{Hunter}, I., {Brott}, I., {Langer}, N., {et~al.} 2009, \aap, 496, 841

\bibitem[{{Hunter} {et~al.}(2008){Hunter}, {Brott}, {Lennon}, {Langer},
  {Dufton}, {Trundle}, {Smartt}, {de Koter}, {Evans}, \& {Ryans}}]{Hunter2008}
{Hunter}, I., {Brott}, I., {Lennon}, D.~J., {et~al.} 2008, \apjl, 676, L29

\bibitem[{{Johnston} {et~al.}(2019){Johnston}, {Tkachenko}, {Aerts},
  {Molenberghs}, {Bowman}, {Pedersen}, {Buysschaert}, \&
  {P{\'a}pics}}]{Johnston2019}
{Johnston}, C., {Tkachenko}, A., {Aerts}, C., {et~al.} 2019, \mnras, 482, 1231

\bibitem[{{Kahraman Ali{\c{c}}avu{\textcommabelow s}} {et~al.}(2016){Kahraman
  Ali{\c{c}}avu{\textcommabelow s}}, {Niemczura}, {De Cat}, {Soydugan},
  {Ko{\l}aczkowski}, {Ostrowski}, {Telting}, {Uytterhoeven}, {Poretti},
  {Rainer}, {Su{\'a}rez}, {Mantegazza}, {Kilmartin}, \&
  {Pollard}}]{Kahraman2016}
{Kahraman Ali{\c{c}}avu{\textcommabelow s}}, F., {Niemczura}, E., {De Cat}, P.,
  {et~al.} 2016, \mnras, 458, 2307

\bibitem[{{Kahraman Ali{\c{c}}avu{\textcommabelow s}} {et~al.}(2020){Kahraman
  Ali{\c{c}}avu{\textcommabelow s}}, {Poretti}, {Catanzaro}, {Smalley},
  {Niemczura}, {Rainer}, \& {Handler}}]{Kahraman2020}
{Kahraman Ali{\c{c}}avu{\textcommabelow s}}, F., {Poretti}, E., {Catanzaro},
  G., {et~al.} 2020, \mnras, 493, 4518

\bibitem[{{K{\i}l{\i}{\c{c}}o{\u{g}}lu}
  {et~al.}(2016){K{\i}l{\i}{\c{c}}o{\u{g}}lu}, {Monier}, {Richer}, {Fossati},
  \& {Albayrak}}]{Kilicoglu2016}
{K{\i}l{\i}{\c{c}}o{\u{g}}lu}, T., {Monier}, R., {Richer}, J., {Fossati}, L.,
  \& {Albayrak}, B. 2016, \aj, 151, 49

\bibitem[{{Kobulnicky} {et~al.}(2014){Kobulnicky}, {Kiminki}, {Lundquist},
  {Burke}, {Chapman}, {Keller}, {Lester}, {Rolen}, {Topel}, {Bhattacharjee},
  {Smullen}, {Vargas {\'A}lvarez}, {Runnoe}, {Dale}, \&
  {Brotherton}}]{Kobulnicky2014}
{Kobulnicky}, H.~A., {Kiminki}, D.~C., {Lundquist}, M.~J., {et~al.} 2014,
  \apjs, 213, 34

\bibitem[{{Kovalev} {et~al.}(2019){Kovalev}, {Bergemann}, {Ting}, \&
  {Rix}}]{Kovalev2019}
{Kovalev}, M., {Bergemann}, M., {Ting}, Y.-S., \& {Rix}, H.-W. 2019, \aap, 628,
  A54

\bibitem[{{Kurtz} {et~al.}(2014){Kurtz}, {Saio}, {Takata}, {Shibahashi},
  {Murphy}, \& {Sekii}}]{Kurtz2014}
{Kurtz}, D.~W., {Saio}, H., {Takata}, M., {et~al.} 2014, \textit{MNRAS}, 444,
  102

\bibitem[{{Kurucz}(1992)}]{Kurucz1992}
{Kurucz}, R.~L. 1992, in The Stellar Populations of Galaxies, ed. B.~{Barbuy}
  \& A.~{Renzini}, Vol. 149, 225

\bibitem[{{Lampens} {et~al.}(2018){Lampens}, {Fr{\'e}mat}, {Vermeylen},
  {S{\'o}dor}, {Skarka}, {De Cat}, {Bogn{\'a}r}, {De Nutte}, {Dumortier},
  {Escorza}, {Oomen}, {Van de Steene}, {Kamath}, {Laverick}, {Samadi},
  {Triana}, \& {Lehmann}}]{Lampens2018}
{Lampens}, P., {Fr{\'e}mat}, Y., {Vermeylen}, L., {et~al.} 2018, \aap, 610, A17

\bibitem[{{Langer}(2012)}]{Langer2012}
{Langer}, N. 2012, \araa, 50, 107

\bibitem[{{Lanz} \& {Hubeny}(2007)}]{Lanz2007}
{Lanz}, T. \& {Hubeny}, I. 2007, \apjs, 169, 83

\bibitem[{{Lebreton} \& {Goupil}(2014)}]{Lebreton2014}
{Lebreton}, Y. \& {Goupil}, M.~J. 2014, \aap, 569, A21

\bibitem[{{Lee} \& {Saio}(1987)}]{LeeSaio1987}
{Lee}, U. \& {Saio}, H. 1987, \textit{MNRAS}, 224, 513

\bibitem[{{Lehmann} {et~al.}(2011){Lehmann}, {Tkachenko}, {Semaan},
  {Guti{\'e}rrez-Soto}, {Smalley}, {Briquet}, {Shulyak}, {Tsymbal}, \& {De
  Cat}}]{Lehmann2011}
{Lehmann}, H., {Tkachenko}, A., {Semaan}, T., {et~al.} 2011, \aap, 526, A124

\bibitem[{{Li} {et~al.}(2020){Li}, {Van Reeth}, {Bedding}, {Murphy}, {Antoci},
  {Ouazzani}, \& {Barbara}}]{Li2020}
{Li}, G., {Van Reeth}, T., {Bedding}, T.~R., {et~al.} 2020, \mnras, 491, 3586

\bibitem[{{Luck}(2017)}]{Luck2017}
{Luck}, R.~E. 2017, \aj, 153, 21

\bibitem[{{Lyubimkov} {et~al.}(2013){Lyubimkov}, {Lambert}, {Poklad},
  {Rachkovskaya}, \& {Rostopchin}}]{Lyubimkov2013}
{Lyubimkov}, L.~S., {Lambert}, D.~L., {Poklad}, D.~B., {Rachkovskaya}, T.~M.,
  \& {Rostopchin}, S.~I. 2013, \mnras, 428, 3497

\bibitem[{{Lyubimkov} {et~al.}(2005){Lyubimkov}, {Rostopchin}, {Rachkovskaya},
  {Poklad}, \& {Lambert}}]{Lyubimkov2005}
{Lyubimkov}, L.~S., {Rostopchin}, S.~I., {Rachkovskaya}, T.~M., {Poklad},
  D.~B., \& {Lambert}, D.~L. 2005, \mnras, 358, 193

\bibitem[{{Maeder} \& {Meynet}(2000)}]{Maeder2000}
{Maeder}, A. \& {Meynet}, G. 2000, \araa, 38, 143

\bibitem[{{Maeder} {et~al.}(2014){Maeder}, {Przybilla}, {Nieva}, {Georgy},
  {Meynet}, {Ekstr{\"o}m}, \& {Eggenberger}}]{Maeder2014}
{Maeder}, A., {Przybilla}, N., {Nieva}, M.-F., {et~al.} 2014, \aap, 565, A39

\bibitem[{{Martins} {et~al.}(2015){Martins}, {Herv{\'e}}, {Bouret},
  {Marcolino}, {Wade}, {Neiner}, {Alecian}, {Grunhut}, \&
  {Petit}}]{Martins2015}
{Martins}, F., {Herv{\'e}}, A., {Bouret}, J.~C., {et~al.} 2015, \aap, 575, A34

\bibitem[{{Mathis}(2009)}]{Mathis2009}
{Mathis}, S. 2009, \textit{Astron. Astrophys.}, 506, 811

\bibitem[{{Mazumdar} {et~al.}(2014){Mazumdar}, {Monteiro}, {Ballot}, {Antia},
  {Basu}, {Houdek}, {Mathur}, {Cunha}, {Silva Aguirre}, {Garc{\'\i}a},
  {Salabert}, {Verner}, {Christensen-Dalsgaard}, {Metcalfe}, {Sanderfer},
  {Seader}, {Smith}, \& {Chaplin}}]{Mazumdar2014}
{Mazumdar}, A., {Monteiro}, M.~J.~P.~F.~G., {Ballot}, J., {et~al.} 2014,
  \textit{Astrophys. J.}, 782, 18

\bibitem[{{Michaud} {et~al.}(2015){Michaud}, {Alecian}, \&
  {Richer}}]{Michaud2015}
{Michaud}, G., {Alecian}, G., \& {Richer}, J. 2015, {Atomic Diffusion in Stars}

\bibitem[{{Mombarg} {et~al.}(2020){Mombarg}, {Dotter}, {Van Reeth},
  {Tkachenko}, {Gebruers}, \& {Aerts}}]{Mombarg2020}
{Mombarg}, J. S.~G., {Dotter}, A., {Van Reeth}, T., {et~al.} 2020, \apj, 895,
  51

\bibitem[{{Mombarg} {et~al.}(2019){Mombarg}, {Van Reeth}, {Pedersen},
  {Molenberghs}, {Bowman}, {Johnston}, {Tkachenko}, \& {Aerts}}]{Mombarg2019}
{Mombarg}, J.~S.~G., {Van Reeth}, T., {Pedersen}, M.~G., {et~al.} 2019, \mnras,
  485, 3248

\bibitem[{{Monier}(2005)}]{Monier2005}
{Monier}, R. 2005, \aap, 442, 563

\bibitem[{{Montgomery} \& {O'Donoghue}(1999)}]{Montgomery1999}
{Montgomery}, M.~H. \& {O'Donoghue}, D. 1999, Delta Scuti Star Newsletter, 13,
  28

\bibitem[{{Moravveji} {et~al.}(2015){Moravveji}, {Aerts}, {P{\'a}pics},
  {Triana}, \& {Vandoren}}]{Moravveji2015}
{Moravveji}, E., {Aerts}, C., {P{\'a}pics}, P.~I., {Triana}, S.~A., \&
  {Vandoren}, B. 2015, \aap, 580, A27

\bibitem[{{Moravveji} {et~al.}(2016){Moravveji}, {Townsend}, {Aerts}, \&
  {Mathis}}]{Moravveji2016}
{Moravveji}, E., {Townsend}, R. H.~D., {Aerts}, C., \& {Mathis}, S. 2016, \apj,
  823, 130

\bibitem[{{Morel} {et~al.}(2008){Morel}, {Hubrig}, \& {Briquet}}]{Morel2008}
{Morel}, T., {Hubrig}, S., \& {Briquet}, M. 2008, \aap, 481, 453

\bibitem[{{Murphy} {et~al.}(2016){Murphy}, {Fossati}, {Bedding}, {Saio},
  {Kurtz}, {Grassitelli}, \& {Wang}}]{Murphy2016}
{Murphy}, S.~J., {Fossati}, L., {Bedding}, T.~R., {et~al.} 2016, \mnras, 459,
  1201

\bibitem[{{Murphy} {et~al.}(2018){Murphy}, {Moe}, {Kurtz}, {Bedding},
  {Shibahashi}, \& {Boffin}}]{Murphy2018}
{Murphy}, S.~J., {Moe}, M., {Kurtz}, D.~W., {et~al.} 2018, \mnras, 474, 4322

\bibitem[{{Neiner} {et~al.}(2012){Neiner}, {Floquet}, {Samadi}, {Espinosa
  Lara}, {Fr{\'e}mat}, {Mathis}, {Leroy}, {de Batz}, {Rainer}, {Poretti},
  {Mathias}, {Guarro Fl{\'o}}, {Buil}, {Ribeiro}, {Alecian}, {Andrade},
  {Briquet}, {Diago}, {Emilio}, {Fabregat}, {Guti{\'e}rrez-Soto}, {Hubert},
  {Janot-Pacheco}, {Martayan}, {Semaan}, {Suso}, \& {Zorec}}]{Neiner2012}
{Neiner}, C., {Floquet}, M., {Samadi}, R., {et~al.} 2012, \textit{Astron.
  Astrophys.}, 546, A47

\bibitem[{{Ness} {et~al.}(2015){Ness}, {Hogg}, {Rix}, {Ho}, \&
  {Zasowski}}]{Ness2015}
{Ness}, M., {Hogg}, D.~W., {Rix}, H.~W., {Ho}, A. Y.~Q., \& {Zasowski}, G.
  2015, \apj, 808, 16

\bibitem[{{Niemczura} {et~al.}(2009){Niemczura}, {Morel}, \&
  {Aerts}}]{Niemczura2009}
{Niemczura}, E., {Morel}, T., \& {Aerts}, C. 2009, \aap, 506, 213

\bibitem[{{Niemczura} {et~al.}(2015){Niemczura}, {Murphy}, {Smalley},
  {Uytterhoeven}, {Pigulski}, {Lehmann}, {Bowman}, {Catanzaro}, {van Aarle},
  {Bloemen}, {Briquet}, {De Cat}, {Drobek}, {Eyer}, {Gameiro}, {Gorlova},
  {Kami{\'n}ski}, {Lampens}, {Marcos-Arenal}, {P{\'a}pics}, {Vandenbussche},
  {Van Winckel}, {St{\c{e}}{\'s}licki}, \& {Fagas}}]{Niemczura2015}
{Niemczura}, E., {Murphy}, S.~J., {Smalley}, B., {et~al.} 2015, \mnras, 450,
  2764

\bibitem[{{Niemczura} {et~al.}(2017){Niemczura}, {Poli{\'n}ska}, {Murphy},
  {Smalley}, {Ko{\l}aczkowski}, {Jessen-Hansen}, {Uytterhoeven}, {Lykke},
  {Trivi{\~n}o Hage}, \& {Michalska}}]{Niemczura2017}
{Niemczura}, E., {Poli{\'n}ska}, M., {Murphy}, S.~J., {et~al.} 2017, \mnras,
  470, 2870

\bibitem[{{Nieva} \& {Przybilla}(2007)}]{Nieva2007}
{Nieva}, M.~F. \& {Przybilla}, N. 2007, \aap, 467, 295

\bibitem[{{Nieva} \& {Przybilla}(2012)}]{Nieva2012}
{Nieva}, M.~F. \& {Przybilla}, N. 2012, \aap, 539, A143

\bibitem[{{Nieva} \& {Sim{\'o}n-D{\'\i}az}(2011)}]{Nieva2011}
{Nieva}, M.~F. \& {Sim{\'o}n-D{\'\i}az}, S. 2011, \aap, 532, A2

\bibitem[{{Ouazzani} {et~al.}(2017){Ouazzani}, {Salmon}, {Antoci}, {Bedding},
  {Murphy}, \& {Roxburgh}}]{Ouazzani2017}
{Ouazzani}, R.-M., {Salmon}, S.~J.~A.~J., {Antoci}, V., {et~al.} 2017, \mnras,
  465, 2294

\bibitem[{{P{\'a}pics} {et~al.}(2012){P{\'a}pics}, {Briquet}, {Baglin},
  {Poretti}, {Aerts}, {Degroote}, {Tkachenko}, {Morel}, {Zima}, {Niemczura},
  {Rainer}, {Hareter}, {Baudin}, {Catala}, {Michel}, {Samadi}, \&
  {Auvergne}}]{Papics2012}
{P{\'a}pics}, P.~I., {Briquet}, M., {Baglin}, A., {et~al.} 2012,
  \textit{Astron. Astrophys.}, 542, A55

\bibitem[{{P{\'a}pics} {et~al.}(2015){P{\'a}pics}, {Tkachenko}, {Aerts}, {Van
  Reeth}, {De Smedt}, {Hillen}, {{\O}stensen}, \& {Moravveji}}]{Papics2015}
{P{\'a}pics}, P.~I., {Tkachenko}, A., {Aerts}, C., {et~al.} 2015, \apjl, 803,
  L25

\bibitem[{{P{\'a}pics} {et~al.}(2013){P{\'a}pics}, {Tkachenko, A.}, {Aerts,
  C.}, {Briquet, M.}, {Marcos-Arenal, P.}, {Beck, P. G.}, {Uytterhoeven, K.},
  {Trivi\~no Hage, A.}, {Southworth, J.}, {Clubb, K. I.}, {Bloemen, S.},
  {Degroote, P.}, {Jackiewicz, J.}, {McKeever, J.}, {Van Winckel, H.},
  {Niemczura, E.}, {Gameiro, J. F.}, \& {Debosscher, J.}}]{Papics2013}
{P{\'a}pics}, P.~I., {Tkachenko, A.}, {Aerts, C.}, {et~al.} 2013, A\&A, 553,
  A127

\bibitem[{{P{\'a}pics} {et~al.}(2017){P{\'a}pics}, {Tkachenko, A.}, {Van Reeth,
  T.}, {Aerts, C.}, {Moravveji, E.}, {Van de Sande, M.}, {De Smedt, K.},
  {Bloemen, S.}, {Southworth, J.}, {Debosscher, J.}, {Niemczura, E.}, \&
  {Gameiro, J. F.}}]{Papics2017}
{P{\'a}pics}, P.~I., {Tkachenko, A.}, {Van Reeth, T.}, {et~al.} 2017, A\&A,
  598, A74

\bibitem[{{Pedersen} {et~al.}(2021){Pedersen}, {Aerts}, {P\'apics},
  {Michielsen}, {Gebruers}, {Rogers}, {Molenberghs}, {Burssens}, {Garcia}, \&
  {Bowman}}]{Pedersen2021}
{Pedersen}, M.~G., {Aerts}, C., {P\'apics}, P.~I., {et~al.} 2021, \textit{Nat.
  Astron.}, in press

\bibitem[{{Pedersen} {et~al.}(2018){Pedersen}, {Aerts}, {P{\'a}pics}, \&
  {Rogers}}]{Pedersen2018}
{Pedersen}, M.~G., {Aerts}, C., {P{\'a}pics}, P.~I., \& {Rogers}, T.~M. 2018,
  \aap, 614, A128

\bibitem[{{Pedersen} {et~al.}(2020){Pedersen}, {Escorza}, {P{\'a}pics}, \&
  {Aerts}}]{Pedersen2020}
{Pedersen}, M.~G., {Escorza}, A., {P{\'a}pics}, P.~I., \& {Aerts}, C. 2020,
  \mnras, 495, 2738

\bibitem[{{Preston}(1974)}]{Preston1974}
{Preston}, G.~W. 1974, \araa, 12, 257

\bibitem[{{Przybilla} {et~al.}(2010){Przybilla}, {Firnstein}, {Nieva},
  {Meynet}, \& {Maeder}}]{Przybilla2010}
{Przybilla}, N., {Firnstein}, M., {Nieva}, M.~F., {Meynet}, G., \& {Maeder}, A.
  2010, \aap, 517, A38

\bibitem[{{Przybilla} {et~al.}(2008){Przybilla}, {Nieva}, \&
  {Butler}}]{Przybilla2008}
{Przybilla}, N., {Nieva}, M.-F., \& {Butler}, K. 2008, \apjl, 688, L103

\bibitem[{{Przybilla} {et~al.}(2011){Przybilla}, {Nieva}, \&
  {Butler}}]{Przybilla2011}
{Przybilla}, N., {Nieva}, M.-F., \& {Butler}, K. 2011, in Journal of Physics
  Conference Series, Vol. 328, Journal of Physics Conference Series, 012015

\bibitem[{{Qian} {et~al.}(2019){Qian}, {Li}, {He}, {Zhang}, {Zhu}, \&
  {Han}}]{Qian2019}
{Qian}, S.-B., {Li}, L.-J., {He}, J.-J., {et~al.} 2019, Research in Astronomy
  and Astrophysics, 19, 001

\bibitem[{{Raskin} {et~al.}(2011){Raskin}, {van Winckel}, {Hensberge},
  {Jorissen}, {Lehmann}, {Waelkens}, {Avila}, {de Cuyper}, {Degroote},
  {Dubosson}, {Dumortier}, {Fr{\'e}mat}, {Laux}, {Michaud}, {Morren}, {Perez
  Padilla}, {Pessemier}, {Prins}, {Smolders}, {van Eck}, \&
  {Winkler}}]{Raskin2011}
{Raskin}, G., {van Winckel}, H., {Hensberge}, H., {et~al.} 2011, \aap, 526, A69

\bibitem[{{Saffe} \& {Levato}(2014)}]{Saffe2014}
{Saffe}, C. \& {Levato}, H. 2014, \aap, 562, A128

\bibitem[{{Saio} {et~al.}(2015){Saio}, {Kurtz}, {Takata}, {Shibahashi},
  {Murphy}, {Sekii}, \& {Bedding}}]{Saio2015}
{Saio}, H., {Kurtz}, D.~W., {Takata}, M., {et~al.} 2015, \textit{MNRAS}, 447,
  3264

\bibitem[{{Salmon} {et~al.}(2014){Salmon}, {Montalb{\'a}n}, {Reese}, {Dupret},
  \& {Eggenberger}}]{Salmon2014}
{Salmon}, S.~J.~A.~J., {Montalb{\'a}n}, J., {Reese}, D.~R., {Dupret}, M.~A., \&
  {Eggenberger}, P. 2014, \aap, 569, A18

\bibitem[{{Shulyak} {et~al.}(2004){Shulyak}, {Tsymbal}, {Ryabchikova},
  {St{\"u}tz}, \& {Weiss}}]{Shulyak2004}
{Shulyak}, D., {Tsymbal}, V., {Ryabchikova}, T., {St{\"u}tz}, C., \& {Weiss},
  W.~W. 2004, \aap, 428, 993

\bibitem[{{Silva Aguirre} {et~al.}(2017){Silva Aguirre}, {Lund}, {Antia},
  {Ball}, {Basu}, {Christensen-Dalsgaard}, {Lebreton}, {Reese}, {Verma},
  {Casagrande}, {Justesen}, {Mosumgaard}, {Chaplin}, {Bedding}, {Davies},
  {Handberg}, {Houdek}, {Huber}, {Kjeldsen}, {Latham}, {White}, {Coelho},
  {Miglio}, \& {Rendle}}]{SilvaAguirre2017}
{Silva Aguirre}, V., {Lund}, M.~N., {Antia}, H.~M., {et~al.} 2017,
  \textit{Astrophys. J.}, 835, 173

\bibitem[{{Sim{\'o}n-D{\'\i}az}(2010)}]{SimonDiaz2010}
{Sim{\'o}n-D{\'\i}az}, S. 2010, \aap, 510, A22

\bibitem[{{Sim{\'o}n-D{\'\i}az} {et~al.}(2017){Sim{\'o}n-D{\'\i}az}, {Godart},
  {Castro}, {Herrero}, {Aerts}, {Puls}, {Telting}, \&
  {Grassitelli}}]{Simon-Diaz2017}
{Sim{\'o}n-D{\'\i}az}, S., {Godart}, M., {Castro}, N., {et~al.} 2017, \aap,
  597, A22

\bibitem[{{Sobol}(1967)}]{Sobol1967}
{Sobol}, I.~M. 1967, USSR Comp. Math. and Math. Phys., 7, 86

\bibitem[{{Szewczuk} \& {Daszy{\'n}ska-Daszkiewicz}(2017)}]{Szewczuk2017}
{Szewczuk}, W. \& {Daszy{\'n}ska-Daszkiewicz}, J. 2017, \mnras, 469, 13

\bibitem[{{Szewczuk} \& {Daszy{\'n}ska-Daszkiewicz}(2018)}]{Szewczuk2018}
{Szewczuk}, W. \& {Daszy{\'n}ska-Daszkiewicz}, J. 2018, \mnras, 478, 2243

\bibitem[{{Ting} {et~al.}(2019){Ting}, {Conroy}, {Rix}, \&
  {Cargile}}]{Ting2019}
{Ting}, Y.-S., {Conroy}, C., {Rix}, H.-W., \& {Cargile}, P. 2019, \apj, 879, 69

\bibitem[{{Tkachenko}(2015)}]{Tkachenko2015}
{Tkachenko}, A. 2015, \aap, 581, A129

\bibitem[{{Tkachenko} {et~al.}(2013){Tkachenko}, {Aerts}, {Yakushechkin},
  {Debosscher}, {Degroote}, {Bloemen}, {P{\'a}pics}, {de Vries}, {Lombaert},
  {Hrudkova}, {Fr{\'e}mat}, {Raskin}, \& {Van Winckel}}]{Tkachenko2013}
{Tkachenko}, A., {Aerts}, C., {Yakushechkin}, A., {et~al.} 2013, \aap, 556, A52

\bibitem[{{Tkachenko} {et~al.}(2012){Tkachenko}, {Lehmann}, {Smalley},
  {Debosscher}, \& {Aerts}}]{Tkachenko2012}
{Tkachenko}, A., {Lehmann}, H., {Smalley}, B., {Debosscher}, J., \& {Aerts}, C.
  2012, \mnras, 422, 2960

\bibitem[{{Tonry} \& {Davis}(1979)}]{Tonry1979}
{Tonry}, J. \& {Davis}, M. 1979, \aj, 84, 1511

\bibitem[{{Torres}(2010)}]{Torres2010b}
{Torres}, G. 2010, \aj, 140, 1158

\bibitem[{{Townsend}(2003)}]{Townsend2003}
{Townsend}, R.~H.~D. 2003, \textit{MNRAS}, 340, 1020

\bibitem[{{Tsymbal}(1996)}]{Tsymbal1996}
{Tsymbal}, V. 1996, in Astronomical Society of the Pacific Conference Series,
  Vol. 108, M.A.S.S., Model Atmospheres and Spectrum Synthesis, ed. S.~J.
  {Adelman}, F.~{Kupka}, \& W.~W. {Weiss}, 198

\bibitem[{{Van Reeth} {et~al.}(2018){Van Reeth}, {Mombarg}, {Mathis},
  {Tkachenko}, {Fuller}, {Bowman}, {Buysschaert}, {Johnston}, {Garc{\'\i}a
  Hern{\'a}ndez}, {Goldstein}, {Townsend}, \& {Aerts}}]{VanReeth2018}
{Van Reeth}, T., {Mombarg}, J.~S.~G., {Mathis}, S., {et~al.} 2018, \aap, 618,
  A24

\bibitem[{{Van Reeth} {et~al.}(2016){Van Reeth}, {Tkachenko}, \&
  {Aerts}}]{VanReeth2016}
{Van Reeth}, T., {Tkachenko}, A., \& {Aerts}, C. 2016, \aap, 593, A120

\bibitem[{{Van Reeth} {et~al.}(2015){Van Reeth}, {Tkachenko}, {Aerts},
  {P{\'a}pics}, {Triana}, {Zwintz}, {Degroote}, {Debosscher}, {Bloemen},
  {Schmid}, {De Smedt}, {Fremat}, {Fuentes}, {Homan}, {Hrudkova},
  {Karjalainen}, {Lombaert}, {Nemeth}, {{\O}stensen}, {Van De Steene}, {Vos},
  {Raskin}, \& {Van Winckel}}]{VanReeth2015}
{Van Reeth}, T., {Tkachenko}, A., {Aerts}, C., {et~al.} 2015, \apjs, 218, 27

\bibitem[{{Varenne} \& {Monier}(1999)}]{Varenne1999}
{Varenne}, O. \& {Monier}, R. 1999, \aap, 351, 247

\bibitem[{{Verma} {et~al.}(2019){Verma}, {Raodeo}, {Basu}, {Silva Aguirre},
  {Mazumdar}, {Mosumgaard}, {Lund}, \& {Ranadive}}]{Verma2019a}
{Verma}, K., {Raodeo}, K., {Basu}, S., {et~al.} 2019, \textit{MNRAS}, 483, 4678

\bibitem[{{Viani} {et~al.}(2018){Viani}, {Basu}, {Ong J.}, {Bonaca}, \&
  {Chaplin}}]{Viani2018}
{Viani}, L.~S., {Basu}, S., {Ong J.}, M.~J., {Bonaca}, A., \& {Chaplin}, W.~J.
  2018, \apj, 858, 28

\bibitem[{{Wu} \& {Li}(2019)}]{WuLi2019-HD50230}
{Wu}, T. \& {Li}, Y. 2019, \textit{Astrophys. J.}, 881, 86

\bibitem[{{Wu} {et~al.}(2020){Wu}, {Li}, {Deng}, {Lin}, {Song}, \&
  {Jiang}}]{Wu2020}
{Wu}, T., {Li}, Y., {Deng}, Z.-m., {et~al.} 2020, \apj, 899, 38

\bibitem[{{Xiang} {et~al.}(2019){Xiang}, {Ting}, {Rix}, {Sand ford}, {Buder},
  {Lind}, {Liu}, {Shi}, \& {Zhang}}]{Xiang2019}
{Xiang}, M., {Ting}, Y.-S., {Rix}, H.-W., {et~al.} 2019, \apjs, 245, 34

\bibitem[{{Zhang} {et~al.}(2018){Zhang}, {Liu}, {Wu}, {Luo}, {Zhang}, {Deng},
  {Fu}, {Zhang}, {Hou}, \& {Wang}}]{Zhang2018}
{Zhang}, C., {Liu}, C., {Wu}, Y., {et~al.} 2018, \apj, 854, 168

\bibitem[{{Zhao} {et~al.}(2012){Zhao}, {Zhao}, {Chu}, {Jing}, \&
  {Deng}}]{Zhao2012}
{Zhao}, G., {Zhao}, Y.-H., {Chu}, Y.-Q., {Jing}, Y.-P., \& {Deng}, L.-C. 2012,
  Research in Astronomy and Astrophysics, 12, 723

\bibitem[{{Zorec} \& {Royer}(2012)}]{Zorec2012}
{Zorec}, J. \& {Royer}, F. 2012, \aap, 537, A120

\end{thebibliography}

\begin{appendix}
\section{Observational logs}
Information about the observations of the $\gamma\,$Dor and SPB stars is given in Tables~\ref{tab:gd_log} and \ref{tab:SPB_log}, including S/N values and radial velocities.

\begin{center}
    \tablefirsthead{
        \hline\hline
        \multicolumn{1}{c}{KIC} & N & Observation times & S/N & RV \\
         &  &  &  & (km\,s$^{-1}$) \\
        \hline}
    \tablehead{
        \textbf{Continued.}\\
        \hline\hline
        \multicolumn{1}{c}{KIC} & N & Observation times & S/N & RV \\
         &  &  &  & (km\,s$^{-1}$) \\
        \hline}
    \topcaption{\label{tab:gd_log} Observation log of the $\gamma\,$Dor stars. N is the total number of spectra and S/N is the combined signal-to-noise ratio in the continuum region between 5440 and 5442 \AA. The last column contains the radial velocity of single stars or the type of spectroscopic binary.} 
    \renewcommand*{\arraystretch}{1.2}
    \begin{supertabular}[H]{r c c c c}
        2575161 & 3 & 2019 & 160 & $-9.3 \pm 0.8$ \\
        2710594 & 4 & 2011 & 80 & $-37 \pm 2$ \\ 
        2846358 & 4 & 2017 & 70 & $2.7 \pm 0.8$ \\
        3331147 & 3 & 2019 & 115 & $-20.0 \pm 0.9$ \\
        3448365 & 4 & 2011, 2014 & 205 & $-30 \pm 2$ \\
        3626325 & 3 & 2019 & 65 & $-24 \pm 1$ \\
        3648936 & 3 & 2019 & 110 & $2 \pm 2$ \\
        3744571 & 4 & 2013 & 35 & $-23 \pm 1$ \\
        3942392 & 3 & 2012 & 170 & $-28.8 \pm 0.9$ \\
        4255166 & 3 & 2019 & 80 & $-25.6 \pm 0.8$ \\
        4567531 & 3 & 2019 & 85 & $-29 \pm 2$ \\
        4659837 & 2 & 2019 & 40 & SB1 \\
        4757184 & 4 & 2011 & 60 & $-24.7 \pm 0.8$ \\
        4846809 & 3 & 2013 & 25 & $0 \pm 1$ \\ 
        5018590 & 4 & 2012 & 150 & $-19 \pm 2$ \\
        5113797 & 2 & 2011 & 170 & $15 \pm 3$ \\
        5114382 & 4 & 2011 & 40 & $15 \pm 2$ \\ 
        5450503 & 3 & 2019 & 115 & $-23 \pm 2$ \\
        5522154 & 3 & 2011 & 105 & $-24 \pm 3$ \\
        5646058 & 3 & 2019 & 55 & $-10 \pm 2$ \\
        5708550 & 4 & 2011 & 75 & $-15 \pm 2$ \\ 
        5788623 & 4 & 2013 & 45 & $-4 \pm 2$ \\ 
        5887983 & 3 & 2019 & 60 & $-9 \pm 3$ \\
        5954264 & 2 & 2010 & 185 & $-33.0 \pm 0.8$ \\
        6064932 & 2 & 2019 & 60 & $-3.5 \pm 0.8$ \\
        6131093 & 4 & 2017 & 110 & $5.6 \pm 0.7$ \\
        6292398 & 8 & 2011-2020 & 105 & SB1\tablefootmark{a} \\
        6301745 & 3 & 2019 & 100 & $-39 \pm 4$ \\
        6468146 & 7 & 2011, 2014, 2019 & 225 & SB1\tablefootmark{b} \\ 
        6468987 & 4 & 2013, 2014 & 30 & $-27.6 \pm 0.7$ \\ 
        6519869 & 2 & 2011 & 85 & $2.5 \pm 0.3$ \\
        6678174 & 4 & 2011 & 60 & $-13.6 \pm 0.9$ \\ 
        6935014 & 5 & 2011 & 80 & $-9 \pm 1$ \\ 
        6953103 & 4 & 2013, 2014 & 30 & $7 \pm 2$ \\ 
        7023122 & 4 & 2011 & 100 & $-18.1 \pm 0.8$ \\ 
        7215607 & 2 & 2019 & 90 & $7.4 \pm 0.3$ \\
        7365537 & 3 & 2011 & 170 & $-30 \pm 2$ \\ 
        7380501 & 4 & 2011 & 40 & $-6 \pm 1$ \\
        7434470 & 4 & 2013 & 35 & SB1\tablefootmark{c} \\ 
        7583663 & 3 & 2013, 2014 & 30 & $-29 \pm 4$ \\ 
        7661054 & 4 & 2012 & 95 & $-17.8 \pm 0.2$ \\
        7694191 & 4 & 2019 & 80 & SB1 \\
        7939065 & 4 & 2013 & 45 & $-22 \pm 3$ \\ 
        8123127 & 3 & 2019 & 60 & $2 \pm 2$ \\
        8197761 & 13 & 2015 & 100 & SB1\tablefootmark{d} \\
        8355130 & 2 & 2010 & 105 & $9.4 \pm 0.6$ \\
        8364249 & 4 & 2011 & 50 & $-23 \pm 4$ \\
        8375138 & 4 & 2011 & 90 & $-4 \pm 2$ \\ 
        8651452 & 3 & 2019 & 80 & $-16 \pm 3$ \\
        8871304 & 16 & 2010 & 195 & SB1 \\
        9210943 & 5 & 2011, 2014 & 80 & $-28 \pm 3$ \\
        9419182 & 4 & 2011, 2019 & 115 & $0 \pm 2$ \\
        9480469 & 4 & 2013, 2014 & 30 & $-18 \pm 3$ \\ 
        9490067 & 2 & 2017 & 40 & $-0.6 \pm 0.2$ \\
        9595743 & 4 & 2013 & 50 & $-9 \pm 1$ \\ 
        9716358 & 4 & 2009 & 60 & $6.1 \pm 0.4$ \\
        9962653 & 19 & 2017, 2019 & 215 & SB1 \\
        10224094 & 4 & 2011 & 50 & $-7.2 \pm 0.5$ \\
        10256787 & 13 & 2012, 2013, 2014 & 50 & $6 \pm 1$ \\
        10317467 & 3 & 2019 & 110 & $6.9 \pm 0.4$ \\
        10467146 & 4 & 2013, 2014 & 35 & SB1\tablefootmark{b} \\ 
        10470294 & 3 & 2019 & 70 & $-17 \pm 7$ \\
        11080103 & 4 & 2013, 2014 & 25 & $4 \pm 1$ \\ 
        11099031 & 4 & 2011 & 155 & $-17 \pm 1$ \\ 
        11294808 & 4 & 2011 & 55 & $-9 \pm 2$ \\
        11456474 & 4 & 2013, 2014 & 25 & $-11 \pm 2$ \\ 
        11607017 & 3 & 2019 & 40 & $-26 \pm 4$ \\
        11612274 & 2 & 2018 & 90 & $1 \pm 2$ \\
        11721304 & 4 & 2011 & 45 & $1.1 \pm 0.7$ \\ 
        11826272 & 11 & 2011, 2015, 2019 & 130 & $-9.6 \pm 0.6$ \\ 
        11907454 & 4 & 2011 & 70 & SB1 \\ 
        11917550 & 4 & 2011 & 75 & $10 \pm 2$ \\ 
        11920505 & 13 & 2011, 2016, 2019 & 245 & $-25 \pm 1$ \\ 
        12066947 & 3 & 2011 & 115 & $9 \pm 2$ \\
        12117689 & 3 & 2019 & 95 & $-24 \pm 6$ \\
        12458189 & 4 & 2011 & 80 & $-35 \pm 1$ \\
        12643786 & 4 & 2011 & 55 & $1 \pm 2$ \\
        \hline 
        \multicolumn{5}{c}{$\gamma$ Dor- $\delta$ Sct hybrids} \\
        \hline
        2168333 & 8 & 2014-2020 & 150 & $-7 \pm 6$ \\
        3241199 & 3 & 2019 & 75 & $-6 \pm 11$ \\
        5294571 & 2 & 2017 & 50 & $-15 \pm 10$ \\
        5608334 & 3 & 2012 & 125 & $-17 \pm 6$ \\
        7106648 & 4 & 2018 & 135 & $-12 \pm 4$ \\
        7748238 & 6 & 2014 & 195 & $-23 \pm 2$ \\
        7770282 & 27 & 2011-2019 & 240 & $7 \pm 1$ \\
        7977996 & 4 & 2019, 2020 & 55 & $13 \pm 11$ \\
        8645874 & 5 & 2011, 2019 & 125 & SB1 \\ 
        8836473 & 4 & 2013, 2014 & 25 & $-26 \pm 1$ \\ 
        9651065 & 2 & 2019 & 50 & $-22 \pm 1$ \\
        9751996 & 8 & 2011, 2019 & 50 & $-51.7 \pm 0.2$ \\ 
        11754232 & 12 & 2013, 2014 & 50 & SB1\tablefootmark{b} \\ 
        12365420 & 5 & 2019, 2020 & 35 & $7 \pm 4$ \\
        \hline
        \multicolumn{5}{c}{SB2, SB3 or higher-order binaries}  \\
        \hline
        3222854 & 10 & 2013, 2014, 2015 & & SB2 \tablefootmark{b} \\
        4480321 & 58 & 2011-2020 & & SB3\tablefootmark{e} (hybrid) \\
        5219533 & 50 & 2010-2019 & & SB3\tablefootmark{e} (hybrid)\\
        6467639 & 12 & 2013, 2014 & & SB3\tablefootmark{b} (hybrid)\\
        6764812 & 3 & 2019 & & SB2 \\
        6778063 & 9 & 2013, 2014 & & SB3\tablefootmark{b} (hybrid)\\
        7385783 & 3 & 2019 & & SB2 \\
        8324305 & 3 & 2011, 2019, 2020 & & SB2 \\
        8975515 & 32 & 2010-2020 & & SB2\tablefootmark{e}  \\
        10080943 & 25 & 2011, 2013, 2014 & & SB2\tablefootmark{f} (hybrid)\\
        \hline
        \multicolumn{5}{c}{low S/N spectra}  \\
        \hline
        5254203 & 4 & 2013, 2014 & 15 & \\
        7746984 & 2 & 2014 & 20 & \\
        9533489 & 3 & 2013, 2014 & 20 & \\
        \hline
    \end{supertabular}
    \tablebib{
    (a) \citet{Niemczura2015}; (b) \citet{VanReeth2015}; (c) \citet{Li2020}; (d) \citet{Murphy2018}; (e) \citet{Lampens2018}; (f) \citet{Tkachenko2013}.}
\end{center}

\renewcommand*{\arraystretch}{1.2}
\begin{table}[ht]
    \caption{Observation log of the SPB stars. N is the total number of spectra and S/N is the combined signal-to-noise ratio in the continuum region between 4960 and 4962 \AA. The last column contains the radial velocity of single stars or the type of spectroscopic binary.}
    \label{tab:SPB_log}
    \centering
    \begin{tabular}{r c c c c}
        \hline \hline
        \multicolumn{1}{c}{KIC} & N & Observation times & S/N & RV \\
         &  &  &  & (km\,s$^{-1}$) \\
        \hline
        3240411 & 6 & 2012, 2016, 2018 & 145 & $-37 \pm 8$ \\
        3756031 & 5 & 2012, 2016 & 90 & $11 \pm 4$ \\
        3839930 & 3 & 2010 & 175 & $-29 \pm 1$ \\
        3865742 & 3 & 2019 & 55 & $-12 \pm 12$ \\
        5941844 & 1 & 2016 & 110 & $-11 \pm 2$ \\
        6462033 & 1 & 2019 & 65 & $3.7 \pm 0.7$ \\
        6780397 & 10 & 2012 & 135 & SB1 \\
        7760680 & 4 & 2012, 2014 & 140 & $-50 \pm 2$ \\
        8057661 & 6 & 2019 & 65 & $-36 \pm 10$ \\
        8087269 & 4 & 2013 & 50 & $96 \pm 46$ \\
        8381949 & 3 & 2010 & 100 & $-111 \pm 10$ \\
        8459899 & 122 & 2010 & 720 & $-10 \pm 3$ \\
        8714886 & 3 & 2010 & 90 & $25.0 \pm 0.8$ \\
        8766405 & 2 & 2010 & 180 & $43 \pm 17$ \\
        9964614 & 6 & 2010, 2016, 2017 & 80 & $-137 \pm 4$ \\
        10285114 & 4 & 2009, 2012 & 135 & $37 \pm 19$ \\
        10536147 & 5 & 2019 & 70 & $-18 \pm 11$ \\
        11360704 & 3 & 2010, 2012 & 105 & $-24 \pm 13$ \\
        11971405 & 5 & 2010, 2015 & 245 & $-27 \pm 7$ \\
        12258330 & 4 & 2010, 2017, 2018 & 170 & $-37 \pm 7$ \\
        \hline
        \multicolumn{5}{c}{SB2} \\
        \hline
        4930889 & 27 & 2010-2015 & & SB2\tablefootmark{a} \\
        6352430 & 29 & 2010, 2011, 2012 & & SB2\tablefootmark{b} \\
        \hline
        \multicolumn{5}{c}{low S/N spectra}  \\
        \hline
        9715425 & 3 & 2019 & 20 & \\
        \hline
    \end{tabular}
    \tablebib{
    (a) \citet{Papics2017}; (b) \citet{Papics2013}.
    }
\end{table}

\section{Test of the performance of the neural network}
\subsection{Size of the training sample}

\begin{figure}[H]
    \centering
    \includegraphics[width=\hsize]{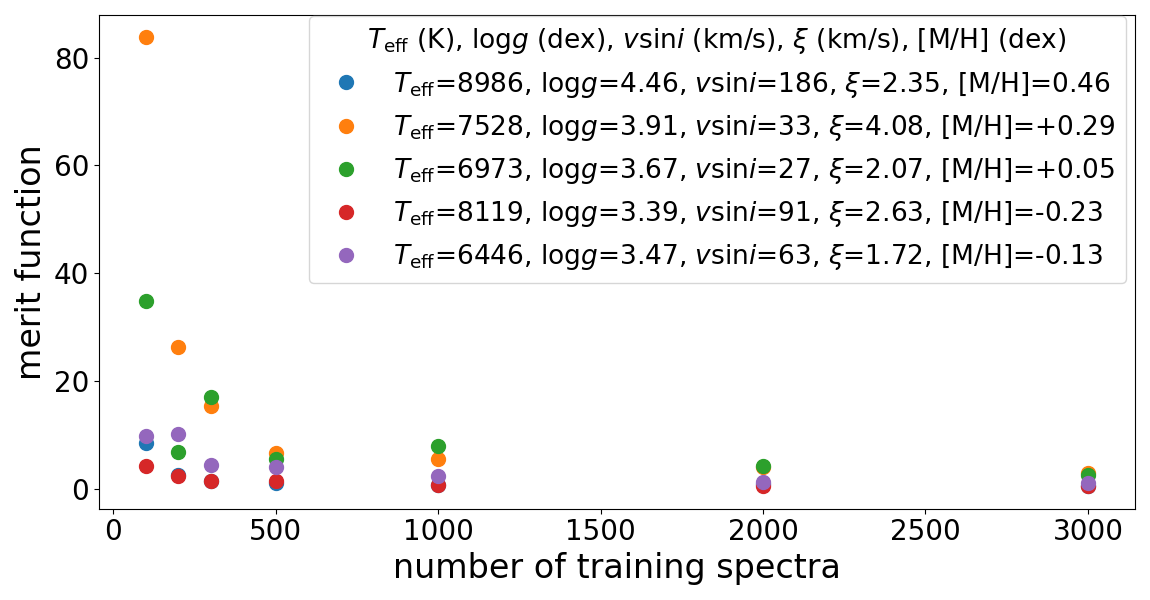}
    \caption{Difference between the spectrum predicted by the NN and the
    synthetic spectrum from GSSP as a function of the number of training spectra for $\gamma\,$Dor stars.}
    \label{fig:size_sample_gDor}
\end{figure}

\begin{figure}[H]
    \centering
    \includegraphics[width=\hsize]{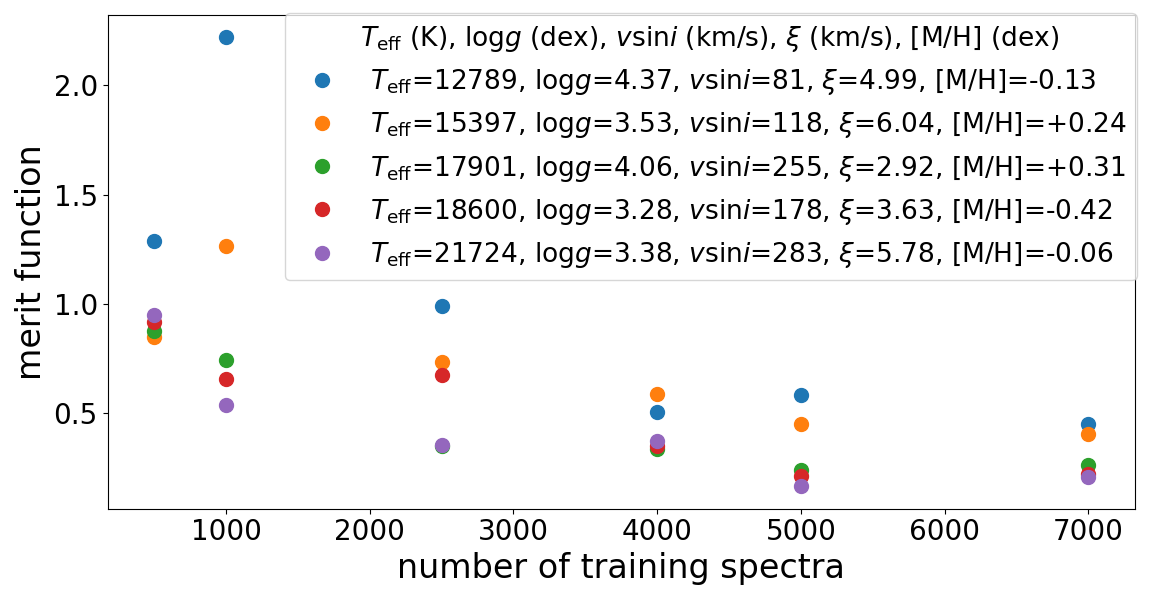}
    \caption{Difference between the spectrum predicted by the NN and the synthetic spectrum from GSSP as function of the number of training spectra for SPB stars.}
    \label{fig:size_sample_SPB}
\end{figure}

In Figs.~\ref{fig:size_sample_gDor} and \ref{fig:size_sample_SPB} we plot the difference between spectra predicted with the NN and synthetic spectra from GSSP for the same stellar parameters in function of the number of spectra used to train the NN. The start of the plateau indicates the optimal training size, which is 1000 and 5000 for $\gamma\,$Dor and SPB stars respectively.

\subsection{Comparison between three methods for parameter determination}

Table~\ref{tab:testNN} summarises the results of the NN performance tests. It contains parameter values for ten synthetic spectra, four with parameters specific for SPB stars and six with typical $\gamma\,$Dor parameters. For each synthetic spectrum we report the real values and values obtained with the three methods described in Sect.~\ref{sec:performance}. Parameter values that deviate from the real value by more than the estimated uncertainties are shown in bold. But for most of them the difference is minimal.

\longtab[1]{
\renewcommand*{\arraystretch}{1.3}
\begin{longtable}{llllll}
    \caption{Results of the NN performance tests. For ten synthetic spectra the real stellar parameter values are given together with the results from methods 1), 2) and 3) (see Sect.~\ref{sec:performance} for more information). Values that deviate from the actual value with (little) more than the uncertainty are marked in bold. The errors are approximately symmetric except when a different upper and lower bound are given.}
    \label{tab:testNN} \\
    \hline\hline
    Method & $T_{\rm{eff}}$ & $\log{g}$  & [M/H] & $v\sin{i}$ & $\xi$ \\
    & (K) & (dex) & (dex) & (km s$^{-1}$) & (km s$^{-1}$)  \\
    \hline
    \endfirsthead
    \caption{continued.} \\
    \hline\hline
    Method & $T_{\rm{eff}}$ & $\log{g}$  & [M/H] & $v\sin{i}$ & $\xi$ \\
    & (K) & (dex) & (dex) & (km s$^{-1}$) & (km s$^{-1}$)  \\
    \hline
    \endhead
    \hline
    \endfoot
    Real & 16500 & 4.0 & -0.2 & 70 & 4.0 \\
    method 1 & \bf{16728 $\pm$ 130} & 4.02 $\pm$ 0.03 & -0.19 $\pm$ 0.06 & 72 $\pm$ 4 & 5 $\pm$ 1 \\
    method 2 & 16521 $\pm$ 108 & 4.00 $\pm$ 0.02 & -0.21 $\pm$ 0.02 & 69.8 $\pm$ 1.9 & 3.9 $\pm$ 0.5 \\
    method 3 & 16424$^{+139}_{-136}$ & 4.00 $\pm$ 0.04 & $\mathbf{-0.09^{+0.05}_{-0.06}}$ & 71 $\pm$ 3 & 3.3$^{+0.8}_{-0.7}$ \\
    \hline
    Real & 16500 & 4.0 & -0.2 & 170 & 4.0 \\
    method 1 & \bf{16631 $\pm$ 110} & 4.00 $\pm$ 0.03 & -0.22 $\pm$ 0.04 & 174 $\pm$ 7 & 4.6 $\pm$ 0.8 \\
    method 2 & 16622$^{+142}_{-156}$ & 4.00 $\pm$ 0.04 & -0.24$^{+0.06}_{-0.07}$ & 170 $\pm$ 5 & 4.0$^{+1.3}_{-1.1}$ \\
    method 3 & 16574$^{+149}_{-170}$ & 3.96 $\pm$ 0.05 & -0.20 $\pm$ 0.06 & 171 $\pm$ 6 & 5.0$^{+1.0}_{-1.1}$ \\
\end{longtable}
}

\section{Stellar parameters and surface abundances of the gamma Dor stars}
The stellar parameters of the $\gamma$ Dor stars are listed in Table~\ref{tab:gd_params} while the surface abundances can be found in Tables~\ref{tab:gd_abund} and \ref{tab:gd_abund2}. 

\longtab[1]{
\renewcommand*{\arraystretch}{1.4}
\begin{longtable}{l|lllll|lll|l}
    \caption{Stellar parameters of the $\gamma\,$Dor stars. The last column contains the luminosities computed from V-band magnitudes or, when indicated, using G-band (bold) or Kepler (italic) magnitudes.}
    \label{tab:gd_params} \\
    \hline\hline
    KIC & $T_{\rm{eff, NN}}$ & $\log{g}_{\rm{NN}}$ & [M/H]$_{\rm{NN}}$ & $v\sin{i}_{\rm{NN}}$ & $\xi_{\rm{NN}}$ & $T_{\rm{eff, GSSP}}$ & $\log{g}_{\rm{GSSP}}$ & $\xi_{\rm{GSSP}}$ & $\log(L/L_\sun)$ \\
    & (K) & (dex) & (dex) & (km s$^{-1}$) & (km s$^{-1}$) & (K) & (dex) & (km s$^{-1}$) &  \\
    \hline
    \endfirsthead
    \caption{continued.} \\
    \hline\hline
    KIC & $T_{\rm{eff, NN}}$ & $\log{g}_{\rm{NN}}$ & [M/H]$_{\rm{NN}}$ & $v\sin{i}_{\rm{NN}}$ & $\xi_{\rm{NN}}$ & $T_{\rm{eff, GSSP}}$ & $\log{g}_{\rm{GSSP}}$ & $\xi_{\rm{GSSP}}$ & $\log(L/L_\sun)$ \\
    & (K) & (dex) & (dex) & (km s$^{-1}$) & (km s$^{-1}$) & (K) & (dex) & (km s$^{-1}$) &  \\
    \hline
    \endhead
    \hline
    \endfoot
    2575161 & 7019 $\pm$ 35 & 3.76 $\pm$ 0.10 & -0.23 $\pm$ 0.04 & 78 $\pm$ 3 & 2.69 $\pm$ 0.19 & 7011$^{+47}_{-49}$ & 3.76$^{+0.16}_{-0.16}$ & 2.64$^{+0.15}_{-0.15}$ & 0.93$^{+0.02}_{-0.02}$ \\
    2710594 & 6937 $\pm$ 40 & 3.41 $\pm$ 0.18 & -0.27 $\pm$ 0.04 & 76 $\pm$ 2 & 3.06 $\pm$ 0.19 & 6916$^{+54}_{-54}$ & 3.41$^{+0.23}_{-0.22}$ & 3.19$^{+0.26}_{-0.23}$ & 0.893$^{+0.008}_{-0.006}$ \\
    2846358 & 6880 $\pm$ 48 & 4.60 $\pm$ 0.11 & -0.18 $\pm$ 0.04 & 37 $\pm$ 2 & 1.87 $\pm$ 0.23 & 6797$^{+74}_{-76}$ & 4.47$^{+0.19}_{-0.21}$ & 1.63$^{+0.23}_{-0.25}$ & 0.73$^{+0.13}_{-0.09}$ \\
    3331147 & 7130 $\pm$ 32 & 4.07 $\pm$ 0.10 & -0.16 $\pm$ 0.03 & 62 $\pm$ 2 & 2.75 $\pm$ 0.16 & 7123$^{+48}_{-49}$ & 3.96$^{+0.17}_{-0.17}$ & 2.83$^{+0.17}_{-0.15}$ & 0.660$^{+0.004}_{-0.006}$ \\
    3448365 & 7065 $\pm$ 30 & 3.98 $\pm$ 0.08 & -0.07 $\pm$ 0.04 & 86 $\pm$ 2 & 2.96 $\pm$ 0.17 & 7079$^{+42}_{-41}$ & 4.11$^{+0.13}_{-0.15}$ & 3.01$^{+0.12}_{-0.13}$ & 0.767$^{+0.002}_{-0.003}$ \\
    3626325 & 7036 $\pm$ 49 & 3.94 $\pm$ 0.13 & -0.11 $\pm$ 0.05 & 51 $\pm$ 2 & 2.78 $\pm$ 0.24 & 7006$^{+61}_{-61}$ & 4.01$^{+0.19}_{-0.22}$ & 2.71$^{+0.20}_{-0.20}$ & 0.763$^{+0.006}_{-0.005}$ \\
    3648936 & 6870 $\pm$ 40 & 3.20 $\pm$ 0.12 & -0.35 $\pm$ 0.04 & 85 $\pm$ 3 & 3.34 $\pm$ 0.16 & 6791$^{+40}_{-39}$ & 3.07$^{+0.15}_{-0.17}$ & 3.48$^{+0.18}_{-0.18}$ & 1.066$^{+0.003}_{-0.003}$ \\
    3744571 & 6987 $\pm$ 63 & 3.40 $\pm$ 0.21 & -0.36 $\pm$ 0.06 & 51 $\pm$ 3 & 2.98 $\pm$ 0.31 & 6938$^{+82}_{-84}$ & 3.52$^{+0.30}_{-0.32}$ & 3.03$^{+0.43}_{-0.33}$ & $\mathbf{0.912^{+0.007}_{-0.006}}$\\
\end{longtable}
}

\longtab[2]{
\begin{landscape}
\renewcommand{\arraystretch}{1.35} 
\begin{longtable}{llllllllllll}
    \caption{\label{tab:gd_abund} Surface abundances of the $\gamma\,$Dor stars with respect to solar values $\left(\rm{[X/H]} = \log_{10}\left(\frac{n_{\rm{x}}}{n_{\rm{tot}}}\right) - \log_{10}\left(\frac{n_{\rm{x}}}{n_{\rm{tot}}}\right)_{\sun}\right)$. The second row in the header contains for the most important elements the mean difference between the NN abundances and the abundances from manual normalisation and its standard deviation for the 37 stars in the sample by \citet{VanReeth2015} (see Sect.~\ref{sec:results_gDor} for more information).} \\
    \hline\hline
    KIC & C & O & Na & Mg & Al & Si & S & Ca & Sc & Ti & V \\
    $\overline{NN-\rm{man}}$ & +0.05 $\pm$ 0.09 & & -0.05 $\pm$ 0.12 & -0.05 $\pm$ 0.10 & & & -0.44$\pm$ 0.06 & -0.22$\pm$ 0.09 & -0.31 $\pm$ 0.17 & -0.16 $\pm$ 0.16 &  \\
    \hline
    \endfirsthead
    \caption{continued.} \\
    \hline\hline
    KIC & C & O & Na & Mg & Al & Si & S & Ca & Sc & Ti & V  \\
    \hline
    \endhead
    \hline
    \endfoot
    2575161 & $+0.08^{+0.11}_{-0.13}$ & $+0.43^{+0.17}_{-0.23}$ & $-0.11^{+0.11}_{-0.12}$ & $+0.09^{+0.11}_{-0.11}$ & $< -0.13$ & $< -0.12$ & $< -0.07$ & $+0.03^{+0.15}_{-0.16}$ & $-0.18^{+0.15}_{-0.15}$ & $-0.18^{+0.10}_{-0.12}$ & $-0.17^{+0.25}_{-0.43}$ \\
    2710594 & $-0.20^{+0.15}_{-0.17}$ & $< +0.07$ & $-0.43^{+0.17}_{-0.21}$ & $+0.09^{+0.11}_{-0.11}$ & $< +0.27$ & $< +0.19$ & $< -0.14$ & $-0.23^{+0.18}_{-0.19}$ & $-0.32^{+0.16}_{-0.17}$ & $-0.25^{+0.13}_{-0.13}$ & $-0.42^{+0.33}_{-0.97}$ \\
    2846358 & $-0.06^{+0.19}_{-0.25}$ & $< +0.46$ & $-0.26^{+0.16}_{-0.17}$ & $-0.19^{+0.10}_{-0.10}$ & $< +0.44$ & $< +0.18$ & $< +0.14$ & $-0.19^{+0.20}_{-0.22}$ & $-0.06^{+0.23}_{-0.25}$ & $-0.16^{+0.17}_{-0.15}$ & $-0.16^{+0.39}_{-0.78}$ \\
    3331147 & $-0.04^{+0.12}_{-0.13}$ & $+0.36^{+0.16}_{-0.23}$ & $-0.08^{+0.13}_{-0.15}$ & $+0.19^{+0.10}_{-0.10}$ & $< -0.18$ & $< +0.12$ & $+0.01^{+0.23}_{-0.36}$ & $-0.04^{+0.15}_{-0.16}$ & $+0.01^{+0.14}_{-0.14}$ & $-0.20^{+0.13}_{-0.13}$ & $-0.07^{+0.22}_{-0.35}$ \\
    3448365 & $+0.24^{+0.09}_{-0.09}$ & $+0.37^{+0.21}_{-0.33}$ & $+0.05^{+0.11}_{-0.12}$ & $+0.18^{+0.10}_{-0.10}$ & $< -0.03$ & $< +0.21$ & $+0.08^{+0.20}_{-0.32}$ & $-0.01^{+0.14}_{-0.15}$ & $+0.24^{+0.13}_{-0.13}$ & $+0.06^{+0.10}_{-0.10}$ & $-0.03^{+0.21}_{-0.31}$ \\
    3626325 & $+0.04^{+0.14}_{-0.16}$ & $+0.44^{+0.18}_{-0.25}$ & $+0.03^{+0.14}_{-0.15}$ & $+0.20^{+0.10}_{-0.10}$ & $< -0.05$ & $< +0.20$ & $-0.06^{+0.27}_{-0.71}$ & $+0.05^{+0.16}_{-0.17}$ & $+0.08^{+0.16}_{-0.17}$ & $-0.10^{+0.14}_{-0.15}$ & $-0.09^{+0.28}_{-0.49}$ \\
    3648936 & $-0.39^{+0.14}_{-0.17}$ & $+0.24^{+0.19}_{-0.27}$ & $-0.06^{+0.12}_{-0.13}$ & $+0.06^{+0.12}_{-0.13}$ & $< -0.37$ & $< +0.04$ & $-0.11^{+0.22}_{-0.35}$ & $-0.19^{+0.15}_{-0.15}$ & $-0.20^{+0.12}_{-0.13}$ & $-0.37^{+0.10}_{-0.11}$ & $-0.28^{+0.20}_{-0.29}$ \\
\end{longtable}
\end{landscape}
}

\longtab[3]{
\begin{landscape}
\renewcommand{\arraystretch}{1.35} 
\begin{longtable}{llllllllllll}
    \caption{\label{tab:gd_abund2} Surface abundances of the $\gamma\,$Dor stars with respect to solar values $\left(\rm{[X/H]} = \log_{10}\left(\frac{n_{\rm{x}}}{n_{\rm{tot}}}\right) - \log_{10}\left(\frac{n_{\rm{x}}}{n_{\rm{tot}}}\right)_{\sun}\right)$. The second row in the header contains for the most important elements the mean difference between the NN abundances and the abundances from manual normalisation and its standard deviation for the 37 stars in the sample by \citet{VanReeth2015} (see Sect.~\ref{sec:results_gDor} for more information).} \\
    \hline\hline
    KIC & Cr & Mn & Fe & Co & Ni & Cu & Zn & Sr & Y & Zr & Ba \\
    $\overline{NN-man}$ & -0.28 $\pm$ 0.14 & -0.73 $\pm$ 0.39 & -0.28 $\pm$ 0.16 & & -0.13 $\pm$ 0.15 & & & & -0.40 $\pm$ 0.20 & & -0.80 $\pm$ 0.33 \\
    \hline
    \endfirsthead
    \caption{continued.} \\
    \hline\hline
    KIC & Cr & Mn & Fe & Co & Ni & Cu & Zn & Sr & Y & Zr & Ba \\
    \hline
    \endhead
    \hline
    \endfoot
    2575161 & $-0.22^{+0.14}_{-0.15}$ & $-0.21^{+0.14}_{-0.16}$ & $-0.27^{+0.07}_{-0.07}$ & $+0.06^{+0.26}_{-0.44}$ & $-0.20^{+0.12}_{-0.14}$ & $< -0.10$ & $< -0.54$ & $-0.26^{+0.29}_{-0.42}$ & $-0.45^{+0.22}_{-0.26}$ & $-0.30^{+0.35}_{-0.85}$ & $+0.13^{+0.17}_{-0.18}$ \\
    2710594 & $-0.28^{+0.16}_{-0.17}$ & $-0.49^{+0.18}_{-0.21}$ & $-0.30^{+0.09}_{-0.09}$ & $< -0.07$ & $-0.35^{+0.17}_{-0.18}$ & $< -0.14$ & $-0.77^{+0.44}_{-1.33}$ & $-0.36^{+0.42}_{-0.61}$ & $-0.45^{+0.22}_{-0.25}$ & $-0.33^{+0.34}_{-0.70}$ & $+0.08^{+0.18}_{-0.19}$ \\
    2846358 & $-0.23^{+0.18}_{-0.20}$ & $-0.38^{+0.24}_{-0.28}$ & $-0.27^{+0.10}_{-0.11}$ & $-0.19^{+0.41}_{-1.86}$ & $-0.32^{+0.18}_{-0.21}$ & $< -0.06$ & $-0.62^{+0.55}_{-0.91}$ & $+0.09^{+0.25}_{-0.33}$ & $-0.20^{+0.36}_{-0.44}$ & $< +0.44$ & $+0.44^{+0.15}_{-0.17}$ \\
    3331147 & $-0.17^{+0.15}_{-0.15}$ & $-0.31^{+0.14}_{-0.16}$ & $-0.25^{+0.09}_{-0.09}$ & $< +0.05$ & $-0.29^{+0.16}_{-0.17}$ & $< -0.01$ & $-0.58^{+0.36}_{-0.63}$ & $+0.23^{+0.19}_{-0.25}$ & $-0.14^{+0.19}_{-0.21}$ & $-0.04^{+0.27}_{-0.44}$ & $+0.22^{+0.15}_{-0.16}$ \\
    3448365 & $-0.10^{+0.14}_{-0.13}$ & $-0.21^{+0.11}_{-0.12}$ & $-0.03^{+0.06}_{-0.07}$ & $< -0.15$ & $+0.01^{+0.10}_{-0.11}$ & $< -0.05^{+0.23}_{-0.34}$ & $-0.57^{+0.32}_{-0.52}$ & $-0.17^{+0.25}_{-0.33}$ & $+0.09^{+0.15}_{-0.16}$ & $+0.10^{+0.25}_{-0.33}$ & $+0.30^{+0.14}_{-0.15}$ \\
    3626325 & $-0.09^{+0.16}_{-0.16}$ & $-0.21^{+0.16}_{-0.19}$ & $-0.19^{+0.08}_{-0.08}$ & $+0.07^{+0.30}_{-0.58}$ & $-0.21^{+0.16}_{-0.17}$ & $< +0.00$ & $-0.54^{+0.43}_{-0.74}$ & $+0.23^{+0.24}_{-0.33}$ & $-0.16^{+0.23}_{-0.26}$ & $-0.07^{+0.35}_{-0.72}$ & $+0.25^{+0.17}_{-0.17}$ \\
    3648936 & $-0.44^{+0.14}_{-0.16}$ & $-0.57^{+0.14}_{-0.17}$ & $-0.52^{+0.07}_{-0.07}$ & $< -0.01$ & $-0.43^{+0.13}_{-0.15}$ & $< -0.15$ & $-0.55^{+0.29}_{-0.38}$ & $+0.01^{+0.21}_{-0.27}$ & $-0.15^{+0.16}_{-0.16}$ & $-0.21^{+0.21}_{-0.28}$ & $+0.02^{+0.15}_{-0.16}$ \\
\end{longtable}
\end{landscape}
}

\section{Stellar parameters and surface abundances of the SPB stars}
The stellar parameters of the SPB stars are listed in Table~\ref{tab:spb_params} while the surface abundances can be found in Table~\ref{tab:spb_abund}.

\longtab[1]{
\small
\begin{longtable}{l|lllll|lll|l}
    \caption{Stellar parameters of the SPB stars.}
    \label{tab:spb_params} \\
    \hline\hline
    KIC & $T_{\rm{eff, NN}}$ & $\log{g}_{\rm{NN}}$ & [M/H]$_{\rm{NN}}$ & $v\sin{i}_{\rm{NN}}$ & $\xi_{\rm{NN}}$ & $T_{\rm{eff, GSSP}}$ & $\log{g}_{\rm{GSSP}}$ & $\xi_{\rm{GSSP}}$ & $\log(L/L_\sun)$ \\
    & (K) & (dex) & (dex) & (km s$^{-1}$) & (km s$^{-1}$) & (K) & (dex) & (km s$^{-1}$) &  \\
    \hline
    \endfirsthead
    \caption{continued.} \\
    \hline\hline
    KIC & $T_{\rm{eff, NN}}$ & $\log{g}_{\rm{NN}}$ & [M/H]$_{\rm{NN}}$ & $v\sin{i}_{\rm{NN}}$ & $\xi_{\rm{NN}}$ & $T_{\rm{eff, GSSP}}$ & $\log{g}_{\rm{GSSP}}$ & $\xi_{\rm{GSSP}}$ & $\log(L/L_\sun)$ \\
    & (K) & (dex) & (dex) & (km s$^{-1}$) & (km s$^{-1}$) & (K) & (dex) & (km s$^{-1}$) &  \\
    \hline
    \endhead
    \hline
    \endfoot
    3240411 & 20910$^{+645}_{-501}$ & 3.99$^{+0.12}_{-0.13}$ & -0.18 $\pm$ 0.10 & 42 $\pm$ 4 & 3.6$^{+2.1}_{-2.8}$ & 20910$^{+645}_{-501}$ & 3.99$^{+0.12}_{-0.13}$ & 3.6$^{+2.1}_{-2.8}$ & 3.43$^{+0.06}_{-0.05}$ \\
    3756031 & 15947$^{+532}_{-583}$ & 3.69$^{+0.11}_{-0.12}$ & -0.44 $\pm$ 0.18 & 32 $\pm$ 4 & $< 3.9$ & 15947$^{+532}_{-583}$ & 3.69$^{+0.11}_{-0.12}$ & $< 3.9$ & 3.12$^{+0.05}_{-0.05}$ \\
    3839930 & 15987$^{+319}_{-333}$ & 4.00$^{+0.08}_{-0.08}$ & -0.23 $\pm$ 0.15 & 33 $\pm$ 4 & $< 3.2$ & 15987$^{+319}_{-333}$ & 4.00$^{+0.08}_{-0.08}$ & $< 3.2$ & 2.71$^{+0.04}_{-0.04}$ \\
    3865742 & 20108 $\pm$ 1820 & 3.86 $\pm$ 0.26 & +0.07 $\pm$ 0.34 & 130 $\pm$ 27 & 4.2 $\pm$ 7.1 & 18798$^{+1682}_{-1584}$ & 3.58$^{+0.28}_{-0.27}$ & 2 (fixed) & 3.68$^{+0.14}_{-0.12}$ \\
    5941844 & 14025$^{+293}_{-305}$ & 4.24$^{+0.08}_{-0.07}$ & -0.13$^{+0.12}_{-0.14}$ & 30 $\pm$ 3 & $< 3.6$ & 13677$^{+237}_{-210}$ & 4.17$^{+0.07}_{-0.08}$ & $< 1.6$ & 2.34$^{+0.03}_{-0.03}$ \\
    6462033 & 18671 $\pm$ 907 & 4.15 $\pm$ 0.12 & -0.32 $\pm$ 0.37 & 75 $\pm$ 21 & 2.2 $\pm$ 4.2 & 17628$^{+951}_{-1021}$ & 3.93$^{+0.19}_{-0.20}$ & $< 5.8$ & 3.29$^{+0.06}_{-0.06}$ \\
    6780397 & 12879 $\pm$ 190 & 3.74 $\pm$ 0.08 & -0.45 $\pm$ 0.11 & 49 $\pm$ 5 & 5.0 $\pm$ 1.4 & 12906$^{+288}_{-245}$ & 3.68$^{+0.11}_{-0.11}$ & $< 2.4$ & 2.69$^{+0.04}_{-0.03}$ \\
    7760680 & 11602 $\pm$ 119 & 3.89 $\pm$ 0.05 & -0.13 $\pm$ 0.06 & 58 $\pm$ 5 & 2.6 $\pm$ 0.6 & 11791$^{+246}_{-249}$ & 3.92$^{+0.10}_{-0.10}$ & 2.0$^{+0.5}_{-0.7}$ & 2.31$^{+0.03}_{-0.03}$ \\
\end{longtable}
}

\longtab[2]{
\small
\begin{longtable}{lllllllllll}
    \caption{Surface abundances of the SPB stars with respect to solar values $\left(\rm{[X/H]} = \log_{10}\left(\frac{n_{\rm{x}}}{n_{\rm{tot}}}\right) - \log_{10}\left(\frac{n_{\rm{x}}}{n_{\rm{tot}}}\right)_{\sun}\right)$.}
    \label{tab:spb_abund} \\
    \hline\hline
    KIC & He & C & N & O & Ne & Mg & Si & S & Ca & Fe \\
    \hline
    \endfirsthead
    \caption{continued.} \\
    \hline\hline
    KIC & He & C & N & O & Ne & Mg & Si & S & Ca & Fe \\
    \hline
    \endhead
    \hline
    \endfoot
    3240411 & $+0.16^{+0.01}_{-0.01}$ & $-0.17^{+0.10}_{-0.11}$ & $-0.08^{+0.12}_{-0.13}$ & $+0.00^{+0.13}_{-0.14}$ & $< +1.37$ & $+0.15^{+0.14}_{-0.15}$ & $-0.45^{+0.14}_{-0.15}$ & $-0.46^{+0.12}_{-0.13}$ & $< +0.99$ & $-0.29^{+0.24}_{-0.36}$ \\
    3756031 & $+0.19^{+0.01}_{-0.01}$ & $+0.10^{+0.12}_{-0.14}$ & $-0.11^{+0.30}_{-0.50}$ & $-0.05^{+0.33}_{-0.53}$ & $< +1.42$ & $+0.01^{+0.16}_{-0.17}$ & $-0.59^{+0.16}_{-0.17}$ & $-0.45^{+0.14}_{-0.15}$ & $< +0.51$ & $-0.77^{+0.28}_{-0.46}$ \\
    3839930 & $+0.21^{+0.01}_{-0.01}$ & $+0.44^{+0.09}_{-0.09}$ & $-0.01^{+0.24}_{-0.35}$ & $-0.06^{+0.25}_{-0.38}$ & $< +1.56$ & $+0.32^{+0.13}_{-0.13}$ & $-0.38^{+0.10}_{-0.10}$ & $-0.26^{+0.13}_{-0.15}$ & $< +0.80$ & $-0.57^{+0.23}_{-0.31}$ \\
    3865742 & $+0.08^{+0.02}_{-0.02}$ & $+0.11^{+0.22}_{-0.25}$ & $-0.11^{+0.43}_{-0.82}$ & $-0.17^{+0.46}_{-0.81}$ & $< +2.44$ & $+0.25^{+0.35}_{-0.46}$ & $-0.47^{+0.47}_{-0.71}$ & $-0.42^{+0.34}_{-0.52}$ & $< +1.59$ & $< +0.19$ \\
    5941844 & $+0.14^{+0.01}_{-0.01}$ & $+0.31^{+0.15}_{-0.17}$ & $+0.50^{+0.38}_{-0.98}$ & $< -0.66$ & $< +1.45$ & $+0.13^{+0.11}_{-0.11}$ & $+0.05^{+0.09}_{-0.09}$ & $-0.24^{+0.17}_{-0.20}$ & $< +0.67$ & $-0.26^{+0.09}_{-0.09}$ \\
    6462033 & $+0.14^{+0.02}_{-0.02}$ & $+0.10^{+0.19}_{-0.22}$ & $< +0.25$ & $< +0.22$ & $< +1.84$ & $-0.01^{+0.25}_{-0.30}$ & $-0.72^{+0.34}_{-0.44}$ & $-0.68^{+0.27}_{-0.41}$ & $< +1.11$ & $< -0.04$ \\
    6780397 & $+0.17^{+0.01}_{-0.01}$ & $+0.20^{+0.19}_{-0.21}$ & $< +0.78$ & $< +0.27$ & $< +1.32$ & $+0.04^{+0.11}_{-0.12}$ & $-0.11^{+0.11}_{-0.12}$ & $-0.32^{+0.19}_{-0.25}$ & $+0.33^{+0.31}_{-0.81}$ & $-0.46^{+0.14}_{-0.16}$ \\
    7760680 & $+0.11^{+0.01}_{-0.01}$ & $+0.47^{+0.22}_{-0.25}$ & $< +1.43$ & $< +0.29$ & $< +1.39$ & $+0.31^{+0.10}_{-0.11}$ & $+0.12^{+0.09}_{-0.10}$ & $-0.24^{+0.27}_{-0.44}$ & $+0.33^{+0.24}_{-0.54}$ & $-0.17^{+0.10}_{-0.11}$ \\
\end{longtable}
}

\end{appendix}
\end{document}